\documentclass[pra,twocolumn,preprintnumbers,amsmath,amssymb,floatfix,longbibliography,superscriptaddress]{revtex4-2}
\usepackage{graphicx}
\usepackage{graphics}
\usepackage{psfrag}
\usepackage{hyperref}
\usepackage{multirow}
\usepackage[sort&compress]{natbib}
\usepackage[export]{adjustbox}
\usepackage{stackrel}
\usepackage{amssymb}
\usepackage{amsmath}
\usepackage{amsthm}
\usepackage{bbold}
\usepackage{bbm}
\usepackage{enumerate}
\usepackage{textcomp}

\newcommand{\vphi}{\varphi}

\newcommand{\rmi}{{\rm i}}

\usepackage{xcolor}

\begin{document}

\hypersetup{pdftitle={title}}
\title{Selberg trace formula in hyperbolic band theory}

\author{Adil Attar}
\email{attar@ualberta.ca}
\affiliation{Department of Physics, University of Alberta, Edmonton, Alberta T6G 2E1, Canada}

\author{Igor Boettcher}
\email{iboettch@ualberta.ca}
\affiliation{Department of Physics, University of Alberta, Edmonton, Alberta T6G 2E1, Canada}
\affiliation{Theoretical Physics Institute, University of Alberta, Edmonton, Alberta T6G 2E1, Canada}

\begin{abstract}
We apply Selberg's trace formula to solve problems in hyperbolic band theory, a recently developed extension of Bloch theory to model band structures on experimentally realized hyperbolic lattices. For this purpose we incorporate the higher-dimensional crystal momentum into the trace formula and evaluate the summation for periodic orbits on the Bolza surface of genus two. We apply the technique to compute partition functions on the Bolza surface and propose an approximate relation between the lowest bands on the Bolza surface and on the $\{8,3\}$ hyperbolic lattice. We discuss the role of automorphism symmetry and its manifestation in the trace formula.
\end{abstract}

\maketitle

\section{Introduction}

Experimental realizations of hyperbolic lattices in both circuit quantum electrodynamics \cite{kollar2019hyperbolic} and topoelectric circuits \cite{lenggenhager2021electric} recently resurged interest in the mathematical properties of hyperbolic space and physical systems in it \cite{BookMagnus,BookCoxeter,BookNumber,Cannon}. A current and experimentally motivated focus of attention is on properties such as band structures \cite{PhysRevX.10.011009,kollar2019line,PhysRevLett.125.053901,PhysRevA.102.032208,kollar2021gap,PhysRevD.102.034511,maciejko2020hyperbolic,maciejko2021automorphic}
and interacting quantum systems \cite{pastawski2015holographic,PhysRevD.103.094507,PhysRevA.103.033703,ZHANG20211967,jahn2021holographic,Zhu2021,boettcher2021crystallography,PhysRevLett.128.013601,Pappalardi,Ikeda,PhysRevResearch.3.L022022,Morice,Malen,Ludewig2021,Lv,Saa,stegmaier2021universality}. 
Historically, however, hyperbolic space served as a crucial platform to investigate theories of both classical and quantum chaos, because key chaotic properties such as ergodicity can be proven mathematically for geodesic flow on hyperbolic surfaces \cite{Hadamard,Mautner,BookAnosov,Bun1,Bun2,SteinerDESY}. One of the most well-studied systems is the Hadamard--Gutzwiller model describing chaotic motion on the Bolza surface \cite{BALAZS1986109,PhysRevLett.61.483,AURICH1988451,AURICH1989169,AURICH199191,PhysRevLett.68.1629,ninnemann,Braun2002}, which is a hyperbolic surface of constant negative curvature and genus two.

In the study of quantum chaos on hyperbolic surfaces, trace formulas play a central role \cite{Bolte1993,BOGOMOLNY1997219,BogAri}. They relate sums over eigenvalues of the quantum Hamiltonian to sums over classical periodic orbits on the surface. While such trace formulas are typically valid semi-classically only, they become exact for hyperbolic surfaces, where they resemble Selberg's celebrated trace formula \cite{selberg,BookAuto,marklof2012ii,BookGrosche}. The latter is pivotal to mathematical fields such as algebraic geometry or number theory. Previous applications of the Selberg trace formula in a physics context, besides investigations of the spectral statistics of quantum chaotic systems, include regularization techniques in quantum field theory or cosmology and partition functions in string theory \cite{PhysRevLett.71.3786,BYTSENKO19961,BookGrosche}. In this work, we show that the trace formula also leads to new insights into the physics of hyperbolic band theory.

In order to define the concept of a hyperbolic lattice, we first define a $\{p,q\}$ lattice, with $p,q\geq 3$ integers, as a lattice constructed from regular $p$-gonal faces with coordination number $q$ for each vertex. The Euclidean square, triangular, and hexagonal lattices correspond to $\{4,4\}$, $\{3,6\}$, and $\{6,3\}$, respectively. These are tessellations of the Euclidean plane. In contrast, for all $\{p,q\}$ such that $(p-2)(q-2)>4$, the ensuing graph is a tessellation of the hyperbolic plane $\mathbb{D}$. These cases will be called hyperbolic lattices. Here we employ the Poincar\'{e} disk model of the hyperbolic plane, whose properties are summarized in App. \ref{AppHyp}. In the following, we summarize both the experimental and mathematical context that motivate the application of the Selberg trace formula to hyperbolic lattices and hyperbolic band theory.

\subsection{Experimental context}

Hyperbolic lattices recently gained relevance in experiments in circuit quantum electrodynamics \cite{kollar2019hyperbolic} and topoelectrical circuits \cite{lenggenhager2021electric}, where the system Hamiltonian is described by a tight-binding Hamiltonian on such lattices. In the former case, for instance, the lattice is implemented through waveguide resonators, with photon propagation on such an arrangement described by the tight-binding Hamiltonian
\begin{align}
 \label{new1} \mathcal{H}_{\rm TB} = -J\sum_{\langle i,j\rangle} \Bigl(\hat{\psi}^\dagger(z_i) \hat{\psi}(z_j) + \text{h.c.}\Bigr)
\end{align}
in second-quantized form. Here $i$ labels the lattice sites with coordinates $z_i\in\mathbb{D}$, $J$ is the hopping amplitude, and $\hat{\psi}^\dagger(z_i)$ is the creation operator for a particle on site $z_i$. The sum extends over all pairs of nearest-neighboring sites. In typical experiments, the number of lattice sites is of order 100. Defining the adjacency matrix $(\mathcal{A}_{ij})$ such that $\mathcal{A}_{ij}=1$ if sites $i,j$ are nearest neighbors, and $\mathcal{A}_{ij}=0$ otherwise, the system Hamiltonian becomes
\begin{align}
 \label{new2} \mathcal{H}_{\rm TB} = -J\sum_{i,j} \mathcal{A}_{ij} \hat{\psi}^\dagger(z_i) \hat{\psi}(z_j).
\end{align}
In topoelectrical circuit networks, $(\mathcal{A}_{ij})$ is realized through the circuit Laplacian, with particles represented by local electric currents.

Finding the eigenenergies and corresponding states of $\mathcal{H}_{\rm TB}$, which is quadratic in the field operators, is equivalent to solving the eigenvalue problem for the adjacency matrix,
\begin{align}
\label{new3} \sum_j \mathcal{A}_{ij} \psi(z_j) = -\mathcal{E} \psi(z_i).
\end{align}
Note that only a finite number of terms on the left-hand side of this equation are nonzero. In experimental situations, the eigenvalues $\mathcal{E}$ will strongly depend on the boundary of the finite graph that is used to model the hyperbolic lattice. (See Ref. \cite{AlbertaWu2021} for a discussion and numerical study of the role of boundary effects on the spectrum.)  Although impossible to implement in experiment, we can assume in theory that the hyperbolic lattice extends infinitely, and ask for all solutions of this eigenvalue problem. This seemingly simple problem, however, remains unsolved for any hyperbolic $\{p,q\}$ lattice so far.

The main reason for our inability to solve the lattice eigenvalue problem in Eq. (\ref{new3}) is that Bloch's theorem in its usual form is not applicable here, as translations---defined below as generators of the Bravais lattice---do not commute on hyperbolic lattices. Instead, a generalized automorphic Bloch theorem needs to be applied \cite{maciejko2021automorphic,RayanHyp}, which contains currently insufficiently understood higher-dimensional representations of the Fuchsian translation group.

One conceptual framework to deal with energy bands in hyperbolic lattices, called hyperbolic band theory, has been developed in Ref. \cite{maciejko2020hyperbolic}. It predicts that eigenfunctions that result from Eq. (\ref{new3}) on the infinite lattice are parameterized by higher-dimensional generalized Brillouin zones. The simplest incarnation are one-dimensional Bloch waves $\psi_{\textbf{k}}(z_i)$ with energy $\mathcal{E}(\textbf{k})$, where $\textbf{k}=(k_1,\dots,k_{2g})$ is a crystal momentum with $2g$ components. Crucially, $g>1$ for hyperbolic lattices, so that momentum space and position space do not have the same dimension.

Through a continuum approximation of Eq. (\ref{new3}), described further in Sec. \ref{SecBand}, the energies $\mathcal{E}(\textbf{k})$ can be mapped to the mathematically well-defined energy spectrum of the Laplacian $\Delta_g$ on a closed hyperbolic surface of genus $g>1$, denoted $E(\textbf{k})$. The Selberg trace formula is a nonperturbative formula to compute sums over all such eigenvalues $E(\textbf{k})$ and in turn, mapping back to $\mathcal{E}(\textbf{k})$, yields predictions about observables on hyperbolic lattices, which are often obtained from functions of the spectrum. While previous application of the Selberg trace formula in quantum chaos only required to consider the case $\textbf{k}=0$, the new application motivated here naturally incorporates the crystal momentum $\textbf{k}\neq 0$. After introducing the corresponding mathematical apparatus and several other applications, we will return to the problem of energy bands on hyperbolic lattices in Sec. \ref{SecBand}.

\subsection{Mathematical context}

 The connection between the hyperbolic band theory developed in Ref. \cite{maciejko2020hyperbolic} and hyperbolic lattices is facilitated by the theory of hyperbolic crystallography \cite{boettcher2021crystallography}. It implies that a large number of hyperbolic lattices can be divided into a unit cell and a hyperbolic Bravais lattice. The simplest hyperbolic Bravais lattice is the $\{8,8\}$ lattice, which is the Bravais lattice of, for instance, the $\{8,3\}$ or $\{8,4\}$ lattices. It is part of a family of self-dual $\{4g,4g\}$ Bravais lattices, whose first member is the square lattice for $g=1$. A single face of these lattices, which is the fundamental domain for the corresponding tessellation of the hyperbolic plane, yields a closed Riemann surface of genus $g$ upon identification of opposite edges  \cite{BookMagnus,BookCoxeter,Edmonds,sausset2007periodic,maciejko2020hyperbolic,boettcher2021crystallography}. For $g>1$, these surfaces are hyperbolic.  In particular, upon imposing such periodic boundary conditions for the fundamental square or octagon of the $\{4,4\}$ or $\{8,8\}$ lattices, we obtain the genus-one torus or genus-two Bolza surface, respectively.

The merit of identifying the Bravais lattice for a given lattice is that the spectrum of the tight-binding model $\mathcal{H}_{\rm TB}$ (or, equivalently, the eigenvalues of its adjacency matrix in Eq. (\ref{new3})) can be classified according to the irreducible representations of the translation group generating the Bravais lattice. This typically simplifies the computational effort immensely. In the Euclidean case, for the $\{4,4\}$ Bravais lattice, say, the Bravais lattice is generated by the Abelian group $\Gamma_1 \simeq \mathbb{Z}^2$ and all irreducible representations are one-dimensional and labelled by the two-dimensional crystal momentum. For fixed crystal momentum, typically a finite number of bands exists. In the hyperbolic cases with $g>1$, the Fuchsian group $\Gamma_g$ generating the Bravais lattice is non-Abelian and the unitary irreducible representations are group homomorphisms
\begin{align}
 \label{intro1} \chi:\ \Gamma_g \to \text{U}(N)
\end{align}
with $N\geq 1$. Automorphic wave functions in these representations transform according to
\begin{align}
 \label{intro2} \psi(\gamma z) = \chi(\gamma) \psi(z),
\end{align}
where $z$ is a site on the lattice and $\gamma\in\Gamma_g$. Importantly, higher-dimensional representations with $N>1$ may occur. Although it would be exciting to study implications of these higher-dimensional representations, the natural first step is to ignore their impact and focus instead on the contributions from one-dimensional ($N=1$) representations of $\Gamma_g$. The corresponding Bloch wave functions $\psi_{\textbf{k}}(z)$ are labelled by a crystal momentum $\textbf{k}=(k_1,\dots,k_{2g})$ with $2g$ components.

In most parts of this work, for concreteness, we restrict our analysis to hyperbolic band theory on the Bolza surface. For one, this model has been covered in the original reference \cite{maciejko2020hyperbolic} and it constitutes the simplest non-Euclidean extension of the usual Bloch wave theory with only a modest number of additional parameters. Furthermore, it is by far the most well-studied model in the context of applications of the Selberg trace formula to Physics and so allows us to make many remarkable connections to previous work. It should be noted that in this case the non-Abelian group $\Gamma_g$ is an arithmetic Fuchsian group, which implies several unique features and simplifies some of the algorithms used. We illuminate this aspect in App. \ref{AppFuchs} but refer to the literature for a detailed exposure \cite{Bolte1993,BOGOMOLNY1997219,BogAri,BENEDITO20161902}.

\begin{figure}[t]
\centering
\includegraphics[width=8.5cm]{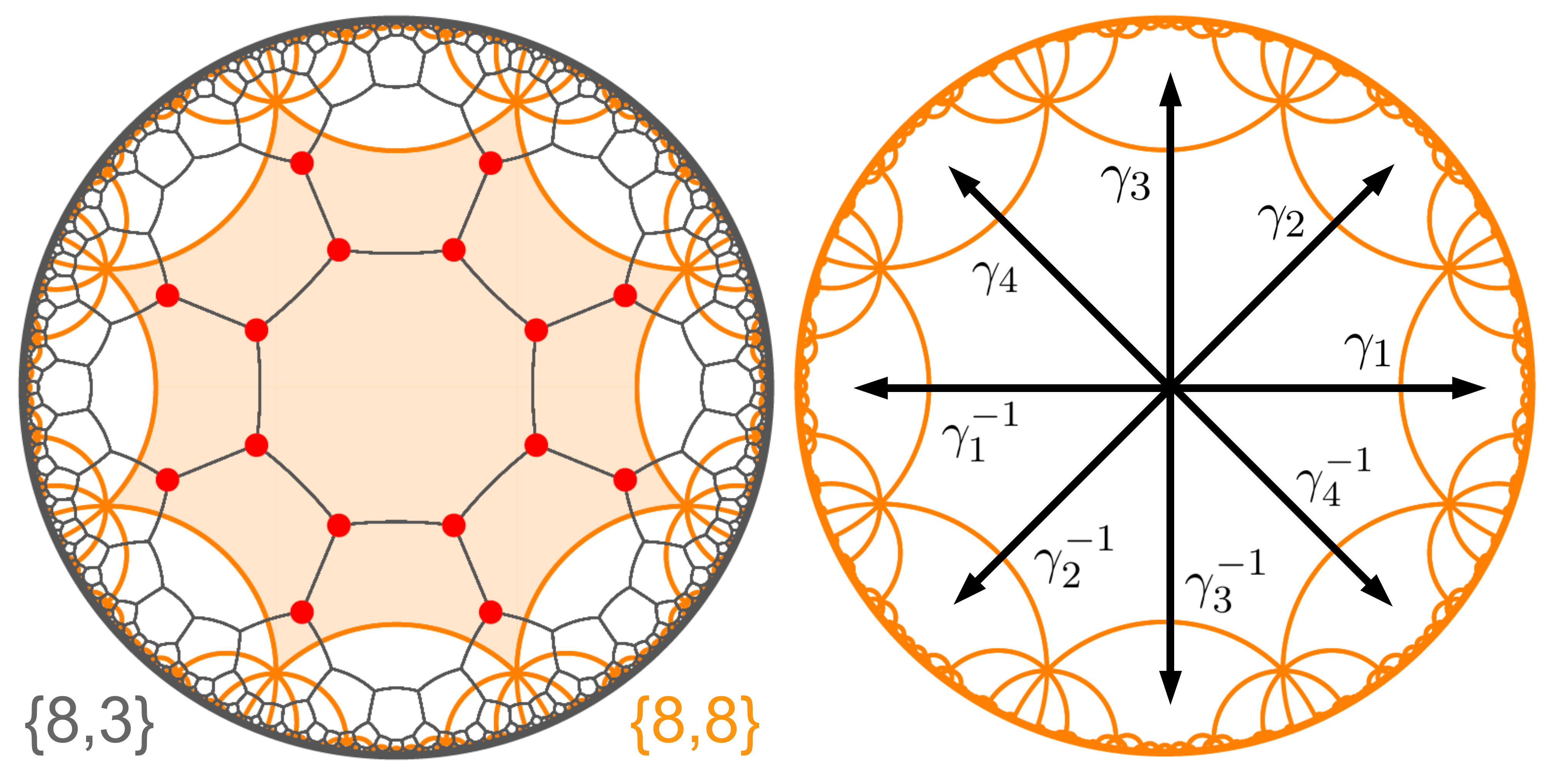}
\caption{The $\{8,8\}$ lattice is a hyperbolic Bravais lattice underlying several $\{p,q\}$ lattices. \emph{Left.} We show the $\{8,3\}$ lattice (gray) with the $\{8,8\}$ Bravais lattice superimposed (orange). The corresponding unit cell of the $\{8,3\}$ lattice has 16 sites (red dots). The fundamental domain $\mathcal{D}$ of the Bolza surface is the central octagon of the $\{8,8\}$ lattice, highlighted in orange shade. \emph{Right.} The Bravais lattice is generated by four generators $\gamma_{1,\dots,4}$ and their inverses through the non-Abelian Fuchsian group $\Gamma$ in Eq. (\ref{intro3}). Note that the $\{8,8\}$ lattice is self-dual, because by placing a vertex into the center of each face of the $\{8,8\}$ lattice and connecting nearest neighbors we obtain another $\{8,8\}$ lattice.}
\label{FigBolza}
\end{figure}

\begin{figure}[t]
\centering
\includegraphics[width=8cm]{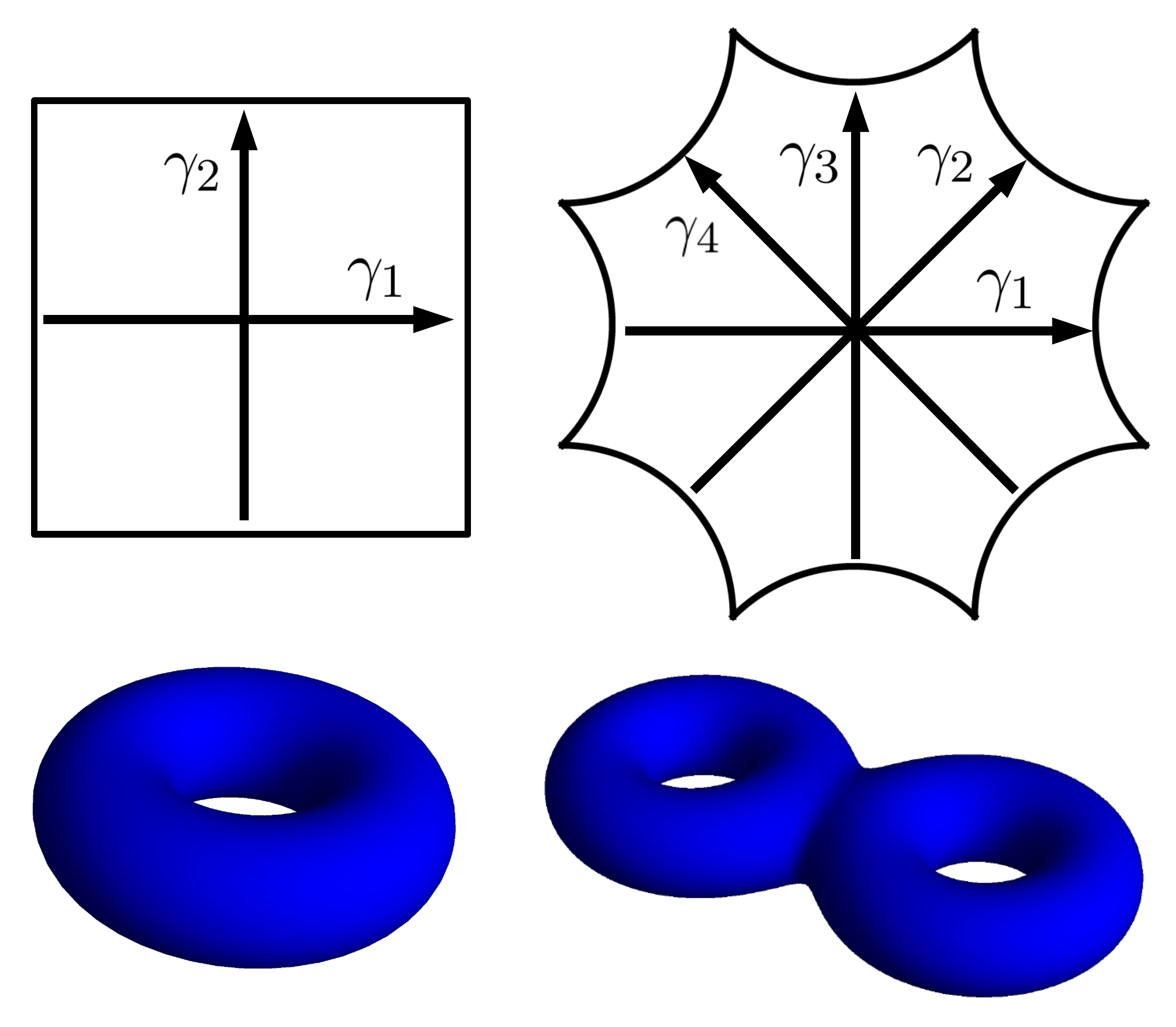}
\caption{\emph{Left.} By identifying opposite sides of a rectangle we obtain a torus of genus one, which is a closed Euclidean manifold. The mappings $\gamma_1$ and $\gamma_2$ that implement the side-pairings are the generators of the $\{4,4\}$ Bravais lattice. Of course, they are simply Euclidean translations in two non-parallel directions. \emph{Right.} By identifying opposite sides of an octagon, we obtain the Bolza surface, a closed hyperbolic manifold of genus two. The four transformations $\gamma_1,\dots,\gamma_4$ that implement the side-pairings are the same as in Fig. \ref{FigBolza} and thus the generators of the $\{8,8\}$ Bravais lattice. }
\label{FigSurfaces}
\end{figure}

The setting for the Bolza surface is depicted in Figs. \ref{FigBolza} and \ref{FigSurfaces}. The fundamental domain $\mathcal{D}$ is the central octagon of the $\{8,8\}$ lattice. We abbreviate the Fuchsian group by $\Gamma:=\Gamma_2$. We have
\begin{align}
 \label{intro3} \Gamma =\langle \gamma_1,\gamma_2,\gamma_3,\gamma_4 | \gamma_1\gamma_2^{-1}\gamma_3\gamma_4^{-1}\gamma_1^{-1}\gamma_2\gamma_3^{-1}\gamma_4=1\rangle.
\end{align}
Every element $\gamma \in\Gamma$ is a product of the four generators $\gamma_{1,2,3,4}$ and their inverses. The transformations $\{\gamma_\mu\}$ generate the Bravais lattice or, alternatively, provide side pairings between opposite sides of $\mathcal{D}$. The Bolza surface is the resulting closed hyperbolic surface of genus $g=2$ and area 
\begin{align}
 \label{intro4} A=4\pi(g-1)=4\pi
\end{align}
in units of the squared curvature radius. Hyperbolic band theory assigns a crystal momentum to each generator according to
\begin{align}
 \label{intro5} \chi_{\textbf{k}}(\gamma_\mu) = \chi_{\textbf{k}}(\gamma_\mu^{-1})^* = e^{\rmi k_\mu}
\end{align}
for $\mu=1,\dots,4$ in Eqs. (\ref{intro1}), (\ref{intro2}). These boundary conditions lead to an infinite, discrete spectrum $\{E_\lambda(\textbf{k})\}$ of eigenvalues of the Laplacian on the Bolza surface. The properties of the latter are studied here with the Selberg trace formula. We note that while for a hyperbolic lattice with $N_{\rm unit}$ sites in the unit cell there are $N_{\rm unit}$ bands for every $\textbf{k}$, the fact that the domain $\mathcal{D}\subset\mathbb{D}$ contains infinitely many points implies that the spectrum $\{E_\lambda(\textbf{k})\}$ is unbounded from above for each $\textbf{k}$.

\subsection{Structure of this work}

This work is organized as follows. In Sec. \ref{SecTrace} we review the Euclidean trace formula and introduce the Selberg trace formula for nonvanishing crystal momentum. In Sec. \ref{SecPart} we turn to a few applications of the so-obtained trace formula in the context of partition functions; in particular, we discuss the ground state energy of the Laplacian, Weyl's law, and the empty lattice approximation on the Bolza surface. In Sec. \ref{SecSymm} we discuss how the automorphism symmetry of the Bolza surface is reflected in the trace formula and how this can be used to organize the enormous amount of data contained in the orbit length spectrum. In Sec. \ref{SecBand} we develop a continuum theory for the lowest band of the adjacency matrix on the $\{8,3\}$ lattice and compute the associated energy band with the Selberg trace formula. We give a summary and outlook in Sec. \ref{SecOut}. The Apps. \ref{AppHyp} and \ref{AppEu} contain a review of hyperbolic geometry, the Euclidean trace formula, and several extensive tables for the Bolza surface. Appendix \ref{AppFuchs} contains a summary of the properties of the arithmetic Fuchsian group $\Gamma$ for the Bolza surface and App. \ref{AppOrb} contains a description of the algorithm to find all primitive periodic orbits.

\section{Trace formulas}\label{SecTrace}

\noindent Trace formulas for the Laplacian $-\Delta_g$ on a closed surface $\mathcal{M}$ allow us to compute expressions of the type
\begin{align}
 \label{int1} \mbox{tr} f(-\Delta_g) = \sum_{\lambda=0}^\infty f(E_\lambda),
\end{align}
where $f(E)$ is a suitable testfunction. The eigenvalues are solutions of
\begin{align}
 \label{int2} -\Delta_g \psi_\lambda(z) = E_\lambda \psi_\lambda(z)
\end{align}
for $z\in\mathcal{M}$. In the following we study trace formula for the torus and the Bolza surface.

\subsection{Warm-up: Euclidean case}

\noindent Before turning to hyperbolic surfaces, it is instructive to review our methodology for the Euclidean case. The results are simplified by the fact that Euclidean translations commute, but there are many similarities to trace formulas on hyperbolic manifolds that are worth pointing out.

The Laplacian in the Euclidean plane is given by
\begin{align}
 \Delta_g = \partial_x^2+\partial_y^2.
\end{align}
We study solutions of Eq. (\ref{int2}), where $z\in \mathcal{S}=[0,L]^2\subset\mathbb{R}^2$ is restricted to one square of the $\{4,4\}$ lattice with lattice constant $L$. Choosing periodic boundary conditions, the eigenvalues are labeled by two integers $\lambda=\textbf{s}\in\mathbb{Z}^2$ and read
\begin{align}
 \label{euc4} E_{\textbf{s}} = \textbf{p}_{\textbf{s}}^2 = \Bigl(\frac{2\pi}{L}\Bigr)^2\textbf{s}^2.
\end{align}
In the following notice the distinct use of $\textbf{p}$ and $p=|\textbf{p}|$.

Using Poisson's summation formula we obtain the trace formula
\begin{align}
 \label{euc7} \sum_{\textbf{s}\in\mathbb{Z}^2} h(p_{\textbf{s}}) = L^2\int \frac{\mbox{d}^2p}{(2\pi)^2} h(p)+ L^2\sum_{\textbf{n}\in\mathbb{Z}^2\backslash(0,0)} \tilde{h}_2(L_{\textbf{n}}),
\end{align}
with the Fourier transform of $h(p)$ in two dimensions given by
\begin{align}
 \label{euc7b} \tilde{h}_2(r) = \frac{1}{(2\pi)^2}\int_0^\infty \mbox{d}p'p'\int_0^{2\pi} \mbox{d}\vphi\ h(p') e^{\rmi p' r\cos\vphi}.
\end{align}
Equation (\ref{euc7}) is derived in Appendix \ref{AppEu}. The sum on the right-hand side extends over all periodic orbits on the torus. These are labeled by two integers $\textbf{n}=(n_1,n_2)\in\mathbb{Z}^2\backslash(0,0)$ and the length of the corresponding closed orbit is $L_{\textbf{n}}=L|\textbf{n}|$. Geometrically, they correspond to straight lines $\{\textbf{x}\}\subset \mathcal{S}$ determined by $\textbf{n}\cdot\textbf{x}=0$ that are periodically continued across the boundaries of $\mathcal{S}$ to yield a closed orbit on the torus.

There are some intriguing number-theoretic aspects to the trace formula (\ref{euc7}). The number of closed orbits of length $Ln$ for a fixed positive $n\in\mathbb{N}$, denoted $\tilde{d}_{\rm E}(n)$, is given by the number of ways of writing $n^2$ as the sum of two squares. Using this function, the last term in Eq. (\ref{euc7}) can be written as
\begin{align}
 \label{euc9} L^2\sum_{n=1}^\infty \tilde{d}_{\rm E}(n) \tilde{h}_2(Ln).
\end{align}
Since any square $n^2$ can be written in at least one way as a sum of two squares ($n^2=n^2+0$), we have $\tilde{d}_{\rm E}(n)>0$ for all positive $n$.

If $\textbf{n}$ yields a closed orbit of length $L_{\textbf{n}}$, then $m\textbf{n}$ for any integer $m>1$ yields a closed orbit of length $mL_{\textbf{n}}$, obtained by traversing the first orbit $m$ times. We may consider the first orbit to be more fundamental or \emph{primitive}. Specifically, we call a Euclidean orbit primitive if it corresponds to an $\textbf{n}$ such that the greatest common divisor of $n_1$ and $n_2$ is unity, i.e. $\text{gcd}(n_1,n_2)=1$. Denote the number of primitive periodic orbits of length $Ln$ for each $n>1$ by $d_{\rm E}(n)$. Then Eq. (\ref{euc9}) becomes
\begin{align}
 \label{euc9b} L^2\sum_{n=1}^\infty d_{\rm E}(n) \sum_{m=1}^\infty \tilde{h}_2(mLn),
\end{align}
where the second sum results from traversing the primitive orbits $m$ times. This expression bears some resemblance to the Selberg trace formula encountered in the next section.

We now consider the domain $\mathcal{S}$ as the fundamental tile of a tessellation of the Euclidean plane, each tile forming a face of the infinite $\{4,4\}$ lattice. Bloch waves are solutions to Eq. (\ref{int2}) for $\textbf{x}\in\mathcal{S}$ that satisfy the twisted boundary condition
\begin{align}
 \label{euc10} \psi_{\textbf{k}\lambda}(\textbf{x}+L\textbf{e}_i) = e^{\rmi k_i} \psi_{\textbf{k}\lambda}(\textbf{x}),
\end{align}
where $\textbf{e}_i$ is the unit vector in $i=1,2$ direction and $\textbf{k}=(k_1,k_2)\in[0,2\pi)^2$ is an external parameter (the \emph{dimensionless} two-dimensional crystal momentum). This boundary condition can also be viewed as threading the torus with nonzero Aharonov--Bohm fluxes. Through Eq. (\ref{euc10}), both eigenfunctions and eigenvalues parametrically depend on $\textbf{k}$. We have $\psi_{\textbf{s},\textbf{k}}(\textbf{x}) = L^{-1}e^{\rmi (\textbf{p}_{\textbf{s}}+\textbf{k}/L)\cdot\textbf{x}}$ for $\textbf{x}\in\mathcal{S}$ and
\begin{align}
 \label{euc11} E_{\textbf{s},\textbf{k}} &= \textbf{p}_{\textbf{s},\textbf{k}}^2 = \Bigl(\textbf{p}_{\textbf{s}}+\frac{1}{L}\textbf{k}\Bigr)^2.
\end{align}
Note how $\textbf{k}\neq 0$ lifts the degeneracy of the eigenvalues $E_{\textbf{s}}\propto \textbf{s}^2$ from Eq. (\ref{euc4}). This characteristic feature is also present in the hyperbolic case. We show in Appendix \ref{AppEu} that the trace formula now reads
\begin{align}
 \nonumber \sum_{\textbf{s}\in\mathbb{Z}^2} h(p_{\textbf{s},\textbf{k}}) &= L^2\int \frac{\mbox{d}^2p}{(2\pi)^2} h(p)\\
 \label{euc12} &+ L^2\sum_{\textbf{n}\in\mathbb{Z}^2\backslash(0,0)} e^{-\rmi \textbf{n}\cdot\textbf{k}}\ \tilde{h}_2(L_{\textbf{n}}).
\end{align}
Importantly, and also foreshadowing the hyperbolic case, the external momentum $\textbf{k}$ only appears in the sum on the right-hand side through a simple, yet characteristic phase factor for each $\textbf{n}$.

\subsection{Selberg trace formula}

\noindent We now consider the trace formula for the Laplacian on the Bolza surface. Basic elements of hyperbolic geo\-metry and or notation are summarized in App. \ref{AppHyp}. The hyperbolic Laplacian on the Poincar\'{e} disk reads
\begin{align}
 \label{hyp1} \Delta_g = \frac{1}{4}(1-|z|^2)^2 (\partial_x^2+\partial_y^2).
\end{align}
We study the eigenvalue problem (\ref{int2}) for $z=x+\rmi y\in \mathcal{D}$, where $\mathcal{D}\subset \mathbb{D}$ is one octagon of the $\{8,8\}$ tessellation. When this fundamental domain $\mathcal{D}$ is equipped with periodic boundary conditions that identify opposite edges, we obtain the Bolza surface, see Fig. \ref{FigSurfaces}.

Periodic boundary conditions are implemented by choosing four generators $\gamma_1,\dots,\gamma_4 \in \text{PSU}(1,1) $ sa\-tisfying
\begin{align}
 \label{hyp2} \gamma_1\gamma_2^{-1}\gamma_3\gamma_4^{-1}\gamma_1^{-1}\gamma_2\gamma_3^{-1}\gamma_4=1
\end{align}
in $\text{PSU}(1,1)$ and constraining solutions of Eq. (\ref{int2}) such that
\begin{align}
 \label{hyp3} \psi_\lambda(\gamma_\mu z) = \psi_\lambda(z)
\end{align}
for $\mu=1,\dots,4$ and all $z\in \mathcal{D}$. The Fuchsian group $\Gamma$ in Eq. (\ref{intro3}), which is a discrete subgroup of $\text{PSU}(1,1)$ made from all possible products of the four generators and their inverses subject to the constraint in Eq. (\ref{hyp2}), is the first homotopy group of the Bolza surface. An alternative point of view is to consider $\Gamma$ as the (non-commutative) translation group of the $\{8,8\}$ Bravais lattice. Then applying the generators $\gamma_\mu$ and their inverses maps one octagon to any of its eight neighboring octagons. Repeating this procedure, every octagon in the $\{8,8\}$ lattice can be uniquely identified with an element $\gamma\in\Gamma$ applied to an arbitrarily chosen central octagon that we identify with $\mathcal{D}$.

In the latter formulation, it appears natural to consider more general boundary conditions than Eq. (\ref{hyp3}). In hyperbolic band theory, Eq. (\ref{hyp3}) is replaced by the twisted boundary condition
\begin{align}
 \label{hyp5} \psi_{\textbf{k}\lambda}(\gamma_\mu z) = e^{\rmi k_\mu} \psi_{\textbf{k}\lambda}(z),
\end{align}
reminiscent of the transformation of Bloch waves in a Euclidean Bravais lattice, with $\textbf{k}=(k_1,k_2,k_3,k_4)\in[0,2\pi)^4$ an external parameter---the hyperbolic crystal momentum. As discussed in the introduction, the factor of automorphy $\chi_{\textbf{k}}(\gamma_\mu)=e^{\rmi k_\mu}$ constitutes a one-dimensional representation of the hyperbolic translation group $\Gamma$.

\begin{figure}[t]
\centering
\includegraphics[width=8.5cm]{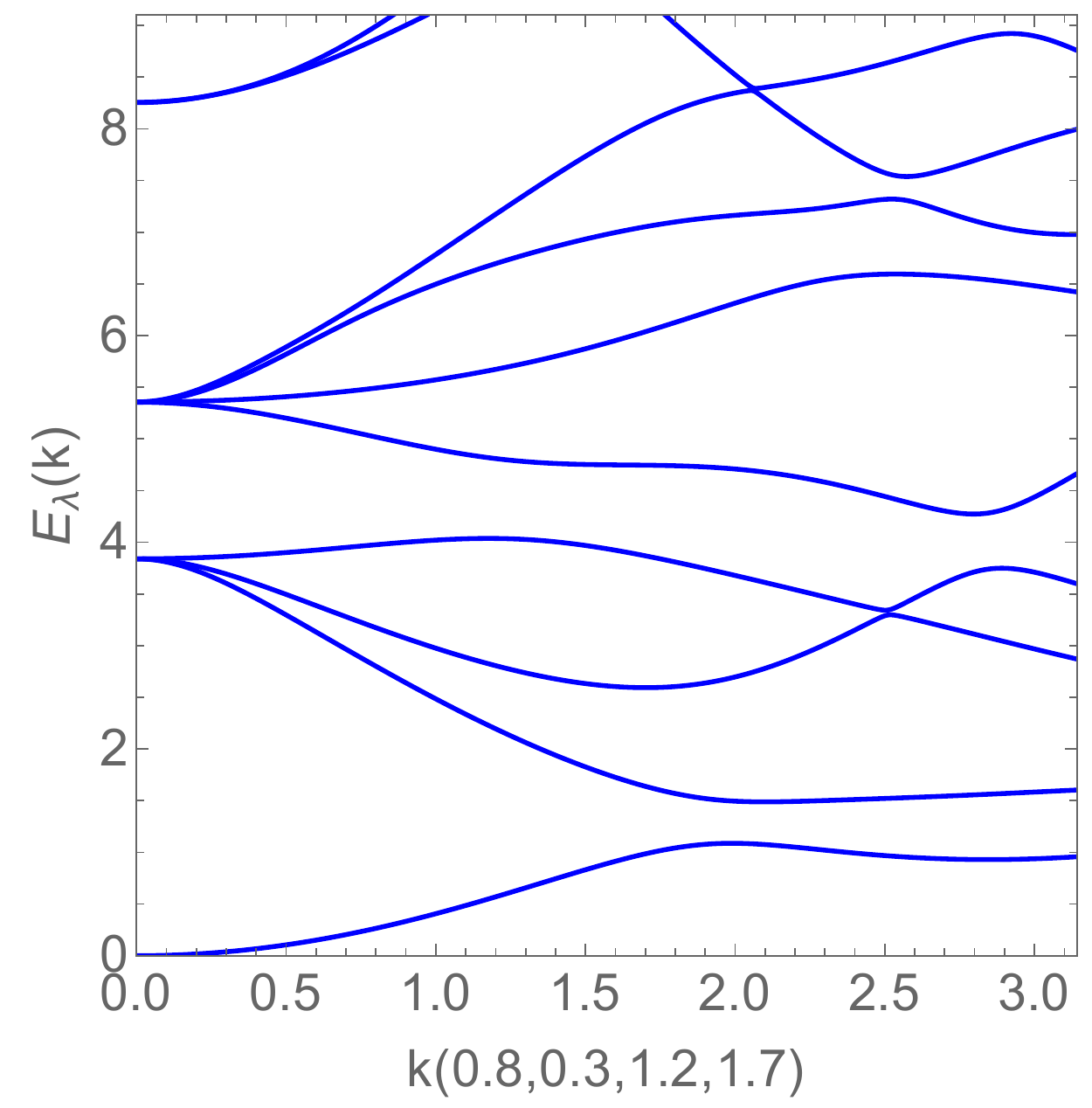}
\caption{Low-lying eigenvalues $E_\lambda(\textbf{k})$ of Eq. (\ref{int2}) on the Bolza surface with twisted boundary condition (\ref{hyp5}) along the generic line $\textbf{k}=(0.8,0.3,1.2,1.7)k$ in four-dimensional momentum space, with $k\in[0,\pi]$. We observe that some de\-generacies present for $\textbf{k}=0$ are lifted for $\textbf{k}\neq0$. Furthermore, re-orderings and crossings of energy bands occur for nonzero crystal momentum, indicating a rich band structure of the model. Data reproduced with kind permission from Ref. \cite{maciejko2020hyperbolic}.}
\label{FigBands}
\end{figure}

The eigenvalues of $-\Delta_g$ on the Bolza surface for $\textbf{k}=0$ are known to high precision \cite{Strohmeier}. The spectrum is discrete, infinite, and unbounded from above. The lowest eigenvalues is $E_0=0$, corresponding to a constant eigenfunction, the next eigenvalues are $E_1=E_2=E_3=3.839$, followed by the fourfold degenerate eigenvalue $5.354$. For $\textbf{k}\neq 0$, the eigenvalues $E_{\lambda}(\textbf{k})$ implicitly depend on the crystal momentum $\textbf{k}$ through the boundary condition in Eq. (\ref{hyp5}), see Fig. \ref{FigBands}. Not much is known about $E_{\lambda}(\textbf{k})$ on the Bolza surface, but a first study in the context of hyperbolic band theory was carried out in Ref. \cite{maciejko2020hyperbolic}, where it was observed that some degeneracies of the eigenvalues are removed for $\textbf{k}\neq 0$ and the invariance under the automorphism group of the surface was studied. Note that the eigenvalues $E_\lambda(\textbf{k})$ are real because $-\Delta_g$ is self-adjoint with respect to the canonical scalar product of functions on $\mathcal{D}$ for any value of $\textbf{k}$.

The eigenvalues of $-\Delta_g$ can be associated with the amplitude $p_\lambda$ of a two-dimensional momentum $\textbf{p}_\lambda\in\mathbb{R}^2$ through the positive root of
\begin{align}
 \label{hyp6} E_\lambda = \frac{1}{4} + p_\lambda^2,
\end{align}
in analogy to Eq. (\ref{euc4}) for the Euclidean case. At this point, it is not obvious why $p=|\textbf{p}|$ should be related to a two-dimensional momentum instead of a one-dimensional one, but this will be suggested by the way in which it appears in the trace formulas below. Of course, Eq. (\ref{hyp6}) can be generalized to $\textbf{k}\neq 0$, expressing the eigenvalues $E_\lambda(\textbf{k})$ through the function $p_\lambda(\textbf{k})=[E_\lambda(\textbf{k})-1/4]^{1/2}$.

In the following, as we did in the Euclidean case, we first discuss the trace formula for $\textbf{k}=0$ and then generalize the setup to arbitrary $\textbf{k}$. The Selberg trace formula on the Bolza surface for $\textbf{k}=0$ reads \cite{BALAZS1986109,PhysRevLett.61.483,marklof2012ii,BookGrosche}
\begin{align}
 \nonumber \sum_{\lambda=0}^\infty h(p_\lambda) = {}& A \int\frac{\mbox{d}^2p}{(2\pi)^2} \tanh(\pi p)h(p)\\
 \label{hyp7} &+\sum_{n=1}^\infty  d_0(n) \sum_{m=1}^\infty  \frac{\ell_n\tilde{h}(m\ell_n)}{2\sinh(m\ell_n/2)},
\end{align}
where $A=4\pi$ is the hyperbolic area of $\mathcal{D}$, $d_0(n)$ is the number of primitive periodic orbits of length $\ell_n$ (defined below), and 
\begin{align}
 \label{hyp8} \tilde{h}(t) &= \int_{-\infty}^\infty \frac{\mbox{d}p}{2\pi} h(p) e^{\rmi p t}
\end{align}
is the Fourier transform of $h(p)$ in one dimension. The first term on the right-hand side of Eq. (\ref{hyp7}) resembles the trace over the continuous spectrum of $-\Delta_g$ in the infinite hyperbolic plane $\mathbb{D}$ \cite{BookAuto,Helgason}. The second term constitutes a sum over all primitive periodic orbits on the Bolza surface. This structure mirrors the Euclidean trace formula in Eq. (\ref{euc7}).

\begin{figure}[t]
\centering
\includegraphics[width=8.5cm]{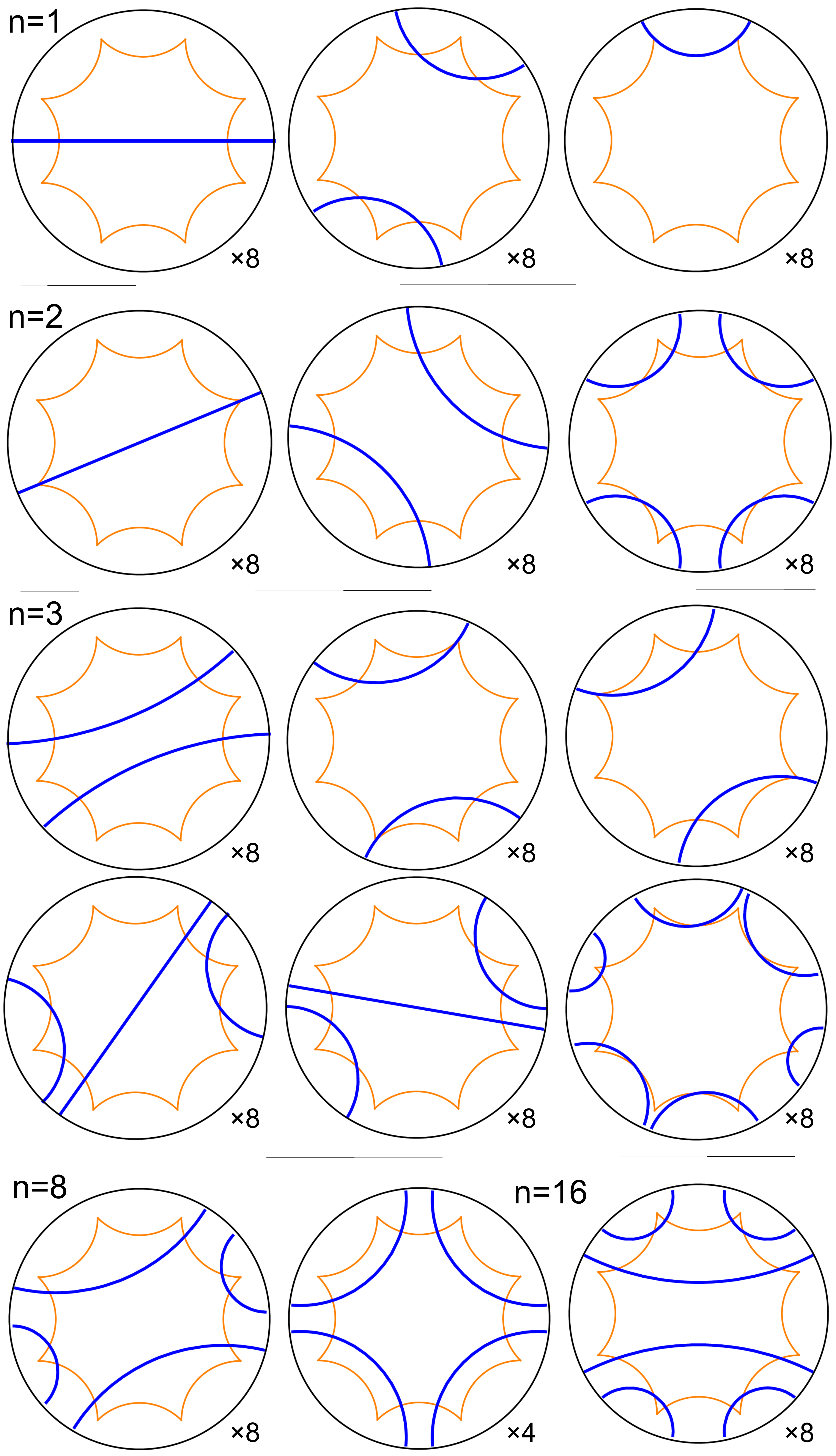}
\caption{All primitive periodic orbits on the Bolza surface of length $\ell_n$ (Eq. (\ref{hyp9})) for some values of $n$. We plot the geodesics in the full Poincar\'{e} disk for better visibility, although they actually need to be continued periodically across the boundaries of the fundamental octagon (orange) through the side-pairings defined in Fig. \ref{FigSurfaces}. The orbits are traversed once in a certain direction: if $\gamma\in\Gamma$ is representative of one direction, then $\gamma^{-1}\in\Gamma$ is representative of the opposite one. Each orbit implies a number of additional orbits obtained by $2\pi/8$-rotations about the origin, indicated at the bottom right; often this number equals 8, but high-symmetry orbits may yield smaller values (as in the case of $n=16$, left).}
\label{FigGeo}
\end{figure}

The determination of the periodic orbits on the Bolza surface or, equivalently, in the Hadamard--Gutzwiller model, has been established in Refs. \cite{BALAZS1986109,PhysRevLett.61.483,AURICH1988451,AURICH1989169,AURICH199191}. We explain the method in some detail in App. \ref{AppOrb}. Here we highlight a few facts that are relevant to understand the second term in the Selberg trace formula. First note that every element $\gamma \neq 1$ from the discrete group $\Gamma$ determines a unique geodesic in $\mathbb{D}$ that is left invariant under the action of $\gamma$. If $\delta\in \Gamma$ is another group element, then the geodesic determined by the conjugate $\delta\gamma\delta^{-1}$ is the one determined by $\gamma$, but the geodesic is shifted by $\delta$. We then only consider those geodesics that pass through the central octagon $\mathcal{D}$. Every geodesic determined by some $\gamma\in\Gamma$ that passes through $\mathcal{D}$ becomes a periodic orbit on the Bolza surface after the opposite edges of $\mathcal{D}$ have been identified and the orbit has been continued accordingly. (This non-obvious fact may be surprising on first encounter.) The closed orbit typi\-cally consists of several geodesic segments when plotted in $\mathcal{D}$, which correspond to group elements that are conjugate to each other, and which need to be counted as one orbit, see Fig. \ref{FigGeo}. All of these considerations result in a one-to-one correspondence between periodic orbits on the closed manifold and conjugacy classes of $\Gamma$, i.e. sets $[\gamma]=\{ \delta\gamma\delta^{-1},\ \delta\in\Gamma\}$. Furthermore, if $[\gamma]$ defines a periodic orbit, then $[\gamma^m]$ with integer $m>1$ defines the same orbit traversed $m$ times. The \emph{primitive} periodic orbits are those conjugacy classes $[\gamma]$, where $\gamma$ cannot be written as $\delta^m$ for some element $\delta\in\Gamma$. The set of lengths of primitive periodic orbits including their degeneracy is called the length spectrum of the surface.

The primitive periodic orbits on the Bolza surface are labeled by a positive integer $n\geq 1$. Their length, which appears in the second term of Eq. (\ref{hyp7}), reads \cite{AURICH199191}
\begin{align}
 \label{hyp9} \ell_n = 2\ \text{arcosh}[m_{\rm o}(n)+n\sqrt{2}],
\end{align}
where $m_{\rm o}(n)$ is the odd integer to best approximate $n\sqrt{2}$, see Eq. (\ref{fuchs8}). The peculiar form of this expression has a number-theoretic origin, related to the fact that $\Gamma$ is an arithmetic Fuchsian group \cite{BOGOMOLNY1997219,BogAri}. The number of primitive periodic orbits of length $\ell_n$ is denoted by $d_0(n)$. The function $d_0(n)$ for $n\leq 1500$ for the Bolza surface has been determined by Aurich, Bogomolny, Steiner \cite{AURICH199191}. With these definitions at hand, it now becomes clear that the second term in the Euclidean and hyperbolic trace formulas, Eqs. (\ref{euc9b}) and (\ref{hyp7}), are structurally very close.

The primitive closed orbits on the torus are labeled by two integers $\textbf{n}\in\mathbb{Z}^2$. For each $n\geq 1$, there are $d_{\rm E}(n)$ such vectors. A similar, yet more subtle, labeling applies to the Bolza surface. We first define the function $\chi_{\textbf{k}} : \Gamma \to \text{U}(1)$ such that $\chi_{\textbf{k}}(\gamma_\mu) = \chi_{\textbf{k}}(\gamma_\mu^{-1})^* = e^{\rmi k_\mu}$ as in Eq. (\ref{intro5}). We have $\chi_{\textbf{k}}(\gamma\gamma')=\chi_{\textbf{k}}(\gamma)\chi_{\textbf{k}}(\gamma')$. Each periodic orbit is associated to a conjugacy class $[\gamma]$ with some representative group element $\gamma$. The latter can be written as a product of generators and their inverses, schematically $\gamma=\gamma_{\nu_1}\gamma_{\nu_2}\cdots\gamma_{\nu_r}$. To this product, we associate a four-component vector $\textbf{v}=(v_1,v_2,v_3,v_4)^T\in\mathbb{Z}^4$ via
\begin{align}
 \label{hyp11} \chi_{\textbf{k}}(\gamma_{\nu_1}\gamma_{\nu_2}\cdots\gamma_{\nu_r}) = e^{\rmi \textbf{v}\cdot\textbf{k}}.
\end{align}
Hence, the component $v_1$ is the number of times $\gamma_1$ appears in the factorization minus the number of times $\gamma_1^{-1}$ appears, etc. For example,
\begin{align}
 \label{hyp12} \chi_{\textbf{k}}(\gamma_3\gamma_4^{-1}\gamma_1\gamma_3) = e^{\rmi(k_1+2k_3-k_4)}\ \Rightarrow\ \textbf{v}= \begin{pmatrix} 1 \\ 0 \\ 2 \\ -1\end{pmatrix}.
\end{align}
The $d_0(n)$ distinct primitive periodic orbits of length $\ell_n$ then give rise to $d_0(n)$ vectors $\textbf{v}$, which we collect in a set denoted $\mathcal{V}_n$.

The Selberg trace formula for $\textbf{k}\neq 0$ was discussed by Selberg in his original paper \cite{selberg} and appears in the literature in the context of nontrivial representations of the Fuchsian group $\Gamma$ \cite{BookGrosche}. However, it does not seem to have been applied in a physical context. It reads
\begin{align}
 \nonumber \sum_{\lambda=0}^\infty h(p_\lambda(\textbf{k})) = {}& A \int\frac{\mbox{d}^2p}{(2\pi)^2} \tanh(\pi p)h(p)\\
 \label{hyp13} &+\sum_{n=1}^\infty  \sum_{\textbf{v}\in\mathcal{V}_n}  \sum_{m=1}^\infty e^{\rmi m\textbf{v}\cdot\textbf{k}} \frac{\ell_n\tilde{h}(m\ell_n)}{2\sinh(m\ell_n/2)}.
\end{align}
This formula is the central tool for our analysis. For $\textbf{k}=0$ we recover Eq. (\ref{hyp7}) because $\sum_{\textbf{v}\in \mathcal{V}_n} 1 = d_0(n)$. It is striking that the $\textbf{k}$-dependence only enters through a simple phase factor in the sum over the primitive periodic orbits, just as in the Euclidean case in Eq. (\ref{euc12}). The corresponding Selberg zeta function in the presence of the factor $\chi_{\textbf{k}}(\gamma)=e^{\rmi \textbf{v}\cdot\textbf{k}}$, and the location of its zeros, have been discussed in Refs. \cite{selberg,BookGrosche}.

To conclude this section we finally specify the regularity conditions that have to be met by the function $h(p)$ entering Eq. (\ref{hyp13}). We analytically continue $h(p)$ to a function defined on the strip of height $\sigma>1/2$ about the real axis given by $\mathcal{S}_\sigma = \{ p\in \mathbb{C},\ |\text{Im}(p)|\leq \sigma\}$. We require the following conditions to be satisfied \cite{marklof2012ii}.
\begin{itemize}
 \item[(I)] $h(p)=h(-p)$ is an even function.
 \item[(II)] $h(p)$ is analytic in $S_\sigma$.
 \item[(III)] There exist $C> 0$ and $\delta>0$ such that $|h(p)|\leq \frac{C}{(1+|\text{Re}(p)|)^{2+\delta}}$ for all $p\in\mathcal{S}_\sigma$.
\end{itemize}

\section{Partition functions}\label{SecPart}

\noindent In this section, we present an applications of Eq. (\ref{hyp13}) to a problem from statistical mechanics: computing partition functions in hyperbolic band theory.

\subsection{Ground state energy}\label{SecE0}

\noindent We consider the partition function or heat kernel
\begin{align}
 \label{gs1} Z(\beta,\textbf{k}) = \sum_{\lambda=0}^\infty e^{-\beta E_\lambda(\textbf{k})}
\end{align}
for the Bolza surface, i.e. we apply the functions 
\begin{align}
 \label{gs2} h(p) &= e^{-\beta(p^2+\frac{1}{4})},\\
 \label{gs3} \tilde{h}(t) &= \frac{e^{-\beta/4}}{\sqrt{4\pi \beta}}  e^{-t^2/(4\beta)}
\end{align}
to Eq. (\ref{hyp13}). Here $\beta$ is a positive parameter that we identify with an inverse temperature. For large $\beta$, the partition function is dominated by the ground state energy $E_0(\textbf{k})$ according to
\begin{align}
 \label{gs4} Z(\beta,\textbf{k}) \stackrel{\beta\to \infty}{\sim} e^{-\beta E_0(\textbf{k})}.
\end{align}
This equation can be used to determine the value of $E_0(\textbf{k})$, i.e. the lowest eigenvalue of the solution of the differential equation (\ref{int2}) with boundary condition (\ref{hyp5}), from a set of purely geometric data determined by the hyperbolic surface.

It is instructive to study whether the relevant contribution to Eq. (\ref{gs4}) comes from the first (integral) or second (sum) part of the trace formula. For $\beta\to \infty$, the momentum integral in the first part is dominated by small momenta and we can approximate $\tanh(\pi p)\approx \pi p$, which leads to
\begin{align}
 \label{gs5} A \int\frac{\mbox{d}^2p}{(2\pi)^2} \tanh(\pi p)h(p) \sim \frac{1}{2}\Bigl(\frac{\pi}{\beta}\Bigr)^{3/2} e^{-\beta/4}.
\end{align}
This exponentially decaying contribution is unimportant for large $\beta$. Since the sum over periodic orbits also contains a factor $e^{-\beta/4}$ in $\tilde{h}(t)$, we recognize that the number of orbits that contributes to the sum needs to increase sufficiently fast to yield a nonzero contribution. How this works is most easily seen for $\textbf{k}=0$, where $E_0(\textbf{0})=0$ and
\begin{align}
 \label{gs6} \lim_{\beta \to \infty} Z(\beta,\textbf{0}) =1.
\end{align}
We use that for large $\ell$ the number of primitive orbits in the interval $[\ell,\ell+\mbox{d}\ell]$ is given by $(e^{\ell}/\ell)\mbox{d}\ell$ \cite{BogAri}. Replacing the sum over $\ell_n$ by an integral, using $\sinh(\ell/2)\sim \frac{1}{2}e^{\ell/2}$, and neglecting the sum over $m>1$ we find from Eqs. (\ref{hyp13}) and (\ref{gs5}) that
\begin{align}
 \label{gs7} Z(\beta,\textbf{0}) &\sim \frac{e^{-\beta/4}}{\sqrt{4\pi\beta}} \int_0^\infty \mbox{d}\ell\ \frac{e^{\ell}}{\ell} \frac{\ell e^{-\ell^2/(4\beta)}}{e^{\ell/2}}\sim 1
\end{align}
as expected.

\begin{figure}[t]
\centering
\includegraphics[width=7.5cm]{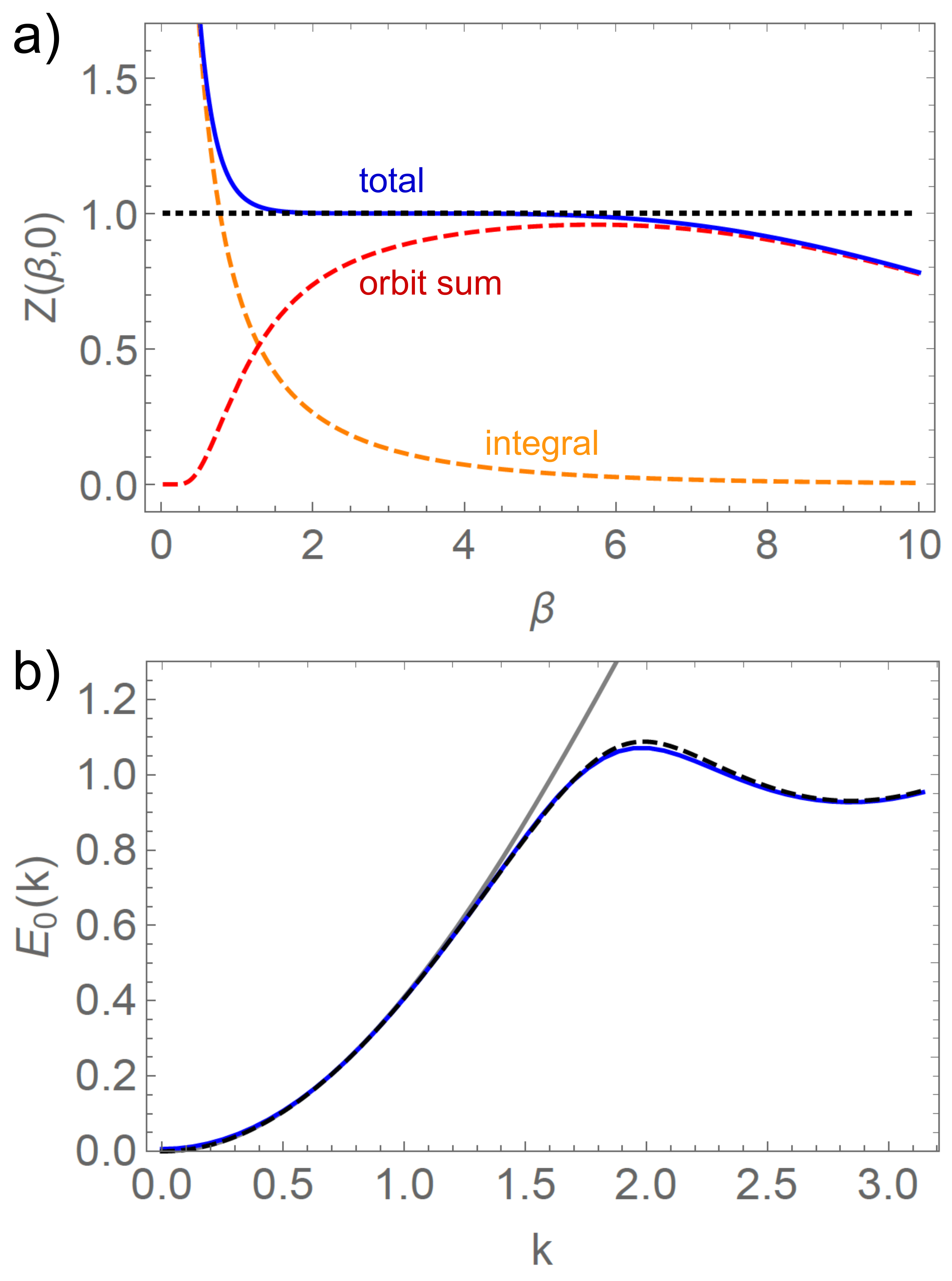}
\caption{Ground state energy $E_0(\textbf{k})$ from the Selberg trace formula (\ref{hyp13}). \emph{Panel a).} Partition function $Z(\beta,\textbf{0})$ for $\textbf{k}=0$ (blue). Since only a finite number of orbits are included ($(n,m)\leq (150,3)$ in this work), there is an optimal range of $\beta$-values from which the asymptotic limit $Z(\beta,0)\sim e^{-\beta E_0} =1$ can be estimated ($\beta \sim 3-5$ in this work). We distinguish integral (dashed orange), sum (dashed red) and total (blue) contributions, together with the limit of unity (dashed horizontal line). \emph{Panel b).} Ground state energy $E_0(\textbf{k})$ along the line $\textbf{k}=(0.8,0.3,1.2,1.7)k$ with $k\in[0,\pi]$ from the trace formula with $(n,m)\leq (150,3)$ (solid blue), compared to the exact value found from numerically solving the Schr\"{o}dinger equation (\ref{int2}) with boundary condition (\ref{hyp5}) (black dashed), see Fig. \ref{FigBands}. The faint gray curve is the quartic expansion at low momenta from Eqs. (\ref{gs12})-(\ref{gs14}).}
\label{FigGS}
\end{figure}

In practice, having only finite information about the length spectrum of the Bolza surface, we terminate the orbit sum at a finite value of $n$. This implies that the truncated sum necessarily vanishes as $\beta\to \infty$. In order to still utilize Eq. (\ref{gs4}) when working with the truncated sum, we need to evaluate it at an optimal value $\beta<\infty$. In this work, we limit the trace formula to terms with $n\leq 150$ and, accordingly, $m\leq 3$, so that $Z(\beta_0,0) \simeq1$ for $\beta_0 \simeq 3$, see Fig. \ref{FigGS}. For an unbiased determination of $E_0(\textbf{k})$ for $\textbf{k}\neq 0$, we define the function
\begin{align}
 \label{gs8} \bar{E}_0(\beta,\textbf{k}) = -\frac{1}{\beta}\log Z(\beta,\textbf{k})
\end{align}
and need to find a suitable criterion to choose the optimal $\beta_0$ with $E_0(\textbf{k})=\bar{E}_0(\beta_0,\textbf{k})$. A sound criterion is $\frac{d}{d\beta}\bar{E}_0(\beta,\textbf{k})=0$, which, if $\bar{E}_0$ has a maximum, implies $E_0(\textbf{k})=\text{max}_\beta \bar{E}_0(\beta,\textbf{k})$. In general, the criterion to find the optimal $\beta_0$ may depend on the particular truncation used for the trace formula. Determining this criterion in general would thus be a problem of asymptotic analysis much beyond the ambition of the present work.

In this work, we determine the ground state energy $E_0(\textbf{k})$ from the trace formula along certain lines in $\textbf{k}$-space. We maximize $\bar{E}_0(\beta,\textbf{k})$ over the interval $\beta\in[3,7]$, which works in our particular case with $n\leq 150$, because typically $\bar{E}_0(\beta,\textbf{k})$ initially increases with $\beta$ and then rapidly decreases for larger $\beta\sim 7$, implying that the best estimate is reached at the maximum. For small $|\textbf{k}|$, the dependence on $\beta\gtrsim 3$ is small. We compare our findings to numerical solutions of the Schr\"{o}dinger equation (\ref{int2}) with the boundary condition (\ref{hyp5}) provided by the authors of Ref. \cite{maciejko2020hyperbolic}. We find good agreement, while largest deviations appear when $E_0(\textbf{k})>0$ is largest. In Fig. \ref{FigGS} we compare both determinations of $E_0(\textbf{k})$ along the generic line $\textbf{k}=(0.8,0.3,1.2,1.7)k$ with $k\in[0,\pi]$. We also compared and found reasonable agreement within the truncation chosen along the lines $\textbf{k}=k(1,1,1,1)$ and $\textbf{k}(1,1,-1,1)$ that are discussed in Sec. \ref{SecBand}. This leads us to suspect that the true value of $E_0(\textbf{k})$ is recovered when including all (infinitely many) primitive periodic orbits in the trace formula.

For nonvanishing crystal momentum $\textbf{k}\neq 0$ we have $E_0(\textbf{k})>0$, because a constant wave function cannot solve the Schr\"{o}dinger equation subject to the boundary condition (\ref{hyp5}). Although a closed formula for $E_0(\textbf{k})$ is absent, an expansion for small $\textbf{k}$ is possible. For this purpose, we introduce the vector $\textbf{K}$ via
\begin{align}
 \label{gs10} \textbf{K} = \begin{pmatrix} k_1+k_4 \\ k_2 -k_1 \\ k_3-k_2\\ k_4-k_3 \end{pmatrix}.
\end{align}
The merit of introducing the variable $\textbf{K}$ is that several formulas involving small $\textbf{k}$ simplify considerably. This is ultimately rooted in the automorphism symmetry of the energy spectrum, discussed in Sec. \ref{SecSymm}, but at this point may be viewed as a mere elegant rewriting.

\begin{figure}[t]
\centering
\includegraphics[width=8.5cm]{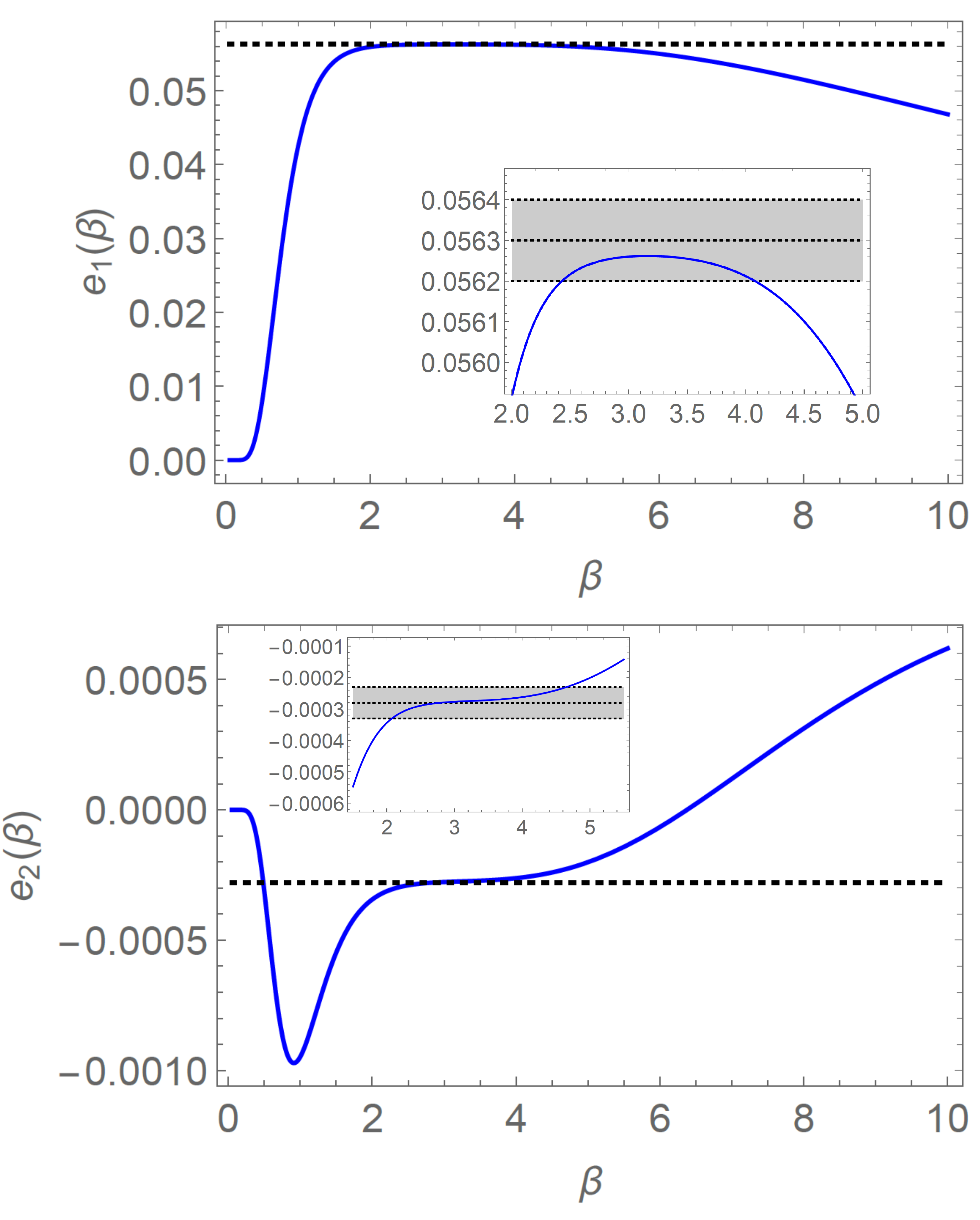}
\caption{The coefficients in the $\textbf{k}$-expansion of $E_0(\textbf{k})$ in Eq. (\ref{gs12}) can be obtained from $Z(\beta,\textbf{k})$ for large values of $\beta$. From the truncated partition function with $(n,m)\leq (150,3)$, we estimate the asymptotic values of $e_1$ and $e_2$ in the range of $\beta$-values where the function has a local plateau, assuming this plateau would extend to $\beta\to \infty$ if all orbits were included. The blue curves are the $\beta$-dependent coefficients and the dashed horizontal lines correspond to the best estimates from Eqs. (\ref{gs13}) and (\ref{gs14}). The insets show the plateau regions together with the estimated errors of $e_1$ and $e_2$ in these equations.}
\label{FigExp}
\end{figure}

Equation \ref{gs10} is an invertible linear transformation, hence $\textbf{k}=0$ if and only if $\textbf{K}=0$. Equations (\ref{inv11}) and (\ref{hur7}) then imply
\begin{align}
 \nonumber \sum_{\textbf{v}\in\mathcal{V}_n}e^{\rmi m \textbf{v}\cdot\textbf{k}} ={}& d_0(n) -3m^2 d_1(n) \textbf{K}^2 \\
  \label{gs10b} &+ \frac{m^4}{8}d_2(n) \textbf{K}^4 + \mathcal{O}(\textbf{k}^6)
\end{align}
with integer coefficients $d_{0,1,2}(n)$ listed in Tab. \ref{TabdTable}. Inserting this expression into the trace formula for the partition function then yields an expansion of $Z(\beta,\textbf{k})$ and $\bar{E}_0(\beta,\textbf{k})$ in powers $\textbf{K}^2$ and $\textbf{K}^4$. We determine the coefficients in these expansions from the truncated trace formula  with $n\leq 150$ and $m\leq 3$ at values of $\beta$ where the functions show a local plateau. These $\beta$-values happen to be around $\beta\sim 3-6$ in accordance with the previous analysis. We also use the variation across the plateau to estimate the error, see Fig. \ref{FigExp}. We obtain
\begin{align}
 \label{gs12} E_0(\textbf{k}) = e_1 \textbf{K}^2 + e_2 \textbf{K}^4+\mathcal{O}(\textbf{k}^6)
\end{align}
with
\begin{align}
 \label{gs13} e_1 &= 0.0563(1),\\
 \label{gs14} e_2 &=-0.00028(5).
\end{align}
Using the data for $E_0(\textbf{k})$ from Ref. \cite{maciejko2020hyperbolic} with $\textbf{k}=(0.8,0.3,1.2,1.7)k$, i.e. $\textbf{K}^2=7.56k^2$, we fit the coefficients in the range $0\leq k \leq 0.40$ to find $e_1=0.0563$ and $e_2=-0.00028$, in excellent agreement with the values obtained from the Selberg trace formula.

\subsection{Weyl law}

\noindent The opposite limit of Eq. (\ref{gs4}), namely $\beta\to0$, corresponds to a high-temperature expansion that allows to probe the asymptotic distribution of large eigenvalues. For this purpose we define the spectral staircase function
\begin{align}
 \label{weyl1} \mathcal{N}_{\textbf{k}}(E) = \sum_\lambda \theta(E-E_\lambda(\textbf{k})),
\end{align}
with $\theta$ the Heaviside step function. Weyl's law states that the eigenvalues of the Laplacian on a Riemann surfaces of area $A$ for $\textbf{k}=0$ satisfy
\begin{align}
 \label{weyl2} \mathcal{N}(E) \sim \frac{A}{4\pi} E
\end{align}
to leading order as $E\to \infty$. In the following, we study the behavior of $Z(\beta,\textbf{k})$ for small $\beta$ to recover this result for the case of the Bolza surface, and identify the contribution from $\textbf{k}\neq 0$.

For $\beta \to 0$, the orbit sum contribution to $Z(\beta,\textbf{k})$ is exponentially suppressed and can be neglected. (We ve\-rified this numerically.) The integral contribution, which is independent of $\textbf{k}$, yields
\begin{align}
 \nonumber \sum_{\lambda=0}^\infty e^{-\beta E_\lambda(\textbf{k})} &\sim e^{-\beta/4} \frac{A}{4\pi\beta}\Bigl(1-\frac{1}{12}\beta+\dots\Bigr)\\
 \label{weyl3} &= \frac{A}{4\pi}\Bigl( \frac{1}{\beta} -\frac{1}{3}+\dots\Bigr).
\end{align}
This implies that
\begin{align}
 \label{weyl14} \mathcal{N}_{\textbf{k}}(E) \sim \frac{A}{4\pi} \Bigl(E -\frac{1}{3}\Bigr)
\end{align}
as $E\to \infty$ from a Tauberian theorem argument, see Ref. \cite{BALAZS1986109}. We find that, in contrast to the low-temperature behavior, the high-temperature asymptotics is not affected by the crystal momentum.

\subsection{Empty lattice approximation}

\noindent In this section, we investigate whether integration over all possible values of $\textbf{k}$ yields the spectrum and trace formula for the infinite system. In solid state physics, this question is related to the empty lattice approximation \cite{BookAshcroft,maciejko2020hyperbolic}.

We first consider the Euclidean case. The spectrum of the Laplacian in the infinite Euclidean plane is para\-meterized by the two-momentum $\textbf{p}$ and energies $E=\textbf{p}^2$. For a suitable testfunction $h(p)$ we then have
\begin{align}
 \label{gs14b} \sum_{\textbf{p}} h(p) = \int\frac{\mbox{d}^2p}{(2\pi)^2} h(p).
\end{align}
On the other hand, the Euclidean plane can be tessellated by squares through the $\{4,4\}$ lattice. The corresponding Bloch wave spectrum on the (dual) square Bravais lattice is given by
\begin{align}
 \label{gs15} E_{\textbf{s},\textbf{k}} = \frac{1}{L^2}\Bigl(2\pi \textbf{s}+ \textbf{k}\Bigr)^2,
\end{align}
which coincides with the spectrum on the torus with $\textbf{k}$-twisted boundary conditions. Write $\textbf{k}=2\pi\hat{\textbf{k}}$. As we vary $\textbf{s}\in\mathbb{Z}^2$ and $\hat{\textbf{k}}\in [0,1)^2$, every $\textbf{p}=\frac{2\pi}{L}(\textbf{s}+ \hat{\textbf{k}})\in\mathbb{R}^2$ is sampled exactly once.  Consequently, tracing over all values of $\textbf{s}$ and $\textbf{k}$ is equivalent to tracing over the spectrum of the infinite Euclidean plane parameterized by $\textbf{p}\in\mathbb{R}^2$. Consider again the testfunction $h(p)$ from Eq. (\ref{gs14b}). Upon integration over the crystal momentum, the $\textbf{k}$-dependent terms in Eq. (\ref{euc12}) vanish due to
\begin{align}
 \label{gs15b} \int_{[0,1)^2}\mbox{d}^2\hat{k}\ e^{-2\pi\rmi \textbf{n}\cdot\hat{\textbf{k}}}=0
\end{align}
for every vector of integers $\textbf{n}\neq 0$. We define the $d$-dimensional torus as $\mathbb{T}^d = [0,2\pi)^d$ and note that
\begin{align}
 \int_{[0,1)^d} \mbox{d}^d\hat{k}\ (\dots) = \int_{\mathbb{T}^d} \frac{\mbox{d}^dk}{(2\pi)^d}\ (\dots).
\end{align}
The nonvanishing contribution to the trace formula is given by
\begin{align}
  \label{gs18}  \int_{\mathbb{T}^2}\frac{\mbox{d}^2k}{(2\pi)^2} \sum_{\textbf{s}\in\mathbb{Z}^2} h(p_{\textbf{s},\textbf{k}}) &=  L^2\int \frac{\mbox{d}^2p}{(2\pi)^2} h(p),
\end{align}
which equals the result in the infinite Euclidean plane up to the prefactor that sets the dimension. We summarize that by sampling over all crystal momenta, the spectrum on the torus yields the spectrum on the infinite Euclidean plane.

Let us now apply the same procedure to the Selberg trace formula on the Bolza surface. The fundamental octagon of the $\{8,8\}$ lattice tessellates the infinite hyperbolic plane $\mathbb{D}$. We explore whether by sampling all crystal momenta $\hat{\textbf{k}}\in[0,1)^4$ of the (dual) $\{8,8\}$ lattice we recover the result obtained from solving the Schr\"{o}dinger problem (\ref{int2}) in $\mathbb{D}$. On the infinite plane, $E=\frac{1}{4}+\textbf{p}^2$ with $\textbf{p}\in\mathbb{R}^2$, and for a suitable testfunction $h(p)$ we have \cite{Helgason}
\begin{align}
 \label{gs19} \sum_{\textbf{p}} h(p) = \int\frac{\mbox{d}^2p}{(2\pi)^2}\ \tanh(\pi p) h(p).
\end{align}
Note that hyperbolic space has a natural length scale, the curvature radius $\kappa$, which can be used to define a momentum with the correct dimension. We continue to work with $\kappa=1$ here.

Integrating the $\textbf{k}$-dependent Selberg trace formula (\ref{hyp13}) over all values of $\textbf{k}\in\mathbb{T}^4$, most terms vanish because for every vector of integers $\textbf{v}\neq \textbf{0}$ we have
\begin{align}
 \label{gs20} \int_{\mathbb{T}^4} \frac{\mbox{d}^4k}{(2\pi)^4}  e^{\rmi m \textbf{v}\cdot\textbf{k}} =0.
\end{align}
On the other hand, this elimination fails if $\textbf{v}=\textbf{0}$, which does occur for certain $n$, see Tab. \ref{TabEven}. We empirically find that if $\textbf{0}\in \mathcal{V}_n$, then all elements of $\mathcal{V}_n$ are equal to $\textbf{0}$, see the next section. Consequently,
\begin{align}
 \label{gs21} \textbf{0}\in \mathcal{V}_n:\ \int_{\mathbb{T}^4} \frac{\mbox{d}^4k}{(2\pi)^4} \sum_{\textbf{v}\in\mathcal{V}_n} e^{\rmi m \textbf{v}\cdot\textbf{k}} = \sum_{\textbf{v}\in\mathcal{V}_n} 1 = d_0(n).
\end{align}
In our investigation of primitive periodic orbits on the Bolza surface with $n\leq 150$, the case $\textbf{0}\in\mathcal{V}_n$ occurs for $n=8, 16, 32, 80, 96, 112$, which are all divisible by 4, with $d_0(n)=8, 12, 32, 48, 48, 48$. Denote the presumably infinite list of $n$-values such that $\textbf{0}\in\mathcal{V}_n$ by
\begin{align}
 \label{gs22} N_0= \{8,\ 16,\ 32,\ 80,\ 96,\ 112,\ \dots\}
\end{align}
and assume that Eq. (\ref{gs21}) holds for all $n\in N_0$. We then arrive at
\begin{align}
 \nonumber \int_{\mathbb{T}^4}\frac{\mbox{d}^4k}{(2\pi)^4} \sum_{\lambda=0}^\infty h(p_\lambda&(\textbf{k})) = A\int\frac{\mbox{d}^2p}{(2\pi)^2} \tanh(\pi p)h(p)\\
 \label{gs23} &+\sum_{n\in N_0} d_0(n) \sum_{m=1}^\infty  \frac{\ell_n\tilde{h}(m\ell_n)}{2\sinh(m\ell_n/2)}.
\end{align}

The first term in Eq. (\ref{gs23}) resembles the trace formula in the infinite hyperbolic plane (\ref{gs19}) up to a prefactor that sets the dimension. The second term, however, precludes the simple notion that by sampling all hyperbolic Bloch momenta we obtain the spectrum of the infinite hyperbolic plane. One interpretation of this finding would be that sampling all hyperbolic Bloch momenta results in an overcounting of eigenstates. It appears challenging though to imagine a procedure whereby omitting certain $\textbf{k}$-values on the left-hand side of Eq. (\ref{gs23}) eliminates the second term on the right-hand side. Another interpretation would be that perhaps the infinite volume limit result in Eq. (\ref{gs19}) is incorrect and receives subleading corrections, precisely of the form of the second term in Eq. (\ref{gs23}). Of course, a third option would be that we were lead on a wrong direction by the success of the Euclidean formula (\ref{gs18}) and that sampling over Bloch momenta has no relation to the spectrum on the infinite plane in hyperbolic space. We leave the solution of this puzzle to future work.

Until here we have not specified the form of $h(p)$ in the empty lattice approximation. For the study of partition functions, with $h(p)$ from Eq. (\ref{gs2}), the second term in Eq. (\ref{gs23}) is small compared to the first term for all $\beta$, because the contribution from the few geodesics with $n\in N_0$ cannot outweigh the exponential decrease due to $\sinh(m\ell_n/2)$ in the denominator. Indeed, it appears that $d_0(n)$ for $n\in N_0$ does not show the characteristic exponential increase of $d_0(n)$ found for generic $n$-values. This means that the partition function in hyperbolic band theory obtained by sampling over all Bloch momenta gives an excellent approximation of the partition function in the infinite hyperbolic plane.

\section{Symmetries}\label{SecSymm}

\noindent The Bolza surface features internal symmetries described by a non-Abelian automorphism group $G$ with 96 elements. It has been shown in Ref. \cite{maciejko2020hyperbolic} that this symmetry results in a $G$-invariance of the eigenvalues $E_\lambda(\textbf{k})$ that solve Eq. (\ref{int2}) on the Bolza surface with the twisted boundary conditions (\ref{hyp5}). In a suitable four-dimensional representation of $G$ introduced below we have
\begin{align}
 \label{inv1} E_\lambda(g\textbf{k}) = E_\lambda(\textbf{k})
\end{align}
for all $g\in G$. This $G$-invariance implies that both sides of the Selberg trace formula (\ref{hyp13}) are invariant under $\textbf{k}\to g\textbf{k}$. From the form of the right-hand side in Eq. (\ref{hyp13}), with sums over $\textbf{v}\in\mathcal{V}_n$ for each $n$, this invariance cannot be deduced directly. We show in the following that these sums can be written in a manifestly $G$-invariant form, which, in turn, yields additional insights into the nature of the $\mathcal{V}_n$.

\subsection{Automorphism invariance}

\noindent The automorphism group $G$ is generated multiplicatively by four elements, denoted $R, S, T, U$. $R$ is a rotation by $2\pi/8$ about the center of the central octagon. Similar geometric interpretation can be given to $S$, $T$ and $U$. The group reads
\begin{align}
 \nonumber G=\Bigl\{ R^iS^jT^kU^l\Bigl|\ &i\in\{0,\dots,7\},\ j\in\{0,1\},\ \\
 \label{inv2} &k\in\{0,1\},\ l\in\{0,1,2\}\Bigr\}.
\end{align}
The four-dimensional representation yielding Eq. (\ref{inv1}) is given by \cite{maciejko2020hyperbolic}
\begin{align}
 \nonumber R &= \begin{pmatrix} 0 & 0 & 0 & -1 \\ 1 & 0 & 0 & 0 \\ 0 & 1 & 0 & 0 \\ 0 & 0 & 1 & 0 \end{pmatrix},\ S =\begin{pmatrix} 0 & 1 & 0 & 0 \\ 1 & 0 & 0 & 0 \\ 0 & 0 & 0 & -1 \\ 0 & 0 & -1 & 0 \end{pmatrix},\\
 \label{inv3} T &=\begin{pmatrix} 0 & -1 & 1 & -1 \\ -1 & 0 & 1 & -1 \\ -1 & 1 & 0 & -1 \\ -1 & 1 & -1 & 0 \end{pmatrix},\ U = \begin{pmatrix} 0 & -1 & 0 & 0 \\ 1 & -1 & 0 & 0 \\ 1 & 0 & -1 & 1 \\ 0 & 1 & -1 & 0 \end{pmatrix}.
\end{align}
The group contains inversion due to $R^4=-\mathbb{1}_4$. Here $\mathbb{1}_4$ is the $4\times 4$ unit matrix.

The group $G$ in the representation of Eqs. (\ref{inv3}) admits exactly two invariants $I(\textbf{k})$ and $Q(\textbf{v})$ that are quadratic in the components of $\textbf{k}$ and $\textbf{v}$, respectively, which satisfy
\begin{align}
 \label{inv4} I(g\textbf{k}) &= I(\textbf{k}),\\
 \label{inv5} Q(g^T\textbf{v}) &= Q(\textbf{v})
\end{align}
for all $g\in G$. From a suitable ansatz for the quadratic form one verifies that
\begin{align}
 \label{inv6} I(\textbf{k}) &= \textbf{k}^T\mathcal{I}\textbf{k},\\
 \label{inv7} Q(\textbf{v}) &= \textbf{v}^T \mathcal{Q} \textbf{v}
\end{align}
with matrices 
\begin{align}
 \label{inv8} \mathcal{I} &=\frac{1}{192}\sum_{g\in G} g^Tg = \mathbb{1}_4+\Sigma,\\
 \label{inv9}\mathcal{Q} &= \frac{1}{192}\sum_{g\in G}gg^T = \mathbb{1}_4-\Sigma,\\
 \label{inv10} \Sigma &= \frac{1}{2} \begin{pmatrix} 0 & -1 & 0 & 1 \\ -1 & 0 & -1 & 0 \\ 0 & -1 & 0 & -1 \\ 1 & 0 & -1 & 0 \end{pmatrix}.
\end{align}
Importantly,
\begin{align}
 \textbf{v}\in\mathbb{Z}^4\ \Rightarrow\ Q(\textbf{v})\in\mathbb{N}_0.
\end{align}
Since $G$ contains inversion, there can be no invariants that contain an odd number of components of $\textbf{k}$ and $\textbf{v}$. The only invariants to quartic order are $I(\textbf{k})^2$ and $Q(\textbf{v})^2$. To sextic order, however, there are nontrivial $G$-invariants that are not powers of $I$ or $Q$. Importantly, when expressed in terms of $\textbf{K}$ from Eq. (\ref{gs10}), we have $\mathcal{I}(\textbf{k})=\frac{1}{2}\textbf{K}^2$. This is derived below in Eq. (\ref{hur7}).

With these definitions we now show that the $G$-invariance of the Selberg trace formula (\ref{hyp13}) is due to the right-hand side being a sum of terms of the form 
\begin{align}
 \label{inv12} \sum_{g\in G} e^{\rmi m \textbf{v}^Tg\textbf{k}},
\end{align}
with $\textbf{v}$ some four-component vector of integers. This expression is manifestly invariant under $\textbf{k}\to g \textbf{k}$. Although we only verified this for $n\leq 150$, we believe the evi\-dence is striking that this pattern persists for larger $n$. Furthermore, we have
\begin{align}
 \nonumber \sum_{g\in G} e^{\rmi m \textbf{v}^Tg\textbf{k}} ={}& 96 -24m^2 Q(\textbf{v})I(\textbf{k}) \\
 \label{inv11}&+2m^4 Q(\textbf{v})^2I(\textbf{k})^2 + \mathcal{O}(\textbf{k}^6).
\end{align}
This allows us to expand the Selberg trace formula for $\sum_{\lambda=0}^\infty h(p_\lambda(\textbf{k}))$ to quartic order in $\textbf{k}$ with coefficients that are universally determined by the primitive periodic orbits. Equation (\ref{inv11}) implies Eq. (\ref{gs10b}), i.e.
\begin{align}
 \nonumber \sum_{\textbf{v}\in\mathcal{V}_n}e^{\rmi m \textbf{v}\cdot\textbf{k}} ={}& d_0(n) -3m^2 d_1(n) \textbf{K}^2 \\
 \label{inv12b} &+ \frac{m^4}{8}d_2(n) \textbf{K}^4 + \mathcal{O}(\textbf{k}^6)
\end{align}
for the Selberg trace formula (\ref{hyp13}) for the Bolza surface. The integer coefficients $d_{0,1,2}(n)$ can be obtained directly from differentiation of Eq. (\ref{inv12b}) or from Eqs. (\ref{inv18c})-(\ref{inv18e}) derived below. We list the coefficients in Table \ref{TabdTable}.

\subsection{Automorphism-invariant sets}

In the following, we introduce an economic way of rewriting the right-hand side of the Selberg trace formula in Eq. (\ref{hyp13}). In general, to evaluate the orbit sum for a given value of $n$, all vectors $\textbf{v}\in\mathcal{V}_n$ need to be known. Since the number of geodesics of length $\ell$ grows as $e^\ell/\ell$, equally many vectors $\textbf{v}$ would need to be listed. However, it turns out that all sets $\mathcal{V}_n$ are made from only a few compound sets $\mathcal{W}=\mathcal{W}_Q^{(a)}$, labelled by the value of $Q=Q(\textbf{v})$ and a finite index $a\geq 1$. Elements within each $\mathcal{W}$ are related to each other through automorphisms from $G^T$. Thus we only need to know how often each $\mathcal{W}$ is contained in each $\mathcal{V}_n$. The resulting data set is very sparse and might hint at some underlying group or number theoretic origin, though we were not able to identify such a connection.

The sets $\mathcal{V}_n$ that enter the Selberg trace formula decompose into $G^T$-invariant sets $\mathcal{W}_Q^{(a)}$ which have the form
\begin{align}
 \label{inv13} \mathcal{W}_Q^{(a)} = G^T \textbf{v}_Q^{(a)} = \Bigl\{\textbf{v}\ \Bigr|\ \textbf{v}= g^T\textbf{v}^{(a)}_Q\ \text{ for some }g\in G\Bigr\}
\end{align}
with $\textbf{v}_Q^{(a)}\in\mathbb{Z}^4$. Obviously, if $\textbf{v}\in\mathcal{W}_Q^{(a)}$ then $g^T\textbf{v}\in\mathcal{W}_Q^{(a)}$. The value of $Q$ is constant among the set $\mathcal{W}_Q^{(a)}$ due to Eq. (\ref{inv5}), $Q(\textbf{v}) = Q(\textbf{v}_Q^{(a)})$\ for all $\textbf{v}\in\mathcal{W}_Q^{(a)}$. This justifies labeling the sets by the integer $Q$. For every fixed $Q=0,1,2,\dots$, there is a finite number of distinct sets $\mathcal{W}_Q^{(a)}$, which is captured by the additional superindex $a$. For example, the first few $\mathcal{V}_n$ are given by 
\begin{align}
 \mathcal{V}_1 &= \mathcal{W}_1,\ \mathcal{V}_2 = \mathcal{W}_2,\ \mathcal{V}_3 = \mathcal{W}_3^{(1)},\ \mathcal{V}_4 = \emptyset,\\
  \mathcal{V}_5 &= \mathcal{W}_3^{(2)}\cup\mathcal{W}_3^{(2)},\ \mathcal{V}_6 = \mathcal{W}_2\cup  \mathcal{W}_2.
\end{align}
For larger $n$, the decompositions become more involved, for instance
\begin{align}
 \label{inv15} \mathcal{V}_{11} = \mathcal{W}_1 \cup \mathcal{W}_1 \cup \mathcal{W}_1 \cup \mathcal{W}_1 \cup \mathcal{W}_5^{(2)} \cup \mathcal{W}_5^{(2)}.
\end{align}
We list all sets $\mathcal{W}_Q^{(a)}$ with $Q\leq 13$  in Tab. \ref{TabW}. These are the cases relevant for $n\leq 150$. Although $G$ contains 96 elements, the number of elements of $G^T\textbf{v}$ for any $\textbf{v}\in\mathbb{Z}^4$ can be smaller than 96 due to repetitions. We find that the numbers of elements in the sets $\mathcal{W}_Q^{(a)}$ varies between 1 (only for $\mathcal{W}_0=\{\textbf{0}\}$), 24, 48, and 96.

\renewcommand{\arraystretch}{1.4}
\begin{table}
\begin{tabular}{|c|c|c|c|c|}
\hline
 $Q$ & $\mathcal{W}_Q^{(a)}$ & $\textbf{v}_Q^{(a)}$ &  \ $|\mathcal{W}_Q^{(a)}|$ \ & $1^{\rm st}$ appearance\\
\hline\hline
 0 & $\mathcal{W}_0$ & $(0,0,0,0)^T$   &    1 & $n=8$ \\
\hline
 1 &$\mathcal{W}_1$ & $(1,0,0,0)^T$    &     24& $n=1$ \\
\hline
 2 & $\mathcal{W}_2$ & $(1,0,1,0)^T$ &      24& $n=2$ \\
\hline
 3 &  $\mathcal{W}_3^{(1)}$ & $(1,1,0,0)^T$ &  48& $n=3$ \\
  &  $\mathcal{W}_3^{(2)}$ & $(1,1,0,1)^T$ & 48& $n=5$ \\
\hline
 4 & $\mathcal{W}_4$ & $2(1,0,0,0)^T$   &     24& $n=12$ \\
\hline 
 5 & $\mathcal{W}_5^{(1)}$ & $(2,0,1,0)^T$ &    96& $n=9$ \\
  & $\mathcal{W}_5^{(2)}$ & $(1,1,1,0)^T$ &       48& $n=11$ \\
\hline
 6 & $\mathcal{W}_6^{(1)}$ & $(1,0,1,2)^T$ &    48& $n=14$ \\
  & $\mathcal{W}_6^{(2)}$ & $(1,1,1,1)^T$ &     48& $n=18$ \\
\hline
 7 & $\mathcal{W}_7^{(1)}$ & $(2,1,0,0)^T$ &    96& $n=15$ \\
  & $\mathcal{W}_7^{(2)}$ & $(1,0,2,1)^T$ &     96& $n=23$ \\
\hline
 8 & $\mathcal{W}_8$ & $2(1,0,1,0)^T$ &   24& $n=40$ \\
\hline
 9 & $\mathcal{W}_9^{(1)}$ & $(1,1,2,0)^T$ &     96& $n=37$ \\
   & $\mathcal{W}_9^{(2)}$ & $(1,1,0,3)^T$ &     48& $n=39$ \\
  & $\mathcal{W}_9^{(3)}$ & $(2,0,2,1)^T$ &    48& $n=41$ \\
  &  $\mathcal{W}_9^{(4)}$ & $(1,1,1,2)^T$ &     96& $n=47$ \\
  & $\mathcal{W}_9^{(5)}$ & $3(1,0,0,0)^T$ &    24& $n=57$ \\
\hline
 10 & $\mathcal{W}_{10}^{(1)}$ & $(1,0,3,0)^T$ &   96& $n=42$ \\
  & $\mathcal{W}_{10}^{(2)}$ & $(1,2,1,0)^T$ &     48& $n=50$ \\
\hline
 11 & $\mathcal{W}_{11}^{(1)}$ & $(1,0,1,3)^T$ &    48& $n=61$ \\
   & $\mathcal{W}_{11}^{(2)}$ & $(1,0,2,2)^T$ &     96& $n=63$ \\
 & $\mathcal{W}_{11}^{(3)}$ & $(1,2,1,1)^T$ &    96& $n=81$ \\
  & $\mathcal{W}_{11}^{(4)}$ & $(2,1,1,2)^T$ &   48& $n=83$ \\
\hline
 12 & $\mathcal{W}_{12}^{(1)}$ & \ $2(1,1,0,0)^T$ \  &  48& $n=68$ \\
 & $\mathcal{W}_{12}^{(2)}$ & $2(1,1,0,1)^T$ &   48& $n=108$ \\
\hline
 \ 13 \    & $\mathcal{W}_{13}^{(1)}$ & $(0,0,1,3)^T$ &  96& $n=69$ \\
 & \ $\mathcal{W}_{13}^{(2)}$ \ & $(2,0,3,0)^T$ &      96& $n=105$ \\
   & $\mathcal{W}_{13}^{(3)}$ & $(1,0,3,1)^T$ &  96& $n=107$ \\
  & $\mathcal{W}_{13}^{(4)}$ & $(2,1,2,0)^T$ &    48& $n=125$ \\
\hline
\end{tabular}
\caption{Invariant sets $\mathcal{W}_Q^{(a)}$ with $Q\leq 13$, which are the sets that appear in $\mathcal{V}_n$ for $n\leq 150$ in the orbit sum. We display the generating vector $\textbf{v}_Q^{(a)}$ in Eq. (\ref{inv13}) and the number of elements in each set, denoted $|\mathcal{W}_Q^{(a)}|$. The number of times that $\mathcal{W}_Q^{(a)}$ appears in $\mathcal{V}_n$, denoted $\nu_{n,Q}^{(a)}$, is listed in Tabs. \ref{TabEven} and \ref{TabOdd} in the appendix.}
\label{TabW}
\end{table}
\renewcommand{\arraystretch}{1}

Having defined the sets $\mathcal{W}_Q^{(a)}$, we next define
\begin{align}
 \label{inv16} \nu_{n,Q}^{(a)} = \begin{cases} \text{number of times }\mathcal{W}_Q^{(a)} \text{ appears}\\
 \text{in the decomposition of }\mathcal{V}_n\end{cases}.
\end{align} 
In the example from Eq. (\ref{inv15}) we have $\nu_{11,1}=4$, $\nu_{11,5}^{(2)}=2$, and all other $\nu_{n,Q}^{(a)}=0$. The Selberg trace formula (\ref{hyp13}) becomes
\begin{align}
 \nonumber &\sum_{\lambda=0}^\infty h(p_\lambda(\textbf{k})) =  A \int\frac{\mbox{d}^2p}{(2\pi)^2} \tanh(\pi p)h(p)\\
 \label{inv18} &+\sum_{n=1}^\infty \sum_{Q=0}^\infty\sum_a \nu_{n,Q}^{(a)}\sum_{\textbf{v}\in\mathcal{W}_Q^{(a)}} \sum_{m=1}^\infty  e^{\rmi m\textbf{v}\cdot\textbf{k}} \frac{\ell_n\tilde{h}(m\ell_n)}{2\sinh(m\ell_n/2)}.
\end{align}

The decomposition of $\mathcal{V}_n$ into $\mathcal{W}_Q^{(a)}$ yields an expression for the integer coefficients in Eq. (\ref{gs10b}). Using
\begin{align}
 \label{inv18b} \sum_{\textbf{v}\in \mathcal{V}_n}e^{\rmi m \textbf{v}\cdot\textbf{k}} =\sum_{Q,a} \nu_{n,Q}^{(a)} \frac{|\mathcal{W}_Q^{(a)}|}{96} \sum_{g\in G} e^{\rmi m \textbf{v}_Q^{(a)}{}^Tg\textbf{k}},
\end{align}
and Eqs. (\ref{inv11}) and (\ref{hur7}), we find
\begin{align}
 \label{inv18c} d_0(n) &= \sum_{Q,a} \nu_{n,Q}^{(a)}|\mathcal{W}_Q^{(a)}|,\\
 \label{inv18d} d_1(n) &= \frac{1}{24}\sum_{Q,a} \nu_{n,Q}^{(a)}|\mathcal{W}_Q^{(a)}| Q,\\
 \label{inv18e} d_2(n) &= \frac{1}{24} \sum_{Q,a} \nu_{n,Q}^{(a)}|\mathcal{W}_Q^{(a)}| Q^2.
\end{align}
These numbers are collected in Tab. \ref{TabdTable}. Since $|\mathcal{W}_Q^{(a)}|$ is divisible by $24$ for $Q>0$, at least for all cases considered here, we conclude that $d_1(n)$ and $d_2(n)$ are integers. The equation for $d_0(n)$ shows that determining the numbers $\nu_{n,Q}^{(a)}$ for $\textbf{k}\neq 0$ might yield new insights into the problem of finding the length spectrum on the Bolza surface even for $\textbf{k}=0$. Note also that Eq. (\ref{inv18b}) is indeed of the form of Eq. (\ref{inv12}).

For a given $n$, only a few terms in the sum over $Q$ are nonzero in Eq. (\ref{inv18}), see Tabs. \ref{TabEven} and \ref{TabOdd}. For $n\leq 150$, we verify the following statement, which we conjecture to be valid for all $n$:\\
\emph{Conjecture. } \text{Assume }$\textbf{v}\in\mathcal{V}_n$.\ \text{Then}
\begin{align}
 \label{inv20} n \text{ even}\ &\Rightarrow\ Q(\textbf{v})\ \text{ even},\\
 \label{inv21} n \text{ odd}\ &\Rightarrow\ Q(\textbf{v})\ \text{ odd}.
\end{align}
Furthermore, for even $n$,
\begin{align}
 \label{inv22} n \equiv 0\text{ (mod 4)}\ &\Rightarrow\ Q(\textbf{v}) \equiv 0\text{ (mod 4)},\\
 \label{inv23} n \equiv 2\text{ (mod 4)}\ &\Rightarrow\ Q(\textbf{v}) \equiv 2\text{ (mod 4)}
\end{align}
Pursuing to proof this conjecture strikes us as a promising task for future investigations. Having a better understanding of the sets $\mathcal{V}_n$ from a number-theoretic point of view bears potential of eventually summing all terms in the Selberg trace formula, which would yield an exact and genuinely nonperturbative tool for studying quantum physics in hyperbolic space.

\subsection{Relation to Hurwitz quaternions}

\noindent Since the group $G$ is finite, all its irreducible representations are unitary. The matrices in Eq. (\ref{inv3}) are not unitary, but there exists a basis in which they are. The corresponding basis change matrix is
\begin{align}
 \label{hur1} \mathcal{U} = \mathbb{1}_4 - R = \begin{pmatrix} 1 & 0 & 0 & 1 \\ -1 & 1 & 0 & 0 \\ 0 & -1 & 1 & 0 \\ 0 & 0 & -1 & 1 \end{pmatrix}.
\end{align}
We denote the unitary representation by an overhat according to
\begin{align}
 \label{hur2} \hat{g} = \mathcal{U} g \mathcal{U}^{-1}.
\end{align}
The generators of $G$ in the unitary representation read
\begin{align}
 \nonumber \hat{R} &= \begin{pmatrix} 0 & 0 & 0 & -1 \\ 1 & 0 & 0 & 0 \\ 0 & 1 & 0 & 0 \\ 0 & 0 & 1 & 0 \end{pmatrix},\ \hat{S} = \begin{pmatrix} 0 & 0 & -1 & 0 \\ 0 & -1 & 0 & 0 \\ -1 & 0 & 0 & 0 \\ 0 & 0 & 0 & 1\end{pmatrix},\\
 \label{hur3} \hat{T} &= \begin{pmatrix} -1 & 0 & 0 & 0 \\ 0 & 1 & 0 & 0 \\ 0 & 0 & -1 & 0 \\ 0 & 0 & 0 & 1 \end{pmatrix},\ \hat{U} = \frac{1}{2}\begin{pmatrix} -1 & -1 & -1 &1 \\ 1 & -1 & -1 & -1 \\ 1 & 1 & -1 & 1 \\ -1 & 1 & -1 & -1 \end{pmatrix}.
\end{align}
For each $\textbf{k}$ and $\textbf{v}$ we define $\textbf{K} = \mathcal{U}\textbf{k}$ and $\textbf{V} =(\mathcal{U}^{-1})^T\textbf{v}$ such that
\begin{align}
 \label{hur4} \textbf{K} &= \begin{pmatrix} k_1+k_4 \\ k_2 -k_1 \\ k_3-k_2\\ k_4-k_3 \end{pmatrix},\ \textbf{V} =\frac{1}{2}\begin{pmatrix} v_1+v_2+v_3+v_4 \\ -v_1+v_2+v_3+v_4 \\ -v_1-v_2+v_3+v_4 \\ -v_1-v_2-v_3+v_4 \end{pmatrix}
\end{align}
and
\begin{align}
 \label{hur6} \textbf{v}^Tg \textbf{k} = \textbf{V}^T\hat{g}\textbf{K}.
\end{align}
The invariants $I$ and $Q$ simply become
\begin{align}
 \label{hur7} I(\textbf{k}) &= \frac{1}{2}\textbf{K}^2,\\
 \label{hur8} Q(\textbf{v}) &= \textbf{V}^2.
\end{align}

The simple form of the invariants $I$ and $Q$ in the new basis suggests that it is a canonical choice for the problem at hand. When written in this new basis, the elements $\textbf{v}\in\mathcal{V}_n$ have an interesting structure as well. For this note that $\mathcal{V}_1$ is given by the 24 elements in Eq. (\ref{orb5}), which translates to
\begin{align}
 \nonumber \hat{\mathcal{V}}_1 ={}&(\mathcal{U}^{-1})^T \mathcal{V}_1\\
 \nonumber ={}&\Bigl\{ (\pm 1,0,0,0)^T,\ (0,\pm 1,0,0)^T,\ (0,0,\pm 1,0)^T,\\
 \label{hur9} & (0,0,0,\pm 1)^T,\ \Bigl(\pm \frac{1}{2}, \pm \frac{1}{2}, \pm \frac{1}{2},\pm \frac{1}{2}\Bigr)^T\Bigr\},
\end{align}
where in the last entry all signs are varied independently of each other. These 24 vectors are precisely the unit Hurwitz quaternions, see e.g. Ref. \cite{BookQuat}. In general, Hurwitz quaternions have the form $\textbf{V}=(a,b,c,d)^T$ with either $a,b,c,d\in\mathbb{Z}$ or $a,b,c,d \in\mathbb{Z}+\frac{1}{2}$, meaning the components are either all integers or all half-integers. The subset with $a,b,c,d\in\mathbb{Z}$ is called Lipshitz quaternions. The reduced norm of a Hurwitz quaternion is $Q=\textbf{V}^2=a^2+b^2+c^3+d^2$, which is simply the squared length of $\textbf{V}$ or the invariant $Q$. Hence, $\hat{\mathcal{W}}_Q^{(a)}$ is a set of Hurwitz quaternions with norm $Q$. It is easy to see that if $\textbf{V}^2$ is even, then $\textbf{V}$ must be a Lipshitz quaternion. Consequently, if the conjecture in Eqs. (\ref{inv20})-(\ref{inv23}) holds true for all $n$, then all sets $\hat{\mathcal{V}}_n$ for $n$ even are Lipshitz quaternions.

Quaternions comprise a division algebra structure that can be conveniently expressed in terms of the basis
\begin{align}
 \label{hur10} \hat{1} = \begin{pmatrix} 1 & 0 \\ 0& 1 \end{pmatrix},\ \hat{\text{i}} = \rmi \sigma_3,\ \hat{\text{j}}=\rmi \sigma_2,\ \hat{\text{k}}=\rmi \sigma_1
\end{align}
with $\sigma_i$ the standard Pauli matrices, so that
\begin{align}
 \label{hur11} \hat{\text{i}}^2 = \hat{\text{j}}^2 = \hat{\text{k}}^2 = \hat{\text{i}}\hat{\text{j}}\hat{\text{k}}=-\hat{1}.
\end{align}
Every Hurwitz quaternion $\textbf{V}$ is then identified with the $2\times 2$ matrix
\begin{align}
 \label{hur12} H(\textbf{V}) = V_1 \hat{1} + V_2 \hat{\text{i}}+ V_3 \hat{\text{j}}+ V_4 \hat{\text{k}}.
\end{align}
Their addition and multiplication is defined through the usual matrix operations and the invariant $Q$ is given by the determinant,
\begin{align}
 \label{hur13} Q = \textbf{V}^2 = \mbox{det} H(\textbf{V}).
\end{align}
In this formulation, Eq. (\ref{hur9}) becomes
\begin{align}
 \label{hur14} \hat{\mathcal{V}}_1=\Bigl\{ \pm \hat{1},\ \pm \hat{\text{i}},\ \pm \hat{\text{j}},\ \pm \hat{\text{k}},\ \frac{\pm \hat{1}\pm\hat{\text{i}}\pm\hat{\text{j}}\pm\hat{\text{k}}}{2}\Bigr\}.
\end{align}
We believe that studying the mapping 
\begin{align}
 \label{hur15} \gamma\in\Gamma \to \textbf{v}(\gamma) \to H(\textbf{V})
\end{align}
in more detail in the future might yield useful insights into higher-dimensional representations of the hyperbolic $\{8,8\}$ Bravais lattice that go beyond the paradigm of $\text{U}(1)$ Bloch waves.

\section{Band structures of hyperbolic lattices}\label{SecBand}

\noindent The hyperbolic band theory and associated trace formulas discussed in this work concern closed Riemann surfaces with crystal momenta imposed through Aharonov--Bohm fluxes or, equivalently, $\textbf{k}$-twisted boundary conditions on the fundamental polygon. In this section, we come back to the problem of band structures on hyperbolic $\{p,q\}$ lattices discussed in the introduction.

Our goal is to solve Eq. (\ref{new3}) for the eigenvalues of the adjacency matrix. It has been shown in Ref. \cite{PhysRevA.102.032208} that the left-hand side of Eq. (\ref{new3}) can be replaced by a continuum approximation according to
\begin{align}
 \label{band0b} \sum_j \mathcal{A}_{ij} \psi(z_j) = \Bigl[q + q h^2\Delta_g +\mathcal{O}(h^3)\Bigr]\psi(z_i),
\end{align}
where the parameter $h=h(p,q)$ depends on the values of $p$ and $q$ and roughly corresponds to a lattice spacing. For the $\{8,3\}$ lattice we have $q=3$ and $h=0.348311$. The eigenvalue problem on the graph can thus be approximated by the eigenvalue problem of the hyperbolic Laplacian.  The continuum approximation (\ref{band0b}) works best for small values of $|z|$ close to the origin, where curvature effects remain small \cite{PhysRevA.102.032208}. Since the hyperbolic lattice is infinite and fills the whole disk $\mathbb{D}$, we unavoidably accumulate errors as $|z|$ increases. It has been found that the first few low-lying eigenvalues $\mathcal{E}$ in Eq. (\ref{new3}) are well-captured by the continuum approximation, whereas energies of highly excited states deviate substantially  \cite{PhysRevA.102.032208}.

Here we propose one way to improve the accuracy of the continuum approximation. For this we employ recent insights into the crystallography of hyperbolic lattices. Sticking to our example of the Bolza surface, we use that the $\{8,3\}$-lattice splits into a 16-site unit cell that is repeated infinitely often within an $\{8,8\}$ hyperbolic Bravais lattice \cite{boettcher2021crystallography}, see Fig. \ref{FigBolza}. Each unit cell is contained in one fundamental domain $\mathcal{D}$, and going from one unit cell of the $\{8,3\}$ lattice to another is equivalent to moving across the faces of the $\{8,8\}$ lattice. Since the Bravais lattice is generated by the Fuchsian group $\Gamma$ from Eq. (\ref{intro3}), we conclude that all eigenstates of the $\{8,3\}$ lattice transform under distinct representations of $\Gamma$. For the $\{8,3\}$ lattice, the one-dimensional representations are found to be particularly relevant \cite{AlbertaWu2021}. They constitute Bloch waves with an associated crystal momentum $\textbf{k}$.

The spectrum of Bloch waves on the $\{8,3\}$ lattice consists of 16 bands $\mathcal{E}_\eta(\textbf{k})$, $\eta=0,\dots,15$, which corresponds to the number of sites in the unit cell. The bands are the eigenvalues of the $16\times 16$ Bloch-Hamiltonian \cite{boettcher2021crystallography}
\begin{align}
 \label{band1} \mathcal{H}(\textbf{k}) &= - \begin{pmatrix} \mathcal{A}_1 & \mathbb{1}_8 \\ \mathbb{1}_8 & \mathcal{A}_2(\textbf{k}) \end{pmatrix},
\end{align}
with $\mathbb{1}_8$ the $8\times8$ unit matrix and 
\begin{widetext}
\begin{align}
 \mathcal{A}_1 &= \begin{pmatrix} 0 & 1 & 0 & 0 & 0 & 0 & 0 & 1 \\ 1 & 0 & 1 & 0 & 0 & 0 & 0 & 0 \\ 0 & 1 & 0 & 1 & 0 & 0 & 0 & 0 \\ 0 & 0 & 1 & 0 & 1 & 0 & 0 & 0 \\ 0 & 0 & 0 & 1 & 0 & 1 & 0 & 0 \\ 0 & 0 & 0 & 0 & 1 & 0 & 1 & 0 \\ 0 & 0 & 0 & 0 & 0 & 1 & 0 & 1 \\ 1 &  0 & 0 & 0 & 0 & 0 & 1 & 0 \end{pmatrix},\
 \mathcal{A}_2(\textbf{k}) = \begin{pmatrix}   
    0 & 0 & 0 & e^{\rmi k_1} & 0 & e^{\rmi k_2} & 0 & 0\\
    0 & 0 & 0 & 0 & e^{\rmi k_2} & 0 & e^{\rmi k_3} & 0\\
    0 & 0 & 0 & 0 & 0 & e^{\rmi k_3} & 0 & e^{\rmi k_4}\\
    e^{-\rmi k_1} & 0 & 0 & 0 & 0 & 0 & e^{\rmi k_4} & 0\\
    0 & e^{-\rmi k_2} & 0 & 0 & 0 & 0 & 0 & e^{-\rmi k_1}\\
    e^{-\rmi k_2} & 0 & e^{-\rmi k_3} & 0 & 0 & 0 & 0 & 0\\
    0 & e^{-\rmi k_3} & 0 & e^{-\rmi k_4} & 0 & 0 & 0 & 0\\
    0 & 0 & e^{-\rmi k_4} & 0 & e^{\rmi k_1} & 0 & 0 & 0 \end{pmatrix}.
\end{align}
\end{widetext}
Note that for $\textbf{k}=0$ the Bloch-wave Hamiltonian constitutes (minus) the adjacency matrix of the 16-site unit cell with periodic boundary conditions, whereas for $\textbf{k}\neq 0$ it features a complex phase $e^{\rmi k_\mu}$ on the bonds that cross the boundary of the fundamental domain. In analogy to Eq. (\ref{inv1}), one finds that the eigenvalues $\{\mathcal{E}_\eta(\textbf{k})\}$ are invariant under the 96 elements of the automorphism group, i.e.
\begin{align}
\label{band2xx}  \mathcal{E}_\eta(g\textbf{k}) = \mathcal{E}_\eta(\textbf{k})
\end{align}
for all $g\in G$. Here we use the four-dimensional representation of $G$ from Eq. (\ref{inv3}). This implies, for instance, that the eigenvalues for small $\textbf{k}$ can be expanded in $\textbf{K}^2$ and $\textbf{K}^4$ to quartic order, with coefficients that depend on the band index $\eta$. Furthermore, this indicates that the 16 sites of the periodic unit cell provide a useful, albeit coarse, discretization of the Bolza surface that incorporates all symmetries of the continuous Riemann surface.

Since Eq. (\ref{band0b}) has been derived from the local properties of the graph, we expect it to comprise a good approximation for $z\in\mathcal{D}$ in the fundamental domain as this limits the value of $|z|$. By gluing together many fundamental domains with boundary conditions defined through crystal momenta, we can expect to have a faithful representation of many low-lying Bloch-wave eigenstates. The best agreement is expected for the lowest band. We thus obtain an approximate identification between the lowest Bloch-wave band on the $\{8,3\}$ lattice, $\mathcal{E}_0(\textbf{k})$, and the lowest band of the hyperbolic Laplacian $-\Delta_g$ on the Bolza surface, $E_0(\textbf{k})$, given by
\begin{align}
 \label{band2} \mathcal{E}_0(\textbf{k}) \approx -3 + 3h^2 E_0(\textbf{k})
\end{align}
with $h=0.348311$. This formula is the central result of this section.

\begin{figure}[t]
\centering
\includegraphics[width=8.8cm]{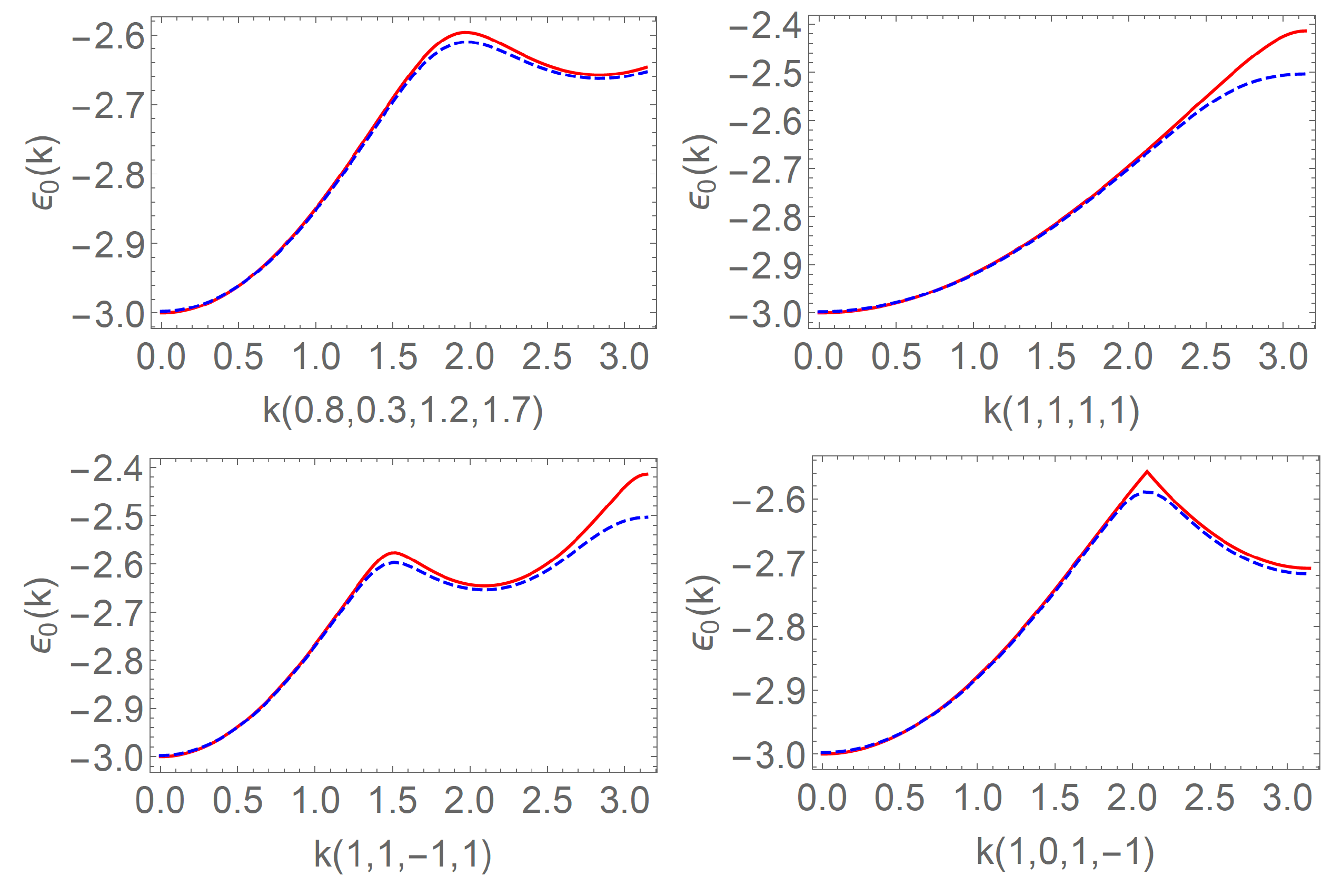}
\caption{Lowest Bloch-wave band $\mathcal{E}_0(\textbf{k})$ on the $\{8,3\}$ lattice, shown by solid red curves, along specific directions in $\textbf{k}$-space. We determine $\mathcal{E}_0(\textbf{k})$ as the lowest eigenvalue of $\mathcal{H}(\textbf{k})=-\mathcal{A}(\textbf{k})$ in Eq. (\ref{band1}). We compare with the lowest band of the Laplacian on the Bolza surface, $E_0(\textbf{k})$, matched through the approximation in Eq. (\ref{band2}), shown as dashed blue curves. We compute $E_0(\textbf{k})$ from the truncated Selberg trace formula for primitive periodic orbits with $(n,m)\leq(150,3)$, which is the same level of accuracy as in Figs. \ref{FigGS} and \ref{FigExp}. Computing $E_0(\textbf{k})$ from the solution of the Schr\"{o}dinger equation yields an improved agreement between the red and blue curves, indicating that some deviations result from the truncation of the trace formula, but no perfect match is obtained in either case. Still, we observe that Eq. (\ref{band2}) provides a decent approximation of the lowest Bloch band in the $\{8,3\}$ lattice.}
\label{Fig83E0}
\end{figure}

In the following we test the validity of the approximation (\ref{band2}). (I) First note that, at least for generic $\textbf{k}\neq 0$, there is no reason to expect the ground state on the lattice or on the Riemann surface to be degenerate, and hence it transforms with a one-dimensional representation under $\Gamma$. The assumption of the ground state being a Bloch wave is therefore justified for generic $\textbf{k}$. We note that certain high-symmetry $\textbf{k}$-points in the Brillouin zone can yield a ground state degeneracy. These points are dictated by the automorphism group $G$ and are isolated. (II) Second, if Eq. (\ref{band2}) is valid, then it should reproduce the expansion of $E_0(\textbf{k})$ in Eq. (\ref{gs12})-(\ref{gs14}), because both expressions are invariant under $G$. We note that we can obtain the expansion of $\mathcal{E}_0(\textbf{k})$ for small $\textbf{k}$ exactly by perturbing along the $\textbf{k}=k(1,0,0,0)^T$ direction with $\textbf{K}^2=2k^2$. We then find
\begin{align}
 \nonumber \mathcal{E}_0(\textbf{k}) &= - 3 + \frac{1}{48}\textbf{K}^2 -\frac{1}{9216}\textbf{K}^4 +\mathcal{O}(k^6),\\
 \label{band3} &\stackrel{!}{\approx} -3 + 3h^2\Bigl(e_1 \textbf{K}^2+e_2\textbf{K}^4\Bigr)+\mathcal{O}(k^6)
\end{align}
with 
\begin{align}
 \label{band4} e_1 &\approx \frac{1}{3h^2}\frac{1}{48} = 0.0572405,\\
 \label{band5} e_2 &\approx -\frac{1}{3h^2}\frac{1}{9216} = -0.000298127.
\end{align}
These are in good quantitative agreement with Eqs. (\ref{gs13}) and (\ref{gs14}), keeping in mind that we neglected higher-order contributions in Eq. (\ref{band2}). (III) Third, we compare $\mathcal{E}_0(\textbf{k})$ and $E_0(\textbf{k})$ along special lines in $\textbf{k}$-space, where the Bolza result $E_0(\textbf{k})$ is obtained from the truncated Selberg trace formula as described in Sec. \ref{SecE0}. The results of this comparison are shown in Fig. \ref{Fig83E0}. We find good agreement for small $|\textbf{k}|$ and small $\mathcal{E}_0(\textbf{k})+3$, but visible deviations at high-symmetry points of the Brillouin zone. Some of these deviations can be traced back to the finite truncation of the trace formula, whereas others are actual shortcomings of Eq. (\ref{band2}) because they persist when computing $E_0(\textbf{k})$ exactly from the Schr\"{o}dinger equation \cite{JosephPrivate}.

Let us summarize the finding of this section. We started from the observation that the continuum approximation in Eq. (\ref{band0b}) allows us to compute the spectrum on hyperbolic lattices from the spectrum of the hyperbolic Laplacian. However, only the first few eigenvalues are reproduced correctly. This is partly due to the fact that the eigenvalue of $-\Delta_g$ in the continuum have the form $\frac{1}{4}+p^2$, and so necessarily grow quadratically in the momentum $p$. As a result, the typical periodicity and cosine-shape of band structures is not reproduced by Eq. (\ref{band0b}). Utilizing the crystallographic properties of the $\{8,3\}$ lattice, however, we were able to extend Eq. (\ref{band0b}) to Eq. (\ref{band2}) to approximate the lowest band on the lattice by the lowest band of the Laplacian on the Bolza surface. The latter correctly reproduces the low-energy band structure qualitatively, at many $\textbf{k}$-points even quantitatively. Hence, we achieved to improve the accuracy of the approximation from a few eigenvalues to a whole band.

\section{Summary and Outlook}\label{SecOut}

\noindent In this work, we have applied the Selberg trace formula to investigate both formal aspects and practical applications in hyperbolic band theory. For this purpose, we considered the trace formula with $\text{U}(1)$ phase factors corresponding  to a nontrivial representation of the Fuchsian group underlying the hyperbolic surface. This extension of the Selberg trace formula naturally appears within hyperbolic band theory but was not relevant to previous applications of the trace formula in studies of, for example, quantum chaotic systems. This is a striking example of new experimental achievements, in this case hyperbolic lattices in circuit quantum electrodynamics and topoelectric circuits, requiring an extended repertoire of mathematical techniques. This development may eventually also influence new directions in mathematics.

We have applied the Selberg trace formula with crystal momentum $\textbf{k}$ to problems concerning the Bolza surface of genus two, which had previously been studied for $\textbf{k}\neq 0$ in the context of hyperbolic band theory without the use of the trace formula, and with the trace formula in the context of quantum chaos for $\textbf{k}=0$. Our work, therefore, comprises a confluence of two directions in physics. We have shown that the geometric data of the classical length spectrum of primitive periodic orbits is sufficient to evaluate the trace formula even for $\textbf{k}\neq 0$. In contrast to the case of $\textbf{k}=0$, however, not only the number of primitive orbits of a certain length $\ell_n$ is relevant (the quantity $d_0(n)$), but instead a representative group element for each orbit is required to compute the vectors $\textbf{v}\in \mathcal{V}_n$. This information, on the other hand, is a by-product of the original algorithm described by Aurich, Bogomolny, and Steiner. The sets $\mathcal{V}_n$ show some intriguing features that we speculate to be of group or number theoretic origin. A better understanding of these relations may help in the future to determine a rule for the sequence of numbers $d_0(n)=|\mathcal{V}_n|$, which, until now, appears to be entirely void of structure.

One practical applications we have considered is the ground state energy $E_0(\textbf{k})$ on the Bolza surface. The presence of the crystal momentum $\textbf{k}\neq 0$ enforces $E_0(\textbf{k})$ to be nonzero. Using the automorphism symmetry of the surface, we determined a universal low-$\textbf{k}$ expansion up to quartic order. Furthermore, we found good agreement between the result deduced from the Selberg trace formula and the result from numerically solving the Schr\"{o}dinger equation; deviations are due to the fact that the trace formula is truncated to orbits with $n\leq 150$ here, and we expect perfect agreement if we were able to include all primitive periodic orbits. We have then shown that the lowest band on the $\{8,3\}$ lattice, $\mathcal{E}_0(\textbf{k})$, can to a decent accuracy be approximated by $E_0(\textbf{k})$, thereby making a connection to the discrete lattices realized in experiment. This approximation is based on recent advances in understanding the crystallography of hyperbolic lattices and improves the accuracy of earlier works on the continuum limit of hyperbolic lattices.

Other applications that we have considered are related to the band structure on the Bolza surface as a whole. Here trace formulas become particularly powerful, since a determination of the high-energy spectrum from the Schr\"{o}dinger equation is computationally challenging. We verified that the asymptotic Weyl law is unaffected by the crystal momentum $\textbf{k}\neq 0$, and have studied the relationship between the empty lattice approximation and averaging over the crystal momenta of Bloch waves. While the latter two procedures were found to not be identical in the hyperbolic case, the small number of terms that violate the identity are quantitatively unimportant in practical applications such as computing partition functions. These findings thus connect the spectrum on closed Riemann surfaces to the spectrum on the infinite hyperbolic plane.

In the following we discuss some extensions and applications of the ideas laid out in this work that we suggest to be addressed in future studies. One such problem is the study of \emph{arithmetical quantum chaos} in hyperbolic band theory. It is understood well that the arithmeticity of the Fuchsian group $\Gamma$ for the Bolza surface results in spectral statistics that does not fall into any of the usual random matrix ensembles of quantum chaos \cite{SteinerDESY,BOGOMOLNY1997219,BogAri}. This behavior is rooted in the distribution of the eigenvalues $\{E_\lambda(\textbf{k})\}$ for $\textbf{k}=0$. For $\textbf{k}\neq 0$, on the other hand, the degeneracies of eigenvalues are lifted, see Fig. \ref{FigBands}, and we conjecture that also the spectral statistics evolves towards that of the usual quantum chaos. This crossover from arithmetic quantum chaos to ordinary quantum chaos in the spectrum $\{E_\lambda(\textbf{k})\}$, and its implications for physical systems in hyperbolic space, strikes us as a very intriguing aspect to be illuminated.

Many of the techniques applied in this work for the Bolza surface, the $\{8,8\}$ and $\{8,3\}$ hyperbolic lattices, can be extended to \emph{other surfaces and other hyperbolic lattices}. The two families of $\{4g,4g\}$ and $\{2(2g+1),2g+1\}$ Bravais lattices  \cite{sausset2007periodic,boettcher2021crystallography} define closed hyperbolic surfaces through identifying the opposite sides of a regular $4g$-gon or $2(2g+1)$-gon, respectively, and result in a crystal momentum $\textbf{k}$ with $2g$ components. At least some of their Fuchsian groups are arithmetic \cite{BENEDITO20161902}, and so the procedure and applications described in this work directly apply. Furthermore, it will be exciting to relate their band structures, such as the ground state energy $E_0(\textbf{k})$, to band structures on $\{p,q\}$ hyperbolic lattices. For instance, the results of Secs. \ref{SecSymm} and \ref{SecBand} can be generalized  to the $\{7,3\}$ hyperbolic lattice that tessellates the Klein quartic of genus three, with $\{14,7\}$ Bravais lattice \cite{boettcher2021crystallography} and an automorphism group $G$ with $2\times168$ elements, satisfying the Hurwitz bound.

In order to apply the Selberg trace formula to hyperbolic lattices beyond ground state or low-energy physics, the mapping of \emph{excited state energy bands} between hyperbolic surfaces and hyperbolic lattices needs to be understood better. For instance, for the Bolza surface, with energies $\{E_\lambda(\textbf{k})\}$, and the $\{8,3\}$ lattice, with energies $\{\mathcal{E}_\eta(\textbf{k})\}$, an extension of the approximate Eq. (\ref{band2}) for excited states should be constructed. There are some immediate obstacles that relate to the degeneracies of bands even at $\textbf{k}=0$, but this might just imply that the mapping is not as simple as Eq. (\ref{band2}). A more direct approach to trace formulas for hyperbolic lattices might start from trace formulas for graph Laplacians such as the Ihara zeta function and related graph-theoretic functions \cite{Ihara,Bass}.

Another exciting extension of the present work amounts to including the \emph{effects of an external magnetic field}. Spectral studies of the Hofstadter butterfly on hyperbolic lattices \cite{PhysRevLett.125.053901,Ikeda,Ikeda2,stegmaier2021universality} reveal an intricate energetic landscape, whose surprising features are deeply rooted in the nature of the quantum hall effect and the Landau level problem in hyperbolic space \cite{Comtet1985,Carey:1997zvh,Ludewig2021}. Also, connections of this problem to the geometric Langlands program have been pointed out \cite{Ikeda3,RayanHyp}. In this context, the trace formula for the Maass Laplacian of nonzero weight discussed in Ref. \cite{BookGrosche} might be useful.

Very recently, ideas of the \emph{conformal bootstrap} have been applied to energy spectra of hyperbolic surfaces such as the Bolza surface and the Klein quartic \cite{Kravchuk,Bonifacio}. This includes a dictionary between conformal field theories and hyperbolic orbifolds. It seems to be a formidable question to turn this around and ask which field theoretic object corresponds to the Selberg trace formula from hyperbolic geometry, and what this implies for our understanding of conformal field theories. Furthermore, it would be exciting to explore which role is played by the crystal momenta and nontrivial representations in this mapping.

\acknowledgements \noindent We thank J. Maciejko for many insightful comments and for kindly providing us with published and unpublished hyperbolic band spectra from the numerical solution of the Schr\"{o}dinger equation on the Bolza surface. We thank R. Aurich for kindly sending us the list of degeneracies of primitive periodic orbits on the Bolza surface. We thank S. Rayan for many helpful comments. AA acknowledges support from a Natural Sciences and Engineering Research Council of Canada (NSERC) Undergraduate Student Research Award. IB acknowledges support from the University of Alberta startup fund UOFAB Startup Boettcher and the Natural Sciences and Engineering Research Council of Canada (NSERC) Discovery Grants RGPIN-2021-02534 and DGECR2021-00043.

\begin{appendix}

\section{Hyperbolic geometry}\label{AppHyp}
\noindent We consider the Poincar\'{e} disk model of hyperbolic space given by the unit disk
\begin{align}
 \label{geo1} \mathbb{D} = \{ z\in \mathbb{C}\ |\ |z|<1\}
\end{align}
equipped with the metric
\begin{align}
 \label{geo2} \mbox{d}s^2 = (2\kappa)^2\frac{\mbox{d}x^2+\mbox{d}y^2}{(1-|z|^2)^2}.
\end{align}
Here $z=re^{\rmi \phi} = x+\rmi y$. Two choices of length scales enter these definitions. First, we could choose the radius of the disk $\mathbb{D}$ to be different from unity, but this is hardly done. Second, the curvature radius $\kappa$ sets the unit of length in the metric. Common choices for $\kappa$ are 1 and $1/2$. In this work we choose $\kappa=1$. Geodesics in $\mathbb{D}$ are arcs of circles that intersect the boundary orthogonally; this includes straight lines through the origin. Integrating Eq. (\ref{geo2}) yields the distance between two points $z,z'\in\mathbb{D}$ as
\begin{align}
 \label{geo3} d(z,z') = \kappa\ \text{arcosh}\Bigl(1+\frac{2|z-z'|^2}{(1-|z|^2)(1-|z'|^2)}\Bigr).
\end{align}
This formula makes it particularly transparent that the hyperbolic plane is infinite, meaning that each point in $\mathbb{D}$ is infinitely far from the boundary $\partial \mathbb{D}$ (points with $|z|=1$) with respect to the metric (\ref{geo2}).

The isometries of the hyperbolic plane, i.e. transformations that preserve the distance $d(z,z')$ between two points, are parameterized by matrices
\begin{align}
 \label{geo4} \begin{pmatrix} a & b \\ b^* & a^*\end{pmatrix} \in\text{PSU}(1,1)
\end{align}
with unit determinant $|a|^2-|b|^2=1$. They act on $z\in\mathbb{D}$ via M\"{o}bius transformations
\begin{align}
 \label{geo5} \begin{pmatrix} a & b \\ b^* & a^*\end{pmatrix}  z := \frac{az +b}{b^*z+a^*}.
\end{align}
Since reversing the sign of $a$ and $b$ simultaneously yields the same transformation, they are elements of $\text{PSU}(1,1)=\text{SU}(1,1)/\{\pm \mathbb{1}\}$. This group is isomorphic to $\text{PSL}(2,\mathbb{R})$. The latter naturally appears as the isometry group when working with the Poincar\'{e} upper half-plane model of the hyperbolic plane.

The Laplace--Beltrami operator on the hyperbolic plane, or simply hyperbolic Laplacian, is given by
\begin{align}
 \label{geo6} \Delta_g = \frac{1}{(2\kappa)^2} (1-|z|^2)^2 (\partial_x^2+\partial_y^2).
\end{align}
It is self-adjoint with respect to the canonical scalar product on the Poincar\'{e} disk,
\begin{align}
 \label{geo7} \langle f_1,f_2\rangle_{\mathbb{D}} := \int_{\mathbb{D}} \frac{\mbox{d}^2z}{(1-|z|^2)^2} f_1(z)^*f_2(z)
\end{align}
or the scalar product in the fundamental octagon $\mathcal{D}$ of the $\{8,8\}$ lattice,
\begin{align}
 \label{geo7b} \langle f_1,f_2\rangle_{\mathcal{D}} := \int_{\mathcal{D}} \frac{\mbox{d}^2z}{(1-|z|^2)^2} f_1(z)^*f_2(z).
\end{align}
The Laplacian commutes with isometries (\ref{geo5}). The continuous spectrum of $-\Delta_g$ on hyperbolic surfaces satisfies $E_\lambda\geq \frac{1}{4}$ and so can be parameterized as $E_\lambda=\frac{1}{4}+p^2$ with $p^2$ real. A discrete spectrum with energies below $\frac{1}{4}$ can exist depending on the surface under consideration, but for the case of the Bolza surface the only eigenvalue that falls into this category  is the ground state energy $E_0=0$.

\section{Euclidean trace formula}\label{AppEu}

\noindent We follow Ref. \cite{BogAri}. We apply Poisson's formula
\begin{align}
 \label{AppEu1} \sum_{\textbf{s}\in\mathbb{Z}^2} f(\textbf{s}) = \sum_{\textbf{n}\in\mathbb{Z}^2} \int_{-\infty}^\infty \mbox{d}s_1 \int_{-\infty}^\infty\mbox{d}s_2\ e^{2\pi\rmi (n_1s_1+n_2s_2)} f(\textbf{s})
\end{align}
to the density of states
\begin{align}
 \label{AppEu2} D_{\textbf{k}}(E) = \sum_{\textbf{s}\in\mathbb{Z}^2} \delta(E-E_{\textbf{s},\textbf{k}}).
\end{align}
Here, $E_{\textbf{s},\textbf{k}}=(\frac{2\pi}{L}\textbf{s}+\frac{1}{L}\textbf{k})^2$ from Eq. (\ref{euc11}) and $\textbf{k}=(k_1,k_2)$ is an external parameter. We obtain
\begin{align}
 \nonumber D_{\textbf{k}}(E) &= \sum_{\textbf{n}\in\mathbb{Z}^2} \int\mbox{d}^2s\ e^{2\pi \rmi \textbf{n}\cdot\textbf{s}} \delta\Bigl(E-\Bigl(\frac{2\pi}{L}\textbf{s}+\frac{1}{L}\textbf{k}\Bigr)^2\Bigr)\\
 \nonumber &= \frac{L^2}{4\pi^2}\sum_{\textbf{n}\in\mathbb{Z}^2}  \int\mbox{d}^2p\ e^{\rmi L\textbf{n}\cdot\textbf{p}} \delta\Bigl(E-\Bigl(\textbf{p}+\frac{1}{L}\textbf{k}\Bigr)^2\Bigr)\\
 \label{AppEu3} &= \frac{L^2}{4\pi} \sum_{\textbf{n}\in\mathbb{Z}^2} e^{-\rmi \textbf{n}\cdot\textbf{k}} J_0(\sqrt{E}L|\textbf{n}|),
\end{align}
where 
\begin{align}
 \label{AppEu4} J_0(y) = \int_0^{2\pi} \frac{\mbox{d}\vphi}{2\pi} e^{\rmi y \cos\vphi}
\end{align}
is a Bessel function. We split off the contribution from $\textbf{n}=(0,0)$ and write
\begin{align}
 \label{AppEu5} D_{\textbf{k}}(E) = \frac{L^2}{4\pi}+ \frac{L^2}{4\pi} \sum_{\textbf{n}\in\mathbb{Z}^2\backslash(0,0)} e^{-\rmi \textbf{n}\cdot\textbf{k}} J_0(\sqrt{E}L|\textbf{n}|).
\end{align}
For a suitable testfunction $H(E)=h(p)$ we have
\begin{align}
 \nonumber &\sum_{\textbf{s}\in\mathbb{Z}^2} H(E_{\textbf{s},\textbf{k}}) = \int_0^\infty\mbox{d}E\ H(E)D_{\textbf{k}}(E)\\
 \nonumber &=\frac{L^2}{2\pi}\int_0^\infty\mbox{d}p'\ p'h(p') \\
  \nonumber &+\sum_{\textbf{n}\in\mathbb{Z}^2\backslash(0,0)} e^{-\rmi \textbf{n}\cdot\textbf{k}} \frac{L^2}{2\pi}\int_0^\infty\mbox{d}p' p' \int_0^{2\pi}\frac{\mbox{d}\vphi}{2\pi} e^{\rmi p' L n \cos\vphi} h(p')\\
 \label{AppEu6} &= L^2\int \frac{\mbox{d}^2p}{(2\pi)^2} h(p)+ L^2\sum_{\textbf{n}\in\mathbb{Z}^2\backslash(0,0)} e^{-\rmi \textbf{n}\cdot\textbf{k}}\ \tilde{h}_2(L_{\textbf{n}}),
\end{align}
which equals Eq. (\ref{euc12}). Equation (\ref{euc7}) is obtained in the limit $\textbf{k}=0$.

\section{Bolza surface}

\noindent In this section we discuss some properties that specifically apply to the Bolza surface and its underlying Fuchsian group $\Gamma$ defined in Eq. (\ref{intro3}). Tables \ref{TabEven} and \ref{TabOdd} collect the numbers $\nu_{n,Q}^{(a)}$ defined in Eq. (\ref{inv16}), and Table \ref{TabdTable} displays the numbers $d_{0,1,2}(n)$ that enter Eq. (\ref{gs10b}) for $n\leq 150$. We describe the algorithm to obtain the primitive periodic orbits on this surface and the associated sets of integer vectors $\mathcal{V}_n$ that enter the Selberg trace formula (\ref{hyp13}).

\renewcommand{\arraystretch}{1.4}
\begin{table*}[t]
\begin{tabular}{|c||c|c|c|c|c|c|c|c|c|c|c|c|c|c|c|c|c|c|c|c|c|}
\hline
 \multicolumn{22}{|c|}{ $\nu_{n,Q}^{(a)}$ for $n$ even} \\
\hline
\multicolumn{20}{c}{} \\
\hline
 $n$ & 2 & 6 & 10 & 38 & 42 & 46 & 50 & 54 & 58 & 82 & 86 & 90 & 94 & 98 & 102 & 130 & 134 & 138 & 142 &  \ 146 \  &  \ 150 \  \\
\hline\hline
 $\mathcal{W}_2$              & 1 & 2 & 2 & 4 & 8 & 8 & 8 & 8 &   & 4 & 10 & 14 & 8 & 8 & 12 & 8 & 24 & 16 & 16 & 16 & 12 \\
\hline
 \ $\mathcal{W}_{10}^{(1)}$ \ &   &   &   &   & 1 &   &   &   &   &   &  2  & 2  & 2   &  2 &    & 2 & 2  &    & 2  & 2  &   \\
\hline
 $\mathcal{W}_{10}^{(2)}$     &   &   &   &   &   &   & 2 &   & 1 &   &    &   &    &   &    & 2 & 2  & 4  & 4  & 2  &   \\
\hline
\multicolumn{22}{c}{} \\
\hline
  \ $n$ \  &  \ 4 \  &  \ 12 \  &  \ 20 \  &  \ 28 \  &  \ 36 \  &  \ 44 \  &  \ 52 \  &  \ 60 \  &  \ 68 \  &  \ 76 \  &  \ 84 \  &  \ 92 \  &  \ 100 \  &  \ 108 \  &  \ 116 \  &  \ 124 \  &  \ 132 \  &  \ 140 \  &  \ 148 \  &  &  \\
\hline\hline
 $\mathcal{W}_4$          &  & 2 & 2 & 4 & 4 & 4 & 4 & 8 &   & 4 & 8 & 4 & 8 & 8 & 4 & 8 & 8 &  & 8 &  &  \\
\hline
 $\mathcal{W}_{12}^{(1)}$ &  &   &   &   &   &   &   &   & 1 &   &   &   &   &   &   &   &   &  &   &  &  \\
\hline 
 $\mathcal{W}_{12}^{(2)}$ &  &   &   &   &   &   &   &   &   &   &   &   &   & 2 &   &   &   &  & 2 &  &  \\
\hline
\multicolumn{22}{c}{} \\
\hline
 $n$ & 8 & 16 & 24 & 32 & 40 & 48 & 56 & 64 & 72 & 80 & 88 & 96 & 104 & 112 & 120 & 128 & 136 & 144 &  &  &  \\
\hline\hline
 $\mathcal{W}_0$ & 8 & 12 &  & 32 &   &  &   &   &  & 48 &   & 48 &   & 48 &   &   &   &   &  &  &  \\
\hline
 $\mathcal{W}_8$ &   &    &  &    & 2 &  & 2 & 4 &  &    & 4 &    & 8 &    & 4 & 4 & 4 & 4 &  &  &  \\
\hline
\multicolumn{22}{c}{} \\
\hline
 $n$ & \ 14 \ & 18 & 22 & 26 & 30 & 34 & 62 & 66 & 70 & 74 & 78 & 106 & 110 & 114 & 118 & 122 & 126 &  &  &  &  \\
\hline\hline
 $\mathcal{W}_{6}^{(1)}$ & 1 &   & 2 & 2 &   &   & 4 & 4 & 4 & 6 & 4 & 2 &   & 8 & 4 & 4 & 7 &  &  &  &  \\
\hline
 $\mathcal{W}_{6}^{(2)}$ &   & 2 &   & 2 & 2 & 2 & 2 & 4 & 4 &   & 4 & 2 & 6 & 6 & 4 & 4 & 4 &  &  &  &  \\
\hline
\end{tabular}
\caption{Number of times $\nu_{n,Q}^{(a)}$ that $\mathcal{W}_Q^{(a)}$ appears in $\mathcal{V}_n$ for even $n$, see Eq. (\ref{inv16}). The nonvanishing entries in the table are those $\nu_{n,Q}^{(a)}$ that are nonzero, whereas all entries not displayed or not included in the table are zero.}
\label{TabEven}
\end{table*}
\renewcommand{\arraystretch}{1}

\renewcommand{\arraystretch}{1.4}
\begin{table*}
\begin{tabular}{|c||c|c|c|c|c|c|c|c|c|c|c|c|c|c|c|c|c|c|c|}
\hline
 \multicolumn{20}{|c|}{ $\nu_{n,Q}^{(a)}$ for $n$ odd} \\
\hline
\multicolumn{20}{c}{} \\
\hline
 $n$ & \ 1 \  & \ 11 \  & \ 17 \  &  \ 27 \  &  \ 29 \  &  \ 39 \  &  \ 41 \  &  \ 57 \  &  \ 67 \  &  \ 69 \  &  \ 79 \  &  \ 85 \  &  \ 95 \  &  \ 97 \  &  \ 107 \  &  \ 125 \  &  \ 135 \  &  \ 137 \  &  \ 147 \  \\
\hline\hline
 $\mathcal{W}_1$               & 1 & 4 & 2 & 4 & 8 & 12 & 2 & 10 & 12 & 8 & 8 & 8 & 16 & 12 & 12 & 24 & 16 & 12 & 16 \\
\hline 
 $\mathcal{W}_{5}^{(2)}$       &   & 2 &   & 2 & 4 & 4  &   &  4 &  4 & 6 & 4 &   &    &  4 &  6 &  6 &  8 &  4 &  4 \\
\hline
 $\mathcal{W}_{9}^{(2)}$       &   &   &   &   &   & 2  &   &  1 &    &   &   & 4 &  4 &    &    &  4 &  4 &  2 &  4 \\
\hline
 $\mathcal{W}_{9}^{(3)}$       &   &   &   &   &   &    & 1 &  2 &  2 &   &   &   &  4 &  4 &  4 &    &    &  1 &  2 \\
\hline
 $\mathcal{W}_{9}^{(5)}$       &   &   &   &   &   &    &   &  4 &    &   &   &   &    &  4 &    &    &  8 &  2 &  4 \\
\hline
 \ $\mathcal{W}_{13}^{(1)}$ \  &   &   &   &   &   &    &   &    &    & 1 &   &   &    &    &    &    &    &    &   \\
\hline
 $\mathcal{W}_{13}^{(3)}$      &   &   &   &   &   &    &   &    &    &   &   &   &    &    &  2 &    &    &    &   \\
\hline
 $\mathcal{W}_{13}^{(4)}$      &   &   &   &   &   &    &   &    &    &   &   &   &    &    &    &  2 &    &    &   \\
\hline
\multicolumn{20}{c}{} \\
\hline
 $n$ & 3 & 13 & 15 & 25 & 43 & 53 & 55 & 65 & 71 & 81 & 83 & 93 & 99 & 109 & 111 & 121 & 123 & 139 & 149 \\
\hline\hline
 $\mathcal{W}_{3}^{(1)}$  & 1 & 2 & 4 & 4 & 5 & 10 & 4 &   &   & 8 & 6 & 8 & 2 & 4 & 8 & 12 & 4 & 12 & 14 \\
\hline
 $\mathcal{W}_{7}^{(1)}$  &   &   & 1 &   & 2 &  2 & 3 & 2 & 4 & 2 & 4 & 2 &   & 4 & 4 &  8 &   &  2 &  2 \\
\hline
 $\mathcal{W}_{11}^{(3)}$ &   &   &   &   &   &    &   &   &   & 2 &   & 2 &   &   &   &  4 & 1 &  1 &    \\
\hline 
 $\mathcal{W}_{11}^{(4)}$ &   &   &   &   &   &    &   &   &   &   & 2 &   & 1 &   &   &    & 2 &  2 &  4 \\
\hline
\multicolumn{20}{c}{} \\
\hline
 $n$ & 5 & 7 & 23 & 33 & 35 & 45 & 51 & 61 & 63 & 73 & 91 & 101 & 103 & 113 & 119 & 129 & 131 & 141 &  \\
\hline\hline
 $\mathcal{W}_{3}^{(2)}$  & 2 & 1 & 3 & 4 & 4 & 6 & 4 & 4 & 5 & 8 & 12 & 6 & 8 & 4 & 10 & 12 & 12 & 12 &  \\
\hline
 $\mathcal{W}_{7}^{(2)}$  &   &   & 2 & 2 & 1 &   & 1 & 2 & 2 & 2 &  3 & 4 & 4 & 2 &  2 &  6 &  3 &  4 &  \\
\hline
 $\mathcal{W}_{11}^{(1)}$ &   &   &   &   &   &   &   & 2 &   &   &    & 2 & 1 &   &  2 &    &    &    &  \\
\hline 
 $\mathcal{W}_{11}^{(2)}$ &   &   &   &   &   &   &   &   & 1 &   &  2 &   & 2 & 2 &    &  2 &    &  2 &  \\
\hline
\multicolumn{20}{c}{} \\
\hline
 $n$ & 9 & 19 & 21 & 31 & 37 & 47 & 49 & 59 & 75 & 77 & 87 & 89 & 105 & 115 & 117 & 127 & 133 & 143 & 145 \\
\hline\hline
 $\mathcal{W}_{5}^{(1)}$  & 1 & 2 & 2 & 2 & 2 & 2 & 4 & 4 & 2 & 6 & 4 & 1 & 3 & 6 & 8 & 4 & 2 & 4 & 6 \\
\hline
 $\mathcal{W}_{9}^{(1)}$  &   &   &   &   & 1 &   & 2 &   &   & 2 & 2 &   & 2 & 4 & 2 & 2 & 2 & 2 & 4 \\
\hline 
 $\mathcal{W}_{9}^{(4)}$  &   &   &   &   &   & 2 &   & 2 & 2 & 1 & 2 & 2 &   & 2 & 3 & 2 & 1 &   & 4 \\
\hline
 $\mathcal{W}_{13}^{(2)}$ &   &   &   &   &   &   &   &   &   &   &   &   & 1 &   &   &   &   &   &   \\
\hline
\end{tabular}
\caption{Number of times $\nu_{n,Q}^{(a)}$ that $\mathcal{W}_Q^{(a)}$ appears in $\mathcal{V}_n$ for odd $n$, see Eq. (\ref{inv16}). The nonvanishing entries in the table are those $\nu_{n,Q}^{(a)}$ that are nonzero, whereas all entries not displayed or not included in the table are zero.}
\label{TabOdd}
\end{table*}
\renewcommand{\arraystretch}{1}

\renewcommand{\arraystretch}{1.4}
\begin{table*}
\begin{tabular}{|c|c|c|c||c|c|c|c||c|c|c|c||c|c|c|c|}
\hline
$n$ &  $d_0(n)$   &   $d_1(n)$   &   $d_2(n)$   &  $n$ &  $d_0(n)$   &   $d_1(n)$   &   $d_2(n)$   & $n$ &  $d_0(n)$   &   $d_1(n)$   &   $d_2(n)$   & $n$ &  $d_0(n)$   &   $d_1(n)$   &   $d_2(n)$  \\
\hline\hline
1	&	24	&	1	&	1	&	39	&	576	&	88	&	536	&	77	&	864	&	228	&	1572	&	115	&	1152	&	336	&	2544	\\
2	&	24	&	2	&	4	&	40	&	48	&	16	&	128	&	78	&	384	&	96	&	576	&	116	&	96	&	16	&	64	\\
3	&	48	&	6	&	18	&	41	&	96	&	20	&	164	&	79	&	384	&	48	&	208	&	117	&	1248	&	340	&	2420	\\
4	&	0	&	0	&	0	&	42	&	288	&	56	&	432	&	80	&	48	&	0	&	0	&	118	&	384	&	96	&	576	\\
5	&	96	&	12	&	36	&	43	&	432	&	86	&	482	&	81	&	768	&	192	&	1504	&	119	&	768	&	160	&	1056	\\
6	&	48	&	4	&	8	&	44	&	96	&	16	&	64	&	82	&	96	&	8	&	16	&	120	&	96	&	32	&	256	\\
7	&	48	&	6	&	18	&	45	&	288	&	36	&	108	&	83	&	768	&	192	&	1376	&	121	&	1728	&	472	&	3720	\\
8	&	8	&	0	&	0	&	46	&	192	&	16	&	32	&	84	&	192	&	32	&	128	&	122	&	384	&	96	&	576	\\
9	&	96	&	20	&	100	&	47	&	384	&	112	&	848	&	85	&	384	&	80	&	656	&	123	&	384	&	112	&	1040	\\
10	&	48	&	4	&	8	&	48	&	0	&	0	&	0	&	86	&	432	&	100	&	840	&	124	&	192	&	32	&	128	\\
11	&	192	&	24	&	104	&	49	&	576	&	152	&	1048	&	87	&	768	&	224	&	1696	&	125	&	1152	&	208	&	1648	\\
12	&	48	&	8	&	32	&	50	&	288	&	56	&	432	&	88	&	96	&	32	&	256	&	126	&	528	&	132	&	792	\\
13	&	96	&	12	&	36	&	51	&	288	&	52	&	268	&	89	&	288	&	92	&	748	&	127	&	768	&	224	&	1696	\\
14	&	48	&	12	&	72	&	52	&	96	&	16	&	64	&	90	&	528	&	108	&	856	&	128	&	96	&	32	&	256	\\
15	&	288	&	52	&	268	&	53	&	672	&	116	&	572	&	91	&	1056	&	244	&	1772	&	129	&	1344	&	328	&	2360	\\
16	&	12	&	0	&	0	&	54	&	192	&	16	&	32	&	92	&	96	&	16	&	64	&	130	&	480	&	136	&	1232	\\
17	&	48	&	2	&	2	&	55	&	480	&	108	&	660	&	93	&	768	&	192	&	1504	&	131	&	864	&	156	&	804	\\
18	&	96	&	24	&	144	&	56	&	48	&	16	&	128	&	94	&	384	&	96	&	832	&	132	&	192	&	32	&	128	\\
19	&	192	&	40	&	200	&	57	&	672	&	140	&	1020	&	95	&	768	&	160	&	1312	&	133	&	480	&	148	&	1172	\\
20	&	48	&	8	&	32	&	58	&	48	&	20	&	200	&	96	&	48	&	0	&	0	&	134	&	864	&	168	&	1296	\\
21	&	192	&	40	&	200	&	59	&	576	&	152	&	1048	&	97	&	768	&	160	&	1184	&	135	&	1152	&	240	&	1712	\\
22	&	96	&	24	&	144	&	60	&	192	&	32	&	128	&	98	&	384	&	96	&	832	&	136	&	96	&	32	&	256	\\
23	&	336	&	74	&	446	&	61	&	480	&	124	&	948	&	99	&	144	&	34	&	278	&	137	&	672	&	124	&	860	\\
24	&	0	&	0	&	0	&	62	&	288	&	72	&	432	&	100	&	192	&	32	&	128	&	138	&	576	&	112	&	864	\\
25	&	192	&	24	&	72	&	63	&	528	&	130	&	966	&	101	&	768	&	192	&	1376	&	139	&	960	&	216	&	1576	\\
26	&	192	&	48	&	288	&	64	&	96	&	32	&	256	&	102	&	288	&	24	&	48	&	140	&	0	&	0	&	0	\\
27	&	192	&	24	&	104	&	65	&	192	&	56	&	392	&	103	&	1008	&	270	&	2138	&	141	&	1152	&	272	&	1968	\\
28	&	96	&	16	&	64	&	66	&	384	&	96	&	576	&	104	&	192	&	64	&	512	&	142	&	768	&	192	&	1664	\\
29	&	384	&	48	&	208	&	67	&	576	&	88	&	536	&	105	&	576	&	184	&	1624	&	143	&	576	&	152	&	1048	\\
30	&	96	&	24	&	144	&	68	&	48	&	24	&	288	&	106	&	192	&	48	&	288	&	144	&	96	&	32	&	256	\\
31	&	192	&	40	&	200	&	69	&	576	&	120	&	984	&	107	&	960	&	248	&	2312	&	145	&	1344	&	408	&	3192	\\
32	&	32	&	0	&	0	&	70	&	384	&	96	&	576	&	108	&	288	&	80	&	704	&	146	&	672	&	152	&	1264	\\
33	&	384	&	80	&	464	&	71	&	384	&	112	&	784	&	109	&	576	&	136	&	856	&	147	&	960	&	200	&	1512	\\
34	&	96	&	24	&	144	&	72	&	0	&	0	&	0	&	110	&	288	&	72	&	432	&	148	&	288	&	80	&	704	\\
35	&	288	&	52	&	268	&	73	&	576	&	104	&	536	&	111	&	768	&	160	&	928	&	149	&	1056	&	228	&	1612	\\
36	&	96	&	16	&	64	&	74	&	288	&	72	&	432	&	112	&	48	&	0	&	0	& \	150 \	&	288	&	24	&	48	\\
37	&	288	&	76	&	524	&	75	&	384	&	112	&	848	&	113	&	576	&	168	&	1432	&	 	&	 	&	 	&	 	\\
 \ 38 \ 	&	96	&	8	&	16	& \	76 \ 	&	96	&	16	&	64	& \	114 \ 	&	672	&	168	&	1008	&	 &	 	&	 	&	 	\\
\hline
\end{tabular}
\caption{Number of primitive periodic orbits of length $\ell_n$, $d_0(n)$, and coefficients $d_1(n)$ and $d_2(n)$ in Eq. (\ref{gs10b}) for $1\leq n \leq 150$. These numbers can be obtained from Eqs. (\ref{inv18c})-(\ref{inv18e}). The values of $d_0(n)$ agree with those determined in Refs. \cite{AURICH1988451,AURICH199191}.}
\label{TabdTable}
\end{table*}
\renewcommand{\arraystretch}{1}

\subsection{Arithmetic Fuchsian group}\label{AppFuchs}

\noindent In this section, we closely follow Ref. \cite{AURICH199191}. The Fuchsian group $\Gamma \subset \text{PSU}(1,1)$ of the Bolza surface is generated by products of four generators and their inverses according to the presentation in Eq. (\ref{intro3}). An explicit representation of the generators is
\begin{align}
 \label{fuchs2} \gamma_\mu = \mathcal{R}(\pi/4)^{\mu-1} \gamma_1 \mathcal{R}(-\pi/4)^{\mu-1},
\end{align}
with
\begin{align}
 \label{fuchs3} \gamma_1 & = \begin{pmatrix} 1+\sqrt{2} & \sqrt{\zeta}(2+\sqrt{2}) \\ \sqrt{\zeta}(2+\sqrt{2}) & 1+\sqrt{2}\end{pmatrix},
\end{align}
where 
\begin{align}
\zeta=\sqrt{2}-1
\end{align}
and
\begin{align}
 \label{fuchs4} \mathcal{R}(\alpha) &= \begin{pmatrix} e^{\rmi  \alpha/2} & 0 \\ 0 & e^{-\rmi  \alpha/2} \end{pmatrix}.
\end{align}
Note that $\gamma_5=\gamma_1^{-1},\ \gamma_6=\gamma_2^{-1}$ and so on. We observe that $\mbox{det}(\gamma_\mu)=1$ for all $\mu$, and so $\mbox{det}(\gamma)=1$ for all $\gamma\in\Gamma$.

The group $\Gamma$ is an arithmetic Fuchsian group, which means that it is generated from a quaternion algebra over an algebraic number field \cite{BOGOMOLNY1997219,BogAri}. In this case, this implies that besides the representation of every group element $\gamma$ in terms of a product of the generators $\{\gamma_\mu\}$, there exists another representation where every $\gamma\in \Gamma$ is identified by four integers $\vec{n}=(n_1,n_2,n_3,n_4)$ and a parity. This representation reads
\begin{align}
 \label{fuchs6} \gamma = \begin{pmatrix} N_1 +\rmi N_2 & \sqrt{\zeta}(N_3+\rmi N_4) \\ \sqrt{\zeta}(N_3-\rmi N_4) &  N_1 -\rmi N_2 \end{pmatrix},
\end{align}
where $N_\mu\in\mathbb{Z}[\sqrt{2}]$ are specific algebraic integers. For example, the matrix $\gamma_1$ from Eq. (\ref{fuchs3}) corresponds to $N_1=1+\sqrt{2}$, $N_2=0$, $N_3=2+\sqrt{2}$, $N_4=0$. To define the relation between $N_\mu$ and $n_\mu$, we introduce $m_{\rm e}(n)$ and $m_{\rm o}(n)$ as the even and odd integers $m$ that best approximate $m\approx n\sqrt{2}$. Thus
\begin{align}
 \label{fuchs7} |m - n \sqrt{2}|_{m=m_{\rm e}(n) \text{ even}} \stackrel{!}{=} \text{min},\\
 \label{fuchs8} |m - n \sqrt{2}|_{m=m_{\rm o}(n) \text{ odd}} \stackrel{!}{=} \text{min}.
\end{align}
We then define the algebraic integers
\begin{align}
 \label{fuchs9} N_{\rm e}(n) &= m_{\rm e}(n) + n \sqrt{2},\\
 \label{fuchs10} N_{\rm o}(n) &= m_{\rm o}(n) + n \sqrt{2}.
\end{align}
We have
\begin{align}
 \label{fuchs11} 2n\sqrt{2}-1 \leq N_{\rm e/o}(n) \leq 2n\sqrt{2}+1
\end{align}
for all $n\geq 1$. We say that the algebraic integers $N_{\rm e}(n)$ and $N_{\rm o}(n)$ have even and odd parity, respectively. The matrices of form $\gamma$ from Eq. (\ref{fuchs6}) that comprise the group $\Gamma$ are such that
\begin{itemize}
 \item[(i)] $N_1$ has odd parity.
 \item[(ii)] $N_2$ has even parity.
 \item[(iii)] $N_3$ and $N_4$ have equal parity.
\end{itemize}
Consequently, matrices $\gamma\in\Gamma$ are specified by four integers $\vec{n}=(n_1,n_2,n_3,n_4)\in\mathbb{Z}^4$ and the parity of the off-diagonal entries, yielding two possibilities
\begin{align} 
 \label{fuchs12} \gamma_{\rm e} &= \begin{pmatrix} N_{\rm o}(n_1) +\rmi N_{\rm e}(n_2) & \sqrt{\zeta}[N_{\rm e}(n_3)+\rmi N_{\rm e}(n_4)] \\ \sqrt{\zeta}[N_{\rm e}(n_3)-\rmi N_{\rm e}(n_4)] &  N_{\rm o}(n_1) -\rmi N_{\rm e}(n_2) \end{pmatrix},\\
 \label{fuchs13} \gamma_{\rm o} &= \begin{pmatrix} N_{\rm o}(n_1) +\rmi N_{\rm e}(n_2) & \sqrt{\zeta}[N_{\rm o}(n_3)+\rmi N_{\rm o}(n_4)] \\ \sqrt{\zeta}[N_{\rm o}(n_3)-\rmi N_{\rm o}(n_4)] &  N_{\rm o}(n_1) -\rmi N_{\rm e}(n_2) \end{pmatrix}.
\end{align}
Only those sets of integers $\vec{n}\in \mathbb{Z}^4$ that yield $\mbox{det}(\gamma)=1$ are allowed. Since the matrices $\gamma$ and $-\gamma$ are identified in $\text{PSU}(1,1)$, we can always assume that $n_1\geq 0$.

Fuchsian group elements acquire a geometric interpretation through their action on the Poincar\'{e} disk in Eq. (\ref{geo5}). Each $1\neq \gamma\in\Gamma$ is hyperbolic, i.e. satisfies $|\mbox{tr}(\gamma)|>2$, and leaves two points on the boundary $\partial\mathbb{D}$ invariant. These two points are connected by a unique geodesic in $\mathbb{D}$ that is left invariant under the action of $\gamma$, i.e. points on the geodesic are mapped to points on the geodesic. Writing $z=x+\rmi y\in\mathbb{D}$, the geodesic is given by the circular arc
\begin{align}
 \label{fuchs14} x^2+y^2-\frac{2\sqrt{\zeta}}{N_2}(N_3y-N_4x)+1=0
\end{align}
if $N_2\neq 0$, or by the straight line
\begin{align}
 \label{fuchs15} N_3y-N_4x=0
\end{align}
if $N_2=0$. Furthermore, given that $|\mbox{tr}(\gamma)|>2$, we call the positive number
\begin{align}
 \label{fuchs16} \ell(\gamma) = 2\kappa\ \text{arcosh}\Bigl(\frac{|\mbox{tr}(\gamma)|}{2}\Bigr)
\end{align}
the length of $\gamma$ ($\kappa=1$). For a group element represented in the form of Eq. (\ref{fuchs6}), the length is given by
\begin{align}
 \label{fuchs17} \ell(\gamma) = \ell_{n_1} = 2\kappa\ \text{arcosh}[m_{\rm o}(n_1)+n_1\sqrt{2}],
\end{align}
which only depends on $n_1$. For every $n\in\mathbb{N}$, we can then determine the number $d_0(n)$ of distinct primitive group elements (modulo conjugacy) of length $\ell_{n}$  and their representative group elements $\gamma\in \Gamma$. In the main text, we therefore write $n$ instead of $n_1$.

\subsection{Primitive periodic orbits}\label{AppOrb}

\noindent In this section we describe the algorithm for finding all distinct primitive periodic orbits on the Bolza surface with length $\ell_n$ for a given value of $n\geq 1$.

We first clarify what the geometric notion of a primitive periodic orbit means algebraically, i.e. when expressed in terms of group elements $\gamma\in\Gamma$. Recall that we only consider those geodesics that go through the fundamental domain or central octagon $\mathcal{D}$. One periodic orbit typically consists of several geodesic segments, see Fig. \ref{FigGeo}. Each such geodesic segment is uniquely associated to a group element $\gamma\in\Gamma$. Consequently, an orbit that consists of $N_O$ segments corresponds to a collection of $N_O$ group elements $\{\gamma^{(1)},\dots,\gamma^{(N_O)}\}$. Each element of this set is conjugate to the others, and so every one of them is a suitable representative of the orbit as a whole. Of course, the length of all elements is $\ell(\gamma^{(1)})=\dots=\ell(\gamma^{(N_O)})=\ell_n$.

Let us give some meaningful examples for $n=1$. The three orbits that are shown for $n=1$ in Fig. \ref{FigGeo}, from left to right, are
\begin{align}
 \label{orb1} \{\gamma_1\},\ \{\gamma_2\gamma_3^{-1},\gamma_3^{-1}\gamma_2\},\ \{\gamma_2\gamma_1^{-1}\gamma_4^{-1}\}.
\end{align}
Hence the first and third orbit consist of one segment, the second one consists of two segments. We readily verify that the length of every segment is $\ell_1=3.05714$. The two segments of the second orbit are conjugate to each other, because
\begin{align}
 \label{orb2} \gamma_2\gamma_3^{-1} = \gamma_3\ (\gamma_3^{-1}\gamma_2)\ \gamma_3^{-1}.
\end{align}
The remaining 21 orbits for $n=1$ are related to the three in Eq. (\ref{orb1}) by successive rotations by $\pi/4$ using $\mathcal{R}(\pi/4)$ from Eq. (\ref{fuchs4}). For instance, from the first orbit $\{\gamma_1\}$ we generate eight orbits given by
\begin{align}
 \nonumber &\{\gamma_1\},\\
 \nonumber &\mathcal{R}(\pi/4)\{\gamma_1\}\mathcal{R}(\pi/4)^{-1}=\{\gamma_2\},\\
 \nonumber &\mathcal{R}(2\pi/4)\{\gamma_1\}\mathcal{R}(2\pi/4)^{-1}=\{\gamma_3\},\\
 \nonumber &\vdots\\
 \label{orb3} &\mathcal{R}(7\pi/4)\{\gamma_1\}\mathcal{R}(7\pi/4)^{-1}=\{\gamma_4^{-1}\}.
\end{align}
Hence the set of representative elements is simply $\gamma_1,\gamma_2,\gamma_3,\gamma_4,\gamma_1^{-1},\gamma_2^{-1},\gamma_3^{-1},\gamma_4^{-1}$, corresponding to eight straight lines connecting opposite sites of the octagon. Repeating this procedure for the other two orbits, a complete set of representative elements of the 24 distinct primitive orbits for $n=1$ is found to be
\begin{align}
 \nonumber &\gamma_1,\gamma_2,\gamma_3,\gamma_4,\gamma_1^{-1},\gamma_2^{-1},\gamma_3^{-1},\gamma_4^{-1},\\
 \nonumber &\gamma_2\gamma_3^{-1}, \gamma_3\gamma_4^{-1}, \gamma_4\gamma_1, \gamma_1^{-1}\gamma_2, \gamma_2^{-1}\gamma_3, \gamma_3^{-1}\gamma_4, \gamma_4^{-1}\gamma_1^{-1}, \gamma_1\gamma_2^{-1},\\
 \nonumber & \gamma_2 \gamma_1^{-1}\gamma_4^{-1}, \gamma_3\gamma_2^{-1}\gamma_1, \gamma_4\gamma_3^{-1}\gamma_2, \gamma_1^{-1}\gamma_4^{-1}\gamma_3, \gamma_2^{-1}\gamma_1\gamma_4,\\
 \label{orb4} & \gamma_3^{-1}\gamma_2\gamma_1^{-1}, \gamma_4^{-1}\gamma_3\gamma_2^{-1}, \gamma_1\gamma_4\gamma_3^{-1}.
\end{align}
The vectors $\textbf{v}$ that are associated to the representative group elements can be read off from the factorization into group generators. We have
\begin{align}
 \nonumber &\mathcal{V}_1 = \mathcal{W}_1 = \Biggl\{{} \begin{pmatrix} \pm 1 \\ 0 \\ 0 \\ 0 \end{pmatrix},\ \begin{pmatrix} 0 \\ \pm 1 \\ 0 \\ 0 \end{pmatrix},\ \begin{pmatrix} 0 \\ 0 \\\pm 1 \\ 0 \end{pmatrix},\ \begin{pmatrix} 0 \\ 0 \\ 0 \\ \pm 1 \end{pmatrix},\\
 \nonumber & \pm \begin{pmatrix} 0 \\ 1 \\ -1 \\ 0 \end{pmatrix},\ \pm \begin{pmatrix} 0 \\ 0 \\ 1 \\ -1 \end{pmatrix},\ \pm \begin{pmatrix} 1 \\ 0 \\ 0 \\ 1 \end{pmatrix},\ \pm \begin{pmatrix} -1 \\ 1 \\ 0 \\ 0 \end{pmatrix},\\
 \label{orb5} & \pm \begin{pmatrix} -1 \\ 1 \\ 0 \\ -1 \end{pmatrix},\ \pm \begin{pmatrix} 1 \\ -1 \\ 1 \\ 0 \end{pmatrix},\ \pm \begin{pmatrix} 0 \\ 1 \\ -1 \\ 1 \end{pmatrix},\ \pm \begin{pmatrix} -1 \\ 0 \\ 1 \\ -1 \end{pmatrix} \Biggr\}.
\end{align}
Paralleling the construction in Eq. (\ref{orb3}), these vectors are related to the three vectors that are obtained from Eq. (\ref{orb1}), namely
\begin{align}
 \label{orb6} \begin{pmatrix} 1 \\ 0 \\ 0 \\ 0 \end{pmatrix},\ \begin{pmatrix} 0 \\ 1 \\ -1 \\ 0 \end{pmatrix},\ \begin{pmatrix} -1 \\ 1 \\ 0 \\ -1 \end{pmatrix}
\end{align}
by successive application of the rotation $R\in G$ in the four-dimensional representation from Eq. (\ref{inv3}).

We now describe how all distinct primitive periodic orbits can be obtained for a given value of $n\geq 1$ and $\ell_n$. This algorithm has been developed in Refs. \cite{AURICH1988451,AURICH199191}. 

(I) In the first step, we generate a large list of candidate matrices $\gamma_{\rm e}(n,n_2,n_3,n_4)$ and $\gamma_{\rm o}(n,n_2,n_3,n_4)$ with the correct value of $n$ and arbitrary values of $n_{2,3,4}$, see Eqs. (\ref{fuchs12}) and (\ref{fuchs13}). One can derive bounds on the number of candidates that need to be sampled, which puts upper limits on $n_{2,3,4}$, but we do not discuss them here. For practical purposes it is sufficient to say that the number of candidates is finite. In our implementation, we choose $0\leq n_4\leq n_3\leq n_{\rm max}$ and $-n_{\rm max}\leq n_2\leq n_{\rm max}$ and choose $n_{\rm max}$ sufficiently large in dependence of $n$. The restriction of $n_{3,4}$ being non-negative can always be achieved by suitable rotations with $\mathcal{R}(\pi/4)$ and so does not constitute a bias. 

(II) Among this initial draft of candidates, which we write as
\begin{align}
 \label{orb7} M=\begin{pmatrix} N_1+\rmi N_2 & \sqrt{\zeta} (N_3+\rmi N_4) \\ \sqrt{\zeta}(N_3-\rmi N_4) & N_1-\rmi N_2\end{pmatrix}, 
\end{align}
we choose those which satisfy two conditions: First, they need to be elements of $\Gamma$ with unit determinant, which yields the condition
\begin{align}
 \label{orb8} N_1^2+N_2^2-\zeta^2(N_3^2+N_4^2) \stackrel{!}{=}1.
\end{align}
Second, their associated geodesics need to pass through the central octagon $\mathcal{D}$. This can be shown to be equivalent to the condition
\begin{align}
 \label{orb9} \rho(M)=\frac{|N_2|}{(2-\sqrt{2})(|N_3|+\zeta|N_4|)}\stackrel{!}{<} 1.
\end{align}
For this equation to be true, we used that $n_3\geq n_4\geq 0$ so that $|N_3|\geq |N_4|$. Note that if $\rho(M)= 1$ occurred in Eq. (\ref{orb9}), then the geodesic would touch a corner of the octagon, but not go through the interior. We can exclude these candidates for the initial draft.

(IIIa) Let us denote the candidate matrices that passed the tests in (I) and (II) by
\begin{align}
 \label{orb10} \mathcal{L}  \subset \Gamma.
\end{align}
By construction we have $\ell(\gamma)=\ell_n$ for all $\gamma\in\mathcal{L}$. The set $\mathcal{L}$ is finite. But not every element $\gamma\in\mathcal{L}$ is representative of a distinct primitive orbit, since orbits can consist of several segments or group elements, such as the second orbit in Eq. (\ref{orb1}), which consists of the segments $\gamma_2\gamma_3^{-1}$ and $\gamma_3^{-1}\gamma_2$. Therefore, the next task is to group those entries of $\mathcal{L}$ together that belong to the same orbit, and then count the number of distinct groups or orbits to obtain $d_0(n)$.

(IIIb) The grouping into orbits is achieved as follows. We chose the first element $\gamma^{(1)}\in\mathcal{L}$ from the list and construct the associated orbit through a method described below. After this is accomplished, we have a list of matrices $\mathcal{L}_1=\{\gamma^{(1)},\dots,\gamma^{(N_O)}\}$ that comprises the $N_O$ segments of this orbit---this list contains at least one element, namely $\gamma^{(1)}$. Importantly though, if the list contains more than one element, then these elements also appear  in $\mathcal{L}$ because they would have passed all the tests in (I) and (II). To proceed, we chose another element from the list $\eta^{(1)}\in\mathcal{L}$. There are two possibilities: Either $\eta^{(1)}$ is one of the elements in $\mathcal{L}_1$, in which case its orbit is already accounted for; we can then discard $\eta^{(1)}$. If $\eta^{(1)}$ is not in $\mathcal{L}_1$, it is part of a distinct orbit and we construct the list $\mathcal{L}_2=\{\eta^{(1)},\dots,\eta^{(N_O')}\}$ of its segments. We update the list of known orbit segments to $\mathcal{L}_1\cup\mathcal{L}_2$. Choosing a third element $\kappa^{(1)}\in\mathcal{L}$, we have to check whether it is already contained in $\mathcal{L}_1\cup\mathcal{L}_2$ or not, and so on. At the end of the procedure, choosing every element of $\mathcal{L}$, we have a list of $d_0(n)$  mutually exclusive sets
\begin{align}
 \label{orb11} \mathcal{L}_1 \cup \mathcal{L}_2\cup\dots\cup \mathcal{L}_{d_0(n)}\subset \Gamma,
\end{align}
where each $\mathcal{L}_i$ represents one periodic orbit. A representative element for each orbit can be chosen arbitrarily from the $\mathcal{L}_i$.

(IIIc) Let us now describe how the orbit for an individual element $\gamma^{(1)}\in\mathcal{L}$ is constructed. We know that $M_1=\gamma^{(1)}$ is the first segment of the orbit. Any other segment connected to $M_1$ needs to be of the form
\begin{align}
 \label{orb12} M_1^{(\mu)} = \gamma_\mu M_1 \gamma_\mu^{-1}
\end{align}
with $\mu=1,\dots,8$, because the connection needs to be through one of the eight side-pairing of the octagon. The matrices $\{M_1^{(\mu)}\}$ are not necessarily in $\mathcal{L}$: Although they have unit determinant, we need to pick those that pass through the central octagon. To do so, we first test if any of the eight matrices satisfies
\begin{align}
 \label{orb13} \rho(M_1^{(\mu)})<1.
\end{align}
(This will be satisfied by none or two of the eight matrices.) If Eq. (\ref{orb13}) is true for any particular matrix, we chose this matrix as $M_2=\gamma^{(2)}$. If Eq. (\ref{orb13}) is not satisfied for any $\mu$, then we test whether 
\begin{align}
 \label{orb14} \rho(M_1^{(\mu)})= 1
\end{align}
is true for any of the eight matrices. If this is true, we choose this matrix as $M_2$. (The segment goes through a corner of the octagon if Eq. (\ref{orb14}) but not Eq. (\ref{orb13}) is satisfied.) In most cases, Eqs. (\ref{orb13}) or (\ref{orb14}) will have yielded a new matrix $M_2$. We then construct the matrix $M_3$ that connects to $M_2$ by considering the eight matrices
\begin{align}
 \label{orb15} M_2^{(\mu)} = \gamma_\mu M_2\gamma_\mu^{-1}
\end{align}
subject to the test $\rho<1$ or $\rho=1$. Importantly, $M_3$ cannot be chosen among the segments $\{M_1,M_2\}$ that we already have. We continue this procedure to eventually obtain a list of distinct matrices $\{M_1,M_2,\dots,M_{N_O}\}$ which is such that considering the eight matrices $M_{N_O}^{(\mu)}$ in Eqs. (\ref{orb13}) and (\ref{orb14}) does not yield any matrices not yet contained in the list. This means that we have found all $N_O$ segments of the orbit, and
\begin{align}
 \label{orb16} \mathcal{L}_1 = \{M_1,M_2,\dots,M_{N_O}\} = \{\gamma^{(1)},\gamma^{(2)},\dots,\gamma^{(N_O)}\}.
\end{align}

(IIId) We comment here on the fact that achieving $N_3\geq N_4\geq 0$ through suitable rotation of $M$ by $\pi/4$ in $\rho(M)$ in Eq. (\ref{orb9}) can be avoided by considering the following modified function: Write $M$ as in Eq. (\ref{orb7}) and define
\begin{widetext}
\begin{align}
 \label{orb17} \bar{\rho}(M)=\min_{\mu=1,\dots,8}\Biggl[ \frac{|N_2|}{(2-\sqrt{2})\Bigl(|\text{Re}[(N_3+\rmi N_4)e^{\rmi \mu\pi/4}]|+\zeta| \text{Im}[(N_3+\rmi N_4)e^{\rmi \mu\pi/4}]|\Bigr)}\Biggr]. 
\end{align}
\end{widetext}
Instead of $\rho(M_1^{(\mu)})<1$ and $\rho(M_1^{(\mu)})= 1$ in Eqs. (\ref{orb13}) and (\ref{orb14}), we can equivalently consider $\bar{\rho}(M_1^{(\mu)})<1$ and $\bar{\rho}(M_1^{(\mu)})= 1$, respectively.

(IV) At last we have to find the factorization of the representatives taken from the $\mathcal{L}_i$ in Eq. (\ref{orb11}) to construct the vectors $\textbf{v}\in\mathcal{V}_n$. The remarkable feature of the algorithm just described is that it yields this factorization in almost all cases. We explain the procedure on the example of $\mathcal{L}_1=\{M_1,\dots,M_{N_O}\}$. The construction in (IIIc) implies that
\begin{align}
 \nonumber M_2 &= \gamma_{\mu_1}M_1\gamma_{\mu_1}^{-1},\\
 \nonumber M_3 &=\gamma_{\mu_2}M_2\gamma_{\mu_2}^{-1}\\
 \nonumber &=\gamma_{\mu_2}\gamma_{\mu_1}M_1(\gamma_{\mu_2}\gamma_{\mu_1})^{-1},\\
 \nonumber &\vdots\\
 \label{orb18} M_{N_O} &= (\gamma_{N_O-1}\cdots \gamma_{\mu_2}\gamma_{\mu_1})M_1 (\gamma_{N_O-1}\cdots \gamma_{\mu_2}\gamma_{\mu_1})^{-1}
\end{align}
for some set of indices $(\mu_1,\mu_2,\dots,\mu_{N_O-1})$ determined  by the algorithm. The fact that the algorithm stops at $M_{N_O}$ implies that there is a $\mu'$ such that 
\begin{align}
 \label{orb19}  M_1 = \gamma_{\mu'} M_{N_O}\gamma_{\mu'}^{-1},
\end{align} 
yielding a closed orbit eventually. Together with Eq. (\ref{orb18}), this implies that $\gamma_{\mu'}\gamma_{N_O-1}\cdots \gamma_{\mu_2}\gamma_{\mu_1}$ commutes with $M_1$. Since $M_1$ is primitive, however, the only group elements commuting with $M_1$ are powers of $M_1$ \cite{marklof2012ii}, and so we obtain
\begin{align}
 \label{orb20} M_1 = \pm (\gamma_{\mu'}\gamma_{N_O-1}\cdots \gamma_{\mu_2}\gamma_{\mu_1})^n
\end{align}
with $n\in\mathbb{Z}$. Testing all possibilities for $\mu'$ and the overall sign  $\pm$, and using small values of $n$, the factorization of $M_1$ is easily found. We found in our numerics that, if none of the segments go through the corners of the octagon, then Eq. (\ref{orb20}) is always solved by $n=1$. If some segments go through the corners, then additionally considering $n=-1$ often gives the correct factorization. In the very few cases where Eq. (\ref{orb20}) with $n=\pm 1$ does not yield the right result, more direct but computationally more expensive methods can be used to find the factorization of $M_1$.

(V) At last let us comment on the role of non-primitive orbits. If $\gamma$ represents a periodic orbit of length $\ell_n$, then $\gamma^m$ represents a periodic orbit of length $m\ell_n$. Consequently, for a given $n\geq 1$, there might be pairs $(n_1,m_1)$ such that
\begin{align}
 \label{orb21} \ell_n = m_1 \ell_{n_1}.
\end{align}
The algorithm described in (I)-(IV) does not distinguish these orbits and so would yield the corresponding matrices of the non-primitive pair $(n_1,m_1)$ as well. Since all primitive orbits are still determined correctly, we can simply eliminate the non-primitive ones by hand. Non-primitive contributions for $n\leq 150$ appear very rarely. They only occur for
\begin{align}
 n&=4:\ (n_1,m_1)=(1,2),\\
 n&= 17:\ (n_1,m_1)=(1,3),\\
 n&=24:\ (n_1,m_1)=(2,2),\\
 n&=60:\ (n_1,m_1)=(3,2),\\
 n&= 80:\ (n_1,m_1)=(1,4),\\
 \label{orb22} n&= 140:\ (n_1,m_1)=(5,2).
\end{align}
The values of $d_0(n)$ in Table \ref{TabdTable} have been corrected for the non-primitive orbits, and thus solely count the primitive contributions.

\end{appendix}

\bibliography{refs_selberg}

\begin{thebibliography}{72}%
\makeatletter
\providecommand \@ifxundefined [1]{%
 \@ifx{#1\undefined}
}%
\providecommand \@ifnum [1]{%
 \ifnum #1\expandafter \@firstoftwo
 \else \expandafter \@secondoftwo
 \fi
}%
\providecommand \@ifx [1]{%
 \ifx #1\expandafter \@firstoftwo
 \else \expandafter \@secondoftwo
 \fi
}%
\providecommand \natexlab [1]{#1}%
\providecommand \enquote  [1]{``#1''}%
\providecommand \bibnamefont  [1]{#1}%
\providecommand \bibfnamefont [1]{#1}%
\providecommand \citenamefont [1]{#1}%
\providecommand \href@noop [0]{\@secondoftwo}%
\providecommand \href [0]{\begingroup \@sanitize@url \@href}%
\providecommand \@href[1]{\@@startlink{#1}\@@href}%
\providecommand \@@href[1]{\endgroup#1\@@endlink}%
\providecommand \@sanitize@url [0]{\catcode `\\12\catcode `\$12\catcode
  `\&12\catcode `\#12\catcode `\^12\catcode `\_12\catcode `\%12\relax}%
\providecommand \@@startlink[1]{}%
\providecommand \@@endlink[0]{}%
\providecommand \url  [0]{\begingroup\@sanitize@url \@url }%
\providecommand \@url [1]{\endgroup\@href {#1}{\urlprefix }}%
\providecommand \urlprefix  [0]{URL }%
\providecommand \Eprint [0]{\href }%
\providecommand \doibase [0]{https://doi.org/}%
\providecommand \selectlanguage [0]{\@gobble}%
\providecommand \bibinfo  [0]{\@secondoftwo}%
\providecommand \bibfield  [0]{\@secondoftwo}%
\providecommand \translation [1]{[#1]}%
\providecommand \BibitemOpen [0]{}%
\providecommand \bibitemStop [0]{}%
\providecommand \bibitemNoStop [0]{.\EOS\space}%
\providecommand \EOS [0]{\spacefactor3000\relax}%
\providecommand \BibitemShut  [1]{\csname bibitem#1\endcsname}%
\let\auto@bib@innerbib\@empty
\bibitem [{\citenamefont {Koll{\'a}r}\ \emph {et~al.}(2019)\citenamefont
  {Koll{\'a}r}, \citenamefont {Fitzpatrick},\ and\ \citenamefont
  {Houck}}]{kollar2019hyperbolic}%
  \BibitemOpen
  \bibfield  {author} {\bibinfo {author} {\bibfnamefont {A.~J.}\ \bibnamefont
  {Koll{\'a}r}}, \bibinfo {author} {\bibfnamefont {M.}~\bibnamefont
  {Fitzpatrick}},\ and\ \bibinfo {author} {\bibfnamefont {A.~A.}\ \bibnamefont
  {Houck}},\ }\bibfield  {title} {\bibinfo {title} {{Hyperbolic lattices in
  circuit quantum electrodynamics}},\ }\href
  {https://doi.org/10.1038/s41586-019-1348-3} {\bibfield  {journal} {\bibinfo
  {journal} {Nature}\ }\textbf {\bibinfo {volume} {571}},\ \bibinfo {pages}
  {45} (\bibinfo {year} {2019})}\BibitemShut {NoStop}%
\bibitem [{\citenamefont {Lenggenhager}\ \emph {et~al.}(2022)\citenamefont
  {Lenggenhager}, \citenamefont {Stegmaier}, \citenamefont {Upreti},
  \citenamefont {Hofmann}, \citenamefont {Helbig}, \citenamefont {Vollhardt},
  \citenamefont {Greiter}, \citenamefont {Lee}, \citenamefont {Imhof},
  \citenamefont {Brand}, \citenamefont {Kiesling}, \citenamefont {Boettcher},
  \citenamefont {Neupert}, \citenamefont {Thomale},\ and\ \citenamefont
  {Bzdusek}}]{lenggenhager2021electric}%
  \BibitemOpen
  \bibfield  {author} {\bibinfo {author} {\bibfnamefont {P.~M.}\ \bibnamefont
  {Lenggenhager}}, \bibinfo {author} {\bibfnamefont {A.}~\bibnamefont
  {Stegmaier}}, \bibinfo {author} {\bibfnamefont {L.~K.}\ \bibnamefont
  {Upreti}}, \bibinfo {author} {\bibfnamefont {T.}~\bibnamefont {Hofmann}},
  \bibinfo {author} {\bibfnamefont {T.}~\bibnamefont {Helbig}}, \bibinfo
  {author} {\bibfnamefont {A.}~\bibnamefont {Vollhardt}}, \bibinfo {author}
  {\bibfnamefont {M.}~\bibnamefont {Greiter}}, \bibinfo {author} {\bibfnamefont
  {C.~H.}\ \bibnamefont {Lee}}, \bibinfo {author} {\bibfnamefont
  {S.}~\bibnamefont {Imhof}}, \bibinfo {author} {\bibfnamefont
  {H.}~\bibnamefont {Brand}}, \bibinfo {author} {\bibfnamefont
  {T.}~\bibnamefont {Kiesling}}, \bibinfo {author} {\bibfnamefont
  {I.}~\bibnamefont {Boettcher}}, \bibinfo {author} {\bibfnamefont
  {T.}~\bibnamefont {Neupert}}, \bibinfo {author} {\bibfnamefont
  {R.}~\bibnamefont {Thomale}},\ and\ \bibinfo {author} {\bibfnamefont
  {T.}~\bibnamefont {Bzdusek}},\ }\bibfield  {title} {\bibinfo {title}
  {{Simulating hyperbolic space on a circuit board}},\ }\href
  {https://doi.org/https://doi.org/10.1038/s41467-022-32042-4} {\bibfield
  {journal} {\bibinfo  {journal} {Nat Commun}\ }\textbf {\bibinfo {volume}
  {13}},\ \bibinfo {pages} {4373} (\bibinfo {year} {2022})}\BibitemShut
  {NoStop}%
\bibitem [{\citenamefont {Magnus}(1974)}]{BookMagnus}%
  \BibitemOpen
  \bibfield  {author} {\bibinfo {author} {\bibfnamefont {W.}~\bibnamefont
  {Magnus}},\ }\href@noop {} {\emph {\bibinfo {title} {{Noneuclidean
  tesselations and their groups}}}}\ (\bibinfo  {publisher} {Academic Press,
  New York},\ \bibinfo {year} {1974})\BibitemShut {NoStop}%
\bibitem [{\citenamefont {Coxeter}\ and\ \citenamefont
  {Moser}(1980)}]{BookCoxeter}%
  \BibitemOpen
  \bibfield  {author} {\bibinfo {author} {\bibfnamefont {H.~S.~M.}\
  \bibnamefont {Coxeter}}\ and\ \bibinfo {author} {\bibfnamefont {W.~O.~J.}\
  \bibnamefont {Moser}},\ }\href@noop {} {\emph {\bibinfo {title} {{Generators
  and Relations for Discrete Groups}}}}\ (\bibinfo  {publisher} {Springer
  Berlin Heidelberg, Berlin, Heidelberg},\ \bibinfo {year} {1980})\BibitemShut
  {NoStop}%
\bibitem [{\citenamefont {Waldschmidt}\ \emph {et~al.}(1992)\citenamefont
  {Waldschmidt}, \citenamefont {Moussa}, \citenamefont {Luck},\ and\
  \citenamefont {Itzykson}}]{BookNumber}%
  \BibitemOpen
  \bibinfo {editor} {\bibfnamefont {M.}~\bibnamefont {Waldschmidt}}, \bibinfo
  {editor} {\bibfnamefont {P.}~\bibnamefont {Moussa}}, \bibinfo {editor}
  {\bibfnamefont {J.-M.}\ \bibnamefont {Luck}},\ and\ \bibinfo {editor}
  {\bibfnamefont {C.}~\bibnamefont {Itzykson}},\ eds.,\ \href@noop {} {\emph
  {\bibinfo {title} {{From Number Theory to Physics}}}}\ (\bibinfo  {publisher}
  {Springer Berlin Heidelberg, Berlin, Heidelberg},\ \bibinfo {year}
  {1992})\BibitemShut {NoStop}%
\bibitem [{\citenamefont {Cannon}\ \emph {et~al.}(1997)\citenamefont {Cannon},
  \citenamefont {Floyd}, \citenamefont {Kenyon},\ and\ \citenamefont
  {Parry}}]{Cannon}%
  \BibitemOpen
  \bibfield  {author} {\bibinfo {author} {\bibfnamefont {J.~W.}\ \bibnamefont
  {Cannon}}, \bibinfo {author} {\bibfnamefont {W.~J.}\ \bibnamefont {Floyd}},
  \bibinfo {author} {\bibfnamefont {R.}~\bibnamefont {Kenyon}},\ and\ \bibinfo
  {author} {\bibfnamefont {W.~R.}\ \bibnamefont {Parry}},\ }\bibfield  {title}
  {\bibinfo {title} {{Hyperbolic Geometry}},\ }in\ \href@noop {} {\emph
  {\bibinfo {booktitle} {{Flavors of Geometry}}}},\ Vol.~\bibinfo {volume}
  {31},\ \bibinfo {editor} {edited by\ \bibinfo {editor} {\bibfnamefont
  {S.}~\bibnamefont {Levy}}}\ (\bibinfo  {publisher} {MSRI Publications},\
  \bibinfo {year} {1997})\ p.~\bibinfo {pages} {59}\BibitemShut {NoStop}%
\bibitem [{\citenamefont {Boyle}\ \emph {et~al.}(2020)\citenamefont {Boyle},
  \citenamefont {Dickens},\ and\ \citenamefont {Flicker}}]{PhysRevX.10.011009}%
  \BibitemOpen
  \bibfield  {author} {\bibinfo {author} {\bibfnamefont {L.}~\bibnamefont
  {Boyle}}, \bibinfo {author} {\bibfnamefont {M.}~\bibnamefont {Dickens}},\
  and\ \bibinfo {author} {\bibfnamefont {F.}~\bibnamefont {Flicker}},\
  }\bibfield  {title} {\bibinfo {title} {{Conformal Quasicrystals and
  Holography}},\ }\href {https://doi.org/10.1103/PhysRevX.10.011009} {\bibfield
   {journal} {\bibinfo  {journal} {Phys. Rev. X}\ }\textbf {\bibinfo {volume}
  {10}},\ \bibinfo {pages} {011009} (\bibinfo {year} {2020})}\BibitemShut
  {NoStop}%
\bibitem [{\citenamefont {Koll{\'a}r}\ \emph {et~al.}(2020)\citenamefont
  {Koll{\'a}r}, \citenamefont {Fitzpatrick}, \citenamefont {Sarnak},\ and\
  \citenamefont {Houck}}]{kollar2019line}%
  \BibitemOpen
  \bibfield  {author} {\bibinfo {author} {\bibfnamefont {A.~J.}\ \bibnamefont
  {Koll{\'a}r}}, \bibinfo {author} {\bibfnamefont {M.}~\bibnamefont
  {Fitzpatrick}}, \bibinfo {author} {\bibfnamefont {P.}~\bibnamefont
  {Sarnak}},\ and\ \bibinfo {author} {\bibfnamefont {A.~A.}\ \bibnamefont
  {Houck}},\ }\bibfield  {title} {\bibinfo {title} {{Line-graph lattices:
  Euclidean and non-Euclidean flat bands, and implementations in circuit
  quantum electrodynamics}},\ }\href
  {https://doi.org/10.1007/s00220-019-03645-8} {\bibfield  {journal} {\bibinfo
  {journal} {Commun. Math. Phys.}\ }\textbf {\bibinfo {volume} {376}},\
  \bibinfo {pages} {1909} (\bibinfo {year} {2020})}\BibitemShut {NoStop}%
\bibitem [{\citenamefont {Yu}\ \emph {et~al.}(2020)\citenamefont {Yu},
  \citenamefont {Piao},\ and\ \citenamefont {Park}}]{PhysRevLett.125.053901}%
  \BibitemOpen
  \bibfield  {author} {\bibinfo {author} {\bibfnamefont {S.}~\bibnamefont
  {Yu}}, \bibinfo {author} {\bibfnamefont {X.}~\bibnamefont {Piao}},\ and\
  \bibinfo {author} {\bibfnamefont {N.}~\bibnamefont {Park}},\ }\bibfield
  {title} {\bibinfo {title} {Topological hyperbolic lattices},\ }\href
  {https://doi.org/10.1103/PhysRevLett.125.053901} {\bibfield  {journal}
  {\bibinfo  {journal} {Phys. Rev. Lett.}\ }\textbf {\bibinfo {volume} {125}},\
  \bibinfo {pages} {053901} (\bibinfo {year} {2020})}\BibitemShut {NoStop}%
\bibitem [{\citenamefont {Boettcher}\ \emph {et~al.}(2020)\citenamefont
  {Boettcher}, \citenamefont {Bienias}, \citenamefont {Belyansky},
  \citenamefont {Koll\'ar},\ and\ \citenamefont
  {Gorshkov}}]{PhysRevA.102.032208}%
  \BibitemOpen
  \bibfield  {author} {\bibinfo {author} {\bibfnamefont {I.}~\bibnamefont
  {Boettcher}}, \bibinfo {author} {\bibfnamefont {P.}~\bibnamefont {Bienias}},
  \bibinfo {author} {\bibfnamefont {R.}~\bibnamefont {Belyansky}}, \bibinfo
  {author} {\bibfnamefont {A.~J.}\ \bibnamefont {Koll\'ar}},\ and\ \bibinfo
  {author} {\bibfnamefont {A.~V.}\ \bibnamefont {Gorshkov}},\ }\bibfield
  {title} {\bibinfo {title} {{Quantum simulation of hyperbolic space with
  circuit quantum electrodynamics: From graphs to geometry}},\ }\href
  {https://doi.org/10.1103/PhysRevA.102.032208} {\bibfield  {journal} {\bibinfo
   {journal} {Phys. Rev. A}\ }\textbf {\bibinfo {volume} {102}},\ \bibinfo
  {pages} {032208} (\bibinfo {year} {2020})}\BibitemShut {NoStop}%
\bibitem [{\citenamefont {Koll{\'a}r}\ and\ \citenamefont
  {Sarnak}(2021)}]{kollar2021gap}%
  \BibitemOpen
  \bibfield  {author} {\bibinfo {author} {\bibfnamefont {A.~J.}\ \bibnamefont
  {Koll{\'a}r}}\ and\ \bibinfo {author} {\bibfnamefont {P.}~\bibnamefont
  {Sarnak}},\ }\bibfield  {title} {\bibinfo {title} {{Gap Sets for the Spectra
  of Cubic Graphs}},\ }\href {https://doi.org/10.1090/cams/3} {\bibfield
  {journal} {\bibinfo  {journal} {Comm. Amer. Math. Soc.}\ }\textbf {\bibinfo
  {volume} {1}},\ \bibinfo {pages} {1} (\bibinfo {year} {2021})}\BibitemShut
  {NoStop}%
\bibitem [{\citenamefont {Asaduzzaman}\ \emph {et~al.}(2020)\citenamefont
  {Asaduzzaman}, \citenamefont {Catterall}, \citenamefont {Hubisz},
  \citenamefont {Nelson},\ and\ \citenamefont
  {Unmuth-Yockey}}]{PhysRevD.102.034511}%
  \BibitemOpen
  \bibfield  {author} {\bibinfo {author} {\bibfnamefont {M.}~\bibnamefont
  {Asaduzzaman}}, \bibinfo {author} {\bibfnamefont {S.}~\bibnamefont
  {Catterall}}, \bibinfo {author} {\bibfnamefont {J.}~\bibnamefont {Hubisz}},
  \bibinfo {author} {\bibfnamefont {R.}~\bibnamefont {Nelson}},\ and\ \bibinfo
  {author} {\bibfnamefont {J.}~\bibnamefont {Unmuth-Yockey}},\ }\bibfield
  {title} {\bibinfo {title} {{Holography on tessellations of hyperbolic
  space}},\ }\href {https://doi.org/10.1103/PhysRevD.102.034511} {\bibfield
  {journal} {\bibinfo  {journal} {Phys. Rev. D}\ }\textbf {\bibinfo {volume}
  {102}},\ \bibinfo {pages} {034511} (\bibinfo {year} {2020})}\BibitemShut
  {NoStop}%
\bibitem [{\citenamefont {Maciejko}\ and\ \citenamefont
  {Rayan}(2021)}]{maciejko2020hyperbolic}%
  \BibitemOpen
  \bibfield  {author} {\bibinfo {author} {\bibfnamefont {J.}~\bibnamefont
  {Maciejko}}\ and\ \bibinfo {author} {\bibfnamefont {S.}~\bibnamefont
  {Rayan}},\ }\bibfield  {title} {\bibinfo {title} {{Hyperbolic band theory}},\
  }\href {https://doi.org/10.1126/sciadv.abe9170} {\bibfield  {journal}
  {\bibinfo  {journal} {Sci. Adv.}\ }\textbf {\bibinfo {volume} {7}},\ \bibinfo
  {pages} {eabe9170} (\bibinfo {year} {2021})}\BibitemShut {NoStop}%
\bibitem [{\citenamefont {Maciejko}\ and\ \citenamefont
  {Rayan}(2022)}]{maciejko2021automorphic}%
  \BibitemOpen
  \bibfield  {author} {\bibinfo {author} {\bibfnamefont {J.}~\bibnamefont
  {Maciejko}}\ and\ \bibinfo {author} {\bibfnamefont {S.}~\bibnamefont
  {Rayan}},\ }\bibfield  {title} {\bibinfo {title} {{Automorphic Bloch theorems
  for hyperbolic lattices}},\ }\href {https://doi.org/10.1073/pnas.2116869119}
  {\bibfield  {journal} {\bibinfo  {journal} {Proc. Natl. Acad. Sci. U.S.A.}\
  }\textbf {\bibinfo {volume} {119}},\ \bibinfo {pages} {e2116869119} (\bibinfo
  {year} {2022})}\BibitemShut {NoStop}%
\bibitem [{\citenamefont {Pastawski}\ \emph {et~al.}(2015)\citenamefont
  {Pastawski}, \citenamefont {Yoshida}, \citenamefont {Harlow},\ and\
  \citenamefont {Preskill}}]{pastawski2015holographic}%
  \BibitemOpen
  \bibfield  {author} {\bibinfo {author} {\bibfnamefont {F.}~\bibnamefont
  {Pastawski}}, \bibinfo {author} {\bibfnamefont {B.}~\bibnamefont {Yoshida}},
  \bibinfo {author} {\bibfnamefont {D.}~\bibnamefont {Harlow}},\ and\ \bibinfo
  {author} {\bibfnamefont {J.}~\bibnamefont {Preskill}},\ }\bibfield  {title}
  {\bibinfo {title} {{Holographic quantum error-correcting codes: Toy models
  for the bulk/boundary correspondence}},\ }\href
  {https://doi.org/10.1007/JHEP06(2015)149} {\bibfield  {journal} {\bibinfo
  {journal} {J. High Energ. Phys.}\ }\textbf {\bibinfo {volume} {2015}},\
  \bibinfo {pages} {149}}\BibitemShut {NoStop}%
\bibitem [{\citenamefont {C.}\ \emph {et~al.}(2021)\citenamefont {C.},
  \citenamefont {Cogburn}, \citenamefont {Fitzpatrick}, \citenamefont
  {Howarth},\ and\ \citenamefont {Tan}}]{PhysRevD.103.094507}%
  \BibitemOpen
  \bibfield  {author} {\bibinfo {author} {\bibfnamefont {R.}~\bibnamefont
  {C.}}, \bibinfo {author} {\bibfnamefont {C.~V.}\ \bibnamefont {Cogburn}},
  \bibinfo {author} {\bibfnamefont {A.~L.}\ \bibnamefont {Fitzpatrick}},
  \bibinfo {author} {\bibfnamefont {D.}~\bibnamefont {Howarth}},\ and\ \bibinfo
  {author} {\bibfnamefont {C.-I.}\ \bibnamefont {Tan}},\ }\bibfield  {title}
  {\bibinfo {title} {{Lattice setup for quantum field theory in
  ${\mathrm{AdS}}_{2}$}},\ }\href {https://doi.org/10.1103/PhysRevD.103.094507}
  {\bibfield  {journal} {\bibinfo  {journal} {Phys. Rev. D}\ }\textbf {\bibinfo
  {volume} {103}},\ \bibinfo {pages} {094507} (\bibinfo {year}
  {2021})}\BibitemShut {NoStop}%
\bibitem [{\citenamefont {Sheng}\ \emph {et~al.}(2021)\citenamefont {Sheng},
  \citenamefont {Huang}, \citenamefont {Yang}, \citenamefont {Gong},
  \citenamefont {Zhu},\ and\ \citenamefont {Liu}}]{PhysRevA.103.033703}%
  \BibitemOpen
  \bibfield  {author} {\bibinfo {author} {\bibfnamefont {C.}~\bibnamefont
  {Sheng}}, \bibinfo {author} {\bibfnamefont {C.}~\bibnamefont {Huang}},
  \bibinfo {author} {\bibfnamefont {R.}~\bibnamefont {Yang}}, \bibinfo {author}
  {\bibfnamefont {Y.}~\bibnamefont {Gong}}, \bibinfo {author} {\bibfnamefont
  {S.}~\bibnamefont {Zhu}},\ and\ \bibinfo {author} {\bibfnamefont
  {H.}~\bibnamefont {Liu}},\ }\bibfield  {title} {\bibinfo {title} {{Simulating
  the escape of entangled photons from the event horizon of black holes in
  nonuniform optical lattices}},\ }\href
  {https://doi.org/10.1103/PhysRevA.103.033703} {\bibfield  {journal} {\bibinfo
   {journal} {Phys. Rev. A}\ }\textbf {\bibinfo {volume} {103}},\ \bibinfo
  {pages} {033703} (\bibinfo {year} {2021})}\BibitemShut {NoStop}%
\bibitem [{\citenamefont {Zhang}\ \emph {et~al.}(2021)\citenamefont {Zhang},
  \citenamefont {Lv}, \citenamefont {Yan},\ and\ \citenamefont
  {Zhou}}]{ZHANG20211967}%
  \BibitemOpen
  \bibfield  {author} {\bibinfo {author} {\bibfnamefont {R.}~\bibnamefont
  {Zhang}}, \bibinfo {author} {\bibfnamefont {C.}~\bibnamefont {Lv}}, \bibinfo
  {author} {\bibfnamefont {Y.}~\bibnamefont {Yan}},\ and\ \bibinfo {author}
  {\bibfnamefont {Q.}~\bibnamefont {Zhou}},\ }\bibfield  {title} {\bibinfo
  {title} {Efimov-like states and quantum funneling effects on synthetic
  hyperbolic surfaces},\ }\href
  {https://doi.org/https://doi.org/10.1016/j.scib.2021.06.017} {\bibfield
  {journal} {\bibinfo  {journal} {Science Bulletin}\ }\textbf {\bibinfo
  {volume} {66}},\ \bibinfo {pages} {1967} (\bibinfo {year}
  {2021})}\BibitemShut {NoStop}%
\bibitem [{\citenamefont {Jahn}\ and\ \citenamefont
  {Eisert}(2021)}]{jahn2021holographic}%
  \BibitemOpen
  \bibfield  {author} {\bibinfo {author} {\bibfnamefont {A.}~\bibnamefont
  {Jahn}}\ and\ \bibinfo {author} {\bibfnamefont {J.}~\bibnamefont {Eisert}},\
  }\bibfield  {title} {\bibinfo {title} {{Holographic tensor network models and
  quantum error correction: a topical review}},\ }\href
  {https://doi.org/10.1088/2058-9565/ac0293} {\bibfield  {journal} {\bibinfo
  {journal} {Quantum Sci. Technol.}\ }\textbf {\bibinfo {volume} {6}},\
  \bibinfo {pages} {033002} (\bibinfo {year} {2021})}\BibitemShut {NoStop}%
\bibitem [{\citenamefont {Zhu}\ \emph {et~al.}(2021)\citenamefont {Zhu},
  \citenamefont {Guo}, \citenamefont {Breuckmann}, \citenamefont {Guo},\ and\
  \citenamefont {Feng}}]{Zhu2021}%
  \BibitemOpen
  \bibfield  {author} {\bibinfo {author} {\bibfnamefont {X.}~\bibnamefont
  {Zhu}}, \bibinfo {author} {\bibfnamefont {J.}~\bibnamefont {Guo}}, \bibinfo
  {author} {\bibfnamefont {N.~P.}\ \bibnamefont {Breuckmann}}, \bibinfo
  {author} {\bibfnamefont {H.}~\bibnamefont {Guo}},\ and\ \bibinfo {author}
  {\bibfnamefont {S.}~\bibnamefont {Feng}},\ }\bibfield  {title} {\bibinfo
  {title} {Quantum phase transitions of interacting bosons on hyperbolic
  lattices},\ }\href {https://doi.org/10.1088/1361-648x/ac0a1a} {\bibfield
  {journal} {\bibinfo  {journal} {J Phys Condens Matter}\ }\textbf {\bibinfo
  {volume} {33}},\ \bibinfo {pages} {335602} (\bibinfo {year}
  {2021})}\BibitemShut {NoStop}%
\bibitem [{\citenamefont {Boettcher}\ \emph {et~al.}(2022)\citenamefont
  {Boettcher}, \citenamefont {Gorshkov}, \citenamefont {Koll\'ar},
  \citenamefont {Maciejko}, \citenamefont {Rayan},\ and\ \citenamefont
  {Thomale}}]{boettcher2021crystallography}%
  \BibitemOpen
  \bibfield  {author} {\bibinfo {author} {\bibfnamefont {I.}~\bibnamefont
  {Boettcher}}, \bibinfo {author} {\bibfnamefont {A.~V.}\ \bibnamefont
  {Gorshkov}}, \bibinfo {author} {\bibfnamefont {A.~J.}\ \bibnamefont
  {Koll\'ar}}, \bibinfo {author} {\bibfnamefont {J.}~\bibnamefont {Maciejko}},
  \bibinfo {author} {\bibfnamefont {S.}~\bibnamefont {Rayan}},\ and\ \bibinfo
  {author} {\bibfnamefont {R.}~\bibnamefont {Thomale}},\ }\bibfield  {title}
  {\bibinfo {title} {Crystallography of hyperbolic lattices},\ }\href
  {https://doi.org/10.1103/PhysRevB.105.125118} {\bibfield  {journal} {\bibinfo
   {journal} {Phys. Rev. B}\ }\textbf {\bibinfo {volume} {105}},\ \bibinfo
  {pages} {125118} (\bibinfo {year} {2022})}\BibitemShut {NoStop}%
\bibitem [{\citenamefont {Bienias}\ \emph {et~al.}(2022)\citenamefont
  {Bienias}, \citenamefont {Boettcher}, \citenamefont {Belyansky},
  \citenamefont {Koll\'ar},\ and\ \citenamefont
  {Gorshkov}}]{PhysRevLett.128.013601}%
  \BibitemOpen
  \bibfield  {author} {\bibinfo {author} {\bibfnamefont {P.}~\bibnamefont
  {Bienias}}, \bibinfo {author} {\bibfnamefont {I.}~\bibnamefont {Boettcher}},
  \bibinfo {author} {\bibfnamefont {R.}~\bibnamefont {Belyansky}}, \bibinfo
  {author} {\bibfnamefont {A.~J.}\ \bibnamefont {Koll\'ar}},\ and\ \bibinfo
  {author} {\bibfnamefont {A.~V.}\ \bibnamefont {Gorshkov}},\ }\bibfield
  {title} {\bibinfo {title} {Circuit quantum electrodynamics in hyperbolic
  space: From photon bound states to frustrated spin models},\ }\href
  {https://doi.org/10.1103/PhysRevLett.128.013601} {\bibfield  {journal}
  {\bibinfo  {journal} {Phys. Rev. Lett.}\ }\textbf {\bibinfo {volume} {128}},\
  \bibinfo {pages} {013601} (\bibinfo {year} {2022})}\BibitemShut {NoStop}%
\bibitem [{\citenamefont {Pappalardi}\ and\ \citenamefont
  {Kurchan}(2022)}]{Pappalardi}%
  \BibitemOpen
  \bibfield  {author} {\bibinfo {author} {\bibfnamefont {S.}~\bibnamefont
  {Pappalardi}}\ and\ \bibinfo {author} {\bibfnamefont {J.}~\bibnamefont
  {Kurchan}},\ }\bibfield  {title} {\bibinfo {title} {{Low temperature quantum
  bounds on simple models}},\ }\href
  {https://doi.org/10.21468/SciPostPhys.13.1.006} {\bibfield  {journal}
  {\bibinfo  {journal} {SciPost Phys.}\ }\textbf {\bibinfo {volume} {13}},\
  \bibinfo {pages} {006} (\bibinfo {year} {2022})}\BibitemShut {NoStop}%
\bibitem [{\citenamefont {Ikeda}\ \emph
  {et~al.}(2021{\natexlab{a}})\citenamefont {Ikeda}, \citenamefont {Matsuki},\
  and\ \citenamefont {Aoki}}]{Ikeda}%
  \BibitemOpen
  \bibfield  {author} {\bibinfo {author} {\bibfnamefont {K.}~\bibnamefont
  {Ikeda}}, \bibinfo {author} {\bibfnamefont {Y.}~\bibnamefont {Matsuki}},\
  and\ \bibinfo {author} {\bibfnamefont {S.}~\bibnamefont {Aoki}},\ }\bibfield
  {title} {\bibinfo {title} {{Algebra of Hyperbolic Band Theory under Magnetic
  Field}},\ }\href {https://arxiv.org/abs/2107.10586} {\bibfield  {journal}
  {\bibinfo  {journal} {arXiv:2107.10586}\ } (\bibinfo {year}
  {2021}{\natexlab{a}})}\BibitemShut {NoStop}%
\bibitem [{\citenamefont {Morice}\ \emph
  {et~al.}(2021{\natexlab{a}})\citenamefont {Morice}, \citenamefont
  {Moghaddam}, \citenamefont {Chernyavsky}, \citenamefont {van Wezel},\ and\
  \citenamefont {van~den Brink}}]{PhysRevResearch.3.L022022}%
  \BibitemOpen
  \bibfield  {author} {\bibinfo {author} {\bibfnamefont {C.}~\bibnamefont
  {Morice}}, \bibinfo {author} {\bibfnamefont {A.~G.}\ \bibnamefont
  {Moghaddam}}, \bibinfo {author} {\bibfnamefont {D.}~\bibnamefont
  {Chernyavsky}}, \bibinfo {author} {\bibfnamefont {J.}~\bibnamefont {van
  Wezel}},\ and\ \bibinfo {author} {\bibfnamefont {J.}~\bibnamefont {van~den
  Brink}},\ }\bibfield  {title} {\bibinfo {title} {Synthetic gravitational
  horizons in low-dimensional quantum matter},\ }\href
  {https://doi.org/10.1103/PhysRevResearch.3.L022022} {\bibfield  {journal}
  {\bibinfo  {journal} {Phys. Rev. Research}\ }\textbf {\bibinfo {volume}
  {3}},\ \bibinfo {pages} {L022022} (\bibinfo {year}
  {2021}{\natexlab{a}})}\BibitemShut {NoStop}%
\bibitem [{\citenamefont {Morice}\ \emph
  {et~al.}(2021{\natexlab{b}})\citenamefont {Morice}, \citenamefont
  {Chernyavsky}, \citenamefont {van Wezel}, \citenamefont {van~den Brink},\
  and\ \citenamefont {Moghaddam}}]{Morice}%
  \BibitemOpen
  \bibfield  {author} {\bibinfo {author} {\bibfnamefont {C.}~\bibnamefont
  {Morice}}, \bibinfo {author} {\bibfnamefont {D.}~\bibnamefont {Chernyavsky}},
  \bibinfo {author} {\bibfnamefont {J.}~\bibnamefont {van Wezel}}, \bibinfo
  {author} {\bibfnamefont {J.}~\bibnamefont {van~den Brink}},\ and\ \bibinfo
  {author} {\bibfnamefont {A.~G.}\ \bibnamefont {Moghaddam}},\ }\bibfield
  {title} {\bibinfo {title} {{Quantum dynamics in 1D lattice models with
  synthetic horizons}},\ }\href {https://arxiv.org/abs/2112.12827} {\bibfield
  {journal} {\bibinfo  {journal} {arXiv:2112.12827}\ } (\bibinfo {year}
  {2021}{\natexlab{b}})}\BibitemShut {NoStop}%
\bibitem [{\citenamefont {Malen}\ \emph {et~al.}(2021)\citenamefont {Malen},
  \citenamefont {Roldan},\ and\ \citenamefont {Toala-Enriquez}}]{Malen}%
  \BibitemOpen
  \bibfield  {author} {\bibinfo {author} {\bibfnamefont {G.}~\bibnamefont
  {Malen}}, \bibinfo {author} {\bibfnamefont {E.}~\bibnamefont {Roldan}},\ and\
  \bibinfo {author} {\bibfnamefont {R.}~\bibnamefont {Toala-Enriquez}},\
  }\bibfield  {title} {\bibinfo {title} {{Extremal $\{p,q\}$-Animals}},\ }\href
  {https://arxiv.org/abs/2109.05331} {\bibfield  {journal} {\bibinfo  {journal}
  {arXiv:2109.05331}\ } (\bibinfo {year} {2021})}\BibitemShut {NoStop}%
\bibitem [{\citenamefont {Ludewig}\ and\ \citenamefont
  {Thiang}(2021)}]{Ludewig2021}%
  \BibitemOpen
  \bibfield  {author} {\bibinfo {author} {\bibfnamefont {M.}~\bibnamefont
  {Ludewig}}\ and\ \bibinfo {author} {\bibfnamefont {G.~C.}\ \bibnamefont
  {Thiang}},\ }\bibfield  {title} {\bibinfo {title} {{{Gaplessness of Landau
  Hamiltonians on Hyperbolic Half-planes via Coarse Geometry}}},\ }\href
  {https://doi.org/10.1007/s00220-021-04068-0} {\bibfield  {journal} {\bibinfo
  {journal} {Commun. Math. Phys.}\ }\textbf {\bibinfo {volume} {386}},\
  \bibinfo {pages} {87} (\bibinfo {year} {2021})}\BibitemShut {NoStop}%
\bibitem [{\citenamefont {Lv}\ \emph {et~al.}(2022)\citenamefont {Lv},
  \citenamefont {Zhang}, \citenamefont {Zhai},\ and\ \citenamefont
  {Zhou}}]{Lv}%
  \BibitemOpen
  \bibfield  {author} {\bibinfo {author} {\bibfnamefont {C.}~\bibnamefont
  {Lv}}, \bibinfo {author} {\bibfnamefont {R.}~\bibnamefont {Zhang}}, \bibinfo
  {author} {\bibfnamefont {Z.}~\bibnamefont {Zhai}},\ and\ \bibinfo {author}
  {\bibfnamefont {Q.}~\bibnamefont {Zhou}},\ }\bibfield  {title} {\bibinfo
  {title} {{Curving the space by non-Hermiticity}},\ }\href
  {https://doi.org/https://doi.org/10.1038/s41467-022-29774-8} {\bibfield
  {journal} {\bibinfo  {journal} {Nat. Commun.}\ }\textbf {\bibinfo {volume}
  {13}},\ \bibinfo {pages} {1} (\bibinfo {year} {2022})}\BibitemShut {NoStop}%
\bibitem [{\citenamefont {Saa}\ \emph {et~al.}(2021)\citenamefont {Saa},
  \citenamefont {Miranda},\ and\ \citenamefont {Rouxinol}}]{Saa}%
  \BibitemOpen
  \bibfield  {author} {\bibinfo {author} {\bibfnamefont {A.}~\bibnamefont
  {Saa}}, \bibinfo {author} {\bibfnamefont {E.}~\bibnamefont {Miranda}},\ and\
  \bibinfo {author} {\bibfnamefont {F.}~\bibnamefont {Rouxinol}},\ }\bibfield
  {title} {\bibinfo {title} {{Higher-dimensional Euclidean and non-Euclidean
  structures in planar circuit quantum electrodynamics}},\ }\href
  {https://arxiv.org/abs/2108.08854} {\bibfield  {journal} {\bibinfo  {journal}
  {arXiv:2108.08854}\ } (\bibinfo {year} {2021})}\BibitemShut {NoStop}%
\bibitem [{\citenamefont {Stegmaier}\ \emph {et~al.}(2022)\citenamefont
  {Stegmaier}, \citenamefont {Upreti}, \citenamefont {Thomale},\ and\
  \citenamefont {Boettcher}}]{stegmaier2021universality}%
  \BibitemOpen
  \bibfield  {author} {\bibinfo {author} {\bibfnamefont {A.}~\bibnamefont
  {Stegmaier}}, \bibinfo {author} {\bibfnamefont {L.~K.}\ \bibnamefont
  {Upreti}}, \bibinfo {author} {\bibfnamefont {R.}~\bibnamefont {Thomale}},\
  and\ \bibinfo {author} {\bibfnamefont {I.}~\bibnamefont {Boettcher}},\
  }\bibfield  {title} {\bibinfo {title} {{Universality of Hofstadter
  Butterflies on Hyperbolic Lattices}},\ }\href
  {https://doi.org/10.1103/PhysRevLett.128.166402} {\bibfield  {journal}
  {\bibinfo  {journal} {Phys. Rev. Lett.}\ }\textbf {\bibinfo {volume} {128}},\
  \bibinfo {pages} {166402} (\bibinfo {year} {2022})}\BibitemShut {NoStop}%
\bibitem [{\citenamefont {Hadamard}(1898)}]{Hadamard}%
  \BibitemOpen
  \bibfield  {author} {\bibinfo {author} {\bibfnamefont {J.}~\bibnamefont
  {Hadamard}},\ }\bibfield  {title} {\bibinfo {title} {{Les surfaces a
  courbures opposees et leurs lignes geodesiques}},\ }\href@noop {} {\bibfield
  {journal} {\bibinfo  {journal} {J. Math. Pures et Appl.}\ }\textbf {\bibinfo
  {volume} {4}},\ \bibinfo {pages} {27} (\bibinfo {year} {1898})}\BibitemShut
  {NoStop}%
\bibitem [{\citenamefont {Mautner}(1957)}]{Mautner}%
  \BibitemOpen
  \bibfield  {author} {\bibinfo {author} {\bibfnamefont {F.~I.}\ \bibnamefont
  {Mautner}},\ }\bibfield  {title} {\bibinfo {title} {{Geodesic Flows on
  Symmetric Riemann Spaces}},\ }\href {https://doi.org/10.2307/1970054}
  {\bibfield  {journal} {\bibinfo  {journal} {Ann. of Math.}\ }\textbf
  {\bibinfo {volume} {65}},\ \bibinfo {pages} {416} (\bibinfo {year}
  {1957})}\BibitemShut {NoStop}%
\bibitem [{\citenamefont {Anosov}(1969)}]{BookAnosov}%
  \BibitemOpen
  \bibfield  {author} {\bibinfo {author} {\bibfnamefont {D.~V.}\ \bibnamefont
  {Anosov}},\ }\href@noop {} {\emph {\bibinfo {title} {{Geodesic flows on
  closed Riemann manifolds with negative curvature}}}},\ \bibinfo {series}
  {Trudy Matematicheskogo instituta imeni V.A. Steklova}, Vol.~\bibinfo
  {volume} {90}\ (\bibinfo  {publisher} {Providence, American Mathematical
  Society},\ \bibinfo {year} {1969})\BibitemShut {NoStop}%
\bibitem [{\citenamefont {Bunimovich}(1979)}]{Bun1}%
  \BibitemOpen
  \bibfield  {author} {\bibinfo {author} {\bibfnamefont {L.~A.}\ \bibnamefont
  {Bunimovich}},\ }\bibfield  {title} {\bibinfo {title} {{On the Ergodic
  Properties of Nowhere Dispersing Billiards}},\ }\href
  {https://doi.org/10.1007/BF01197884} {\bibfield  {journal} {\bibinfo
  {journal} {Commun. Math. Phys.}\ }\textbf {\bibinfo {volume} {65}},\ \bibinfo
  {pages} {295} (\bibinfo {year} {1979})}\BibitemShut {NoStop}%
\bibitem [{\citenamefont {Bunimovich}\ \emph {et~al.}(1991)\citenamefont
  {Bunimovich}, \citenamefont {Sinai},\ and\ \citenamefont {Chernov}}]{Bun2}%
  \BibitemOpen
  \bibfield  {author} {\bibinfo {author} {\bibfnamefont {L.~A.}\ \bibnamefont
  {Bunimovich}}, \bibinfo {author} {\bibfnamefont {Y.~G.}\ \bibnamefont
  {Sinai}},\ and\ \bibinfo {author} {\bibfnamefont {N.~I.}\ \bibnamefont
  {Chernov}},\ }\bibfield  {title} {\bibinfo {title} {{Statistical properties
  of two-dimensional hyperbolic billiards}},\ }\href@noop {} {\bibfield
  {journal} {\bibinfo  {journal} {Russian Math. Surveys}\ }\textbf {\bibinfo
  {volume} {46:4}},\ \bibinfo {pages} {P47} (\bibinfo {year}
  {1991})}\BibitemShut {NoStop}%
\bibitem [{\citenamefont {Steiner}(1994)}]{SteinerDESY}%
  \BibitemOpen
  \bibfield  {author} {\bibinfo {author} {\bibfnamefont {F.}~\bibnamefont
  {Steiner}},\ }\bibfield  {title} {\bibinfo {title} {{Quantum Chaos}},\ }\href
  {https://arxiv.org/abs/chao-dyn/9402001} {\bibfield  {journal} {\bibinfo
  {journal} {arXiv:chao-dyn/9402001}\ } (\bibinfo {year} {1994})}\BibitemShut
  {NoStop}%
\bibitem [{\citenamefont {Balazs}\ and\ \citenamefont
  {Voros}(1986)}]{BALAZS1986109}%
  \BibitemOpen
  \bibfield  {author} {\bibinfo {author} {\bibfnamefont {N.}~\bibnamefont
  {Balazs}}\ and\ \bibinfo {author} {\bibfnamefont {A.}~\bibnamefont {Voros}},\
  }\bibfield  {title} {\bibinfo {title} {{Chaos on the pseudosphere}},\ }\href
  {https://doi.org/https://doi.org/10.1016/0370-1573(86)90159-6} {\bibfield
  {journal} {\bibinfo  {journal} {Phys. Rep.}\ }\textbf {\bibinfo {volume}
  {143}},\ \bibinfo {pages} {109} (\bibinfo {year} {1986})}\BibitemShut
  {NoStop}%
\bibitem [{\citenamefont {Aurich}\ \emph {et~al.}(1988)\citenamefont {Aurich},
  \citenamefont {Sieber},\ and\ \citenamefont {Steiner}}]{PhysRevLett.61.483}%
  \BibitemOpen
  \bibfield  {author} {\bibinfo {author} {\bibfnamefont {R.}~\bibnamefont
  {Aurich}}, \bibinfo {author} {\bibfnamefont {M.}~\bibnamefont {Sieber}},\
  and\ \bibinfo {author} {\bibfnamefont {F.}~\bibnamefont {Steiner}},\
  }\bibfield  {title} {\bibinfo {title} {Quantum chaos of the
  hadamard-gutzwiller model},\ }\href
  {https://doi.org/10.1103/PhysRevLett.61.483} {\bibfield  {journal} {\bibinfo
  {journal} {Phys. Rev. Lett.}\ }\textbf {\bibinfo {volume} {61}},\ \bibinfo
  {pages} {483} (\bibinfo {year} {1988})}\BibitemShut {NoStop}%
\bibitem [{\citenamefont {Aurich}\ and\ \citenamefont
  {Steiner}(1988)}]{AURICH1988451}%
  \BibitemOpen
  \bibfield  {author} {\bibinfo {author} {\bibfnamefont {R.}~\bibnamefont
  {Aurich}}\ and\ \bibinfo {author} {\bibfnamefont {F.}~\bibnamefont
  {Steiner}},\ }\bibfield  {title} {\bibinfo {title} {On the periodic orbits of
  a strongly chaotic system},\ }\href
  {https://doi.org/https://doi.org/10.1016/0167-2789(88)90068-1} {\bibfield
  {journal} {\bibinfo  {journal} {Physica D: Nonlinear Phenomena}\ }\textbf
  {\bibinfo {volume} {32}},\ \bibinfo {pages} {451} (\bibinfo {year}
  {1988})}\BibitemShut {NoStop}%
\bibitem [{\citenamefont {Aurich}\ and\ \citenamefont
  {Steiner}(1989)}]{AURICH1989169}%
  \BibitemOpen
  \bibfield  {author} {\bibinfo {author} {\bibfnamefont {R.}~\bibnamefont
  {Aurich}}\ and\ \bibinfo {author} {\bibfnamefont {F.}~\bibnamefont
  {Steiner}},\ }\bibfield  {title} {\bibinfo {title} {Periodic-orbit sum rules
  for the hadamard-gutzwiller model},\ }\href
  {https://doi.org/https://doi.org/10.1016/0167-2789(89)90003-1} {\bibfield
  {journal} {\bibinfo  {journal} {Physica D: Nonlinear Phenomena}\ }\textbf
  {\bibinfo {volume} {39}},\ \bibinfo {pages} {169} (\bibinfo {year}
  {1989})}\BibitemShut {NoStop}%
\bibitem [{\citenamefont {Aurich}\ \emph {et~al.}(1991)\citenamefont {Aurich},
  \citenamefont {Bogomolny},\ and\ \citenamefont {Steiner}}]{AURICH199191}%
  \BibitemOpen
  \bibfield  {author} {\bibinfo {author} {\bibfnamefont {R.}~\bibnamefont
  {Aurich}}, \bibinfo {author} {\bibfnamefont {E.}~\bibnamefont {Bogomolny}},\
  and\ \bibinfo {author} {\bibfnamefont {F.}~\bibnamefont {Steiner}},\
  }\bibfield  {title} {\bibinfo {title} {{Periodic orbits on the regular
  hyperbolic octagon}},\ }\href
  {https://doi.org/https://doi.org/10.1016/0167-2789(91)90053-C} {\bibfield
  {journal} {\bibinfo  {journal} {Physica D: Nonlinear Phenomena}\ }\textbf
  {\bibinfo {volume} {48}},\ \bibinfo {pages} {91} (\bibinfo {year}
  {1991})}\BibitemShut {NoStop}%
\bibitem [{\citenamefont {Aurich}\ \emph {et~al.}(1992)\citenamefont {Aurich},
  \citenamefont {Matthies}, \citenamefont {Sieber},\ and\ \citenamefont
  {Steiner}}]{PhysRevLett.68.1629}%
  \BibitemOpen
  \bibfield  {author} {\bibinfo {author} {\bibfnamefont {R.}~\bibnamefont
  {Aurich}}, \bibinfo {author} {\bibfnamefont {C.}~\bibnamefont {Matthies}},
  \bibinfo {author} {\bibfnamefont {M.}~\bibnamefont {Sieber}},\ and\ \bibinfo
  {author} {\bibfnamefont {F.}~\bibnamefont {Steiner}},\ }\bibfield  {title}
  {\bibinfo {title} {Novel rule for quantizing chaos},\ }\href
  {https://doi.org/10.1103/PhysRevLett.68.1629} {\bibfield  {journal} {\bibinfo
   {journal} {Phys. Rev. Lett.}\ }\textbf {\bibinfo {volume} {68}},\ \bibinfo
  {pages} {1629} (\bibinfo {year} {1992})}\BibitemShut {NoStop}%
\bibitem [{\citenamefont {Ninnemann}(1995)}]{ninnemann}%
  \BibitemOpen
  \bibfield  {author} {\bibinfo {author} {\bibfnamefont {H.}~\bibnamefont
  {Ninnemann}},\ }\bibfield  {title} {\bibinfo {title} {{Gutzwiller's octagon
  and the triangular billiard T*(2,3,8) as models for the quantization of
  chaotic systems by Selberg's trace formula}},\ }\href
  {https://doi.org/10.1142/S0217979295000719} {\bibfield  {journal} {\bibinfo
  {journal} {International Journal of Modern Physics B}\ }\textbf {\bibinfo
  {volume} {09}},\ \bibinfo {pages} {1647} (\bibinfo {year}
  {1995})}\BibitemShut {NoStop}%
\bibitem [{\citenamefont {Braun}\ \emph {et~al.}(2002)\citenamefont {Braun},
  \citenamefont {Heusler}, \citenamefont {Mueller},\ and\ \citenamefont
  {Haake}}]{Braun2002}%
  \BibitemOpen
  \bibfield  {author} {\bibinfo {author} {\bibfnamefont {P.~A.}\ \bibnamefont
  {Braun}}, \bibinfo {author} {\bibfnamefont {S.}~\bibnamefont {Heusler}},
  \bibinfo {author} {\bibfnamefont {S.}~\bibnamefont {Mueller}},\ and\ \bibinfo
  {author} {\bibfnamefont {F.}~\bibnamefont {Haake}},\ }\bibfield  {title}
  {\bibinfo {title} {Statistics of self-crossings and avoided crossings of
  periodic orbits in the hadamard-gutzwiller model},\ }\href
  {https://doi.org/10.1140/epjb/e2002-00374-7} {\bibfield  {journal} {\bibinfo
  {journal} {Eur. Phys. J. B}\ }\textbf {\bibinfo {volume} {30}},\ \bibinfo
  {pages} {189} (\bibinfo {year} {2002})}\BibitemShut {NoStop}%
\bibitem [{\citenamefont {Bolte}(1993)}]{Bolte1993}%
  \BibitemOpen
  \bibfield  {author} {\bibinfo {author} {\bibfnamefont {J.}~\bibnamefont
  {Bolte}},\ }\bibfield  {title} {\bibinfo {title} {Some studies on
  arithmetical chaos in classical and quantum mechanics},\ }\href
  {https://doi.org/10.1142/S0217979293003759} {\bibfield  {journal} {\bibinfo
  {journal} {International Journal of Modern Physics B}\ }\textbf {\bibinfo
  {volume} {07}},\ \bibinfo {pages} {4451} (\bibinfo {year}
  {1993})}\BibitemShut {NoStop}%
\bibitem [{\citenamefont {Bogomolny}\ \emph {et~al.}(1997)\citenamefont
  {Bogomolny}, \citenamefont {Georgeot}, \citenamefont {Giannoni},\ and\
  \citenamefont {Schmit}}]{BOGOMOLNY1997219}%
  \BibitemOpen
  \bibfield  {author} {\bibinfo {author} {\bibfnamefont {E.}~\bibnamefont
  {Bogomolny}}, \bibinfo {author} {\bibfnamefont {B.}~\bibnamefont {Georgeot}},
  \bibinfo {author} {\bibfnamefont {M.-J.}\ \bibnamefont {Giannoni}},\ and\
  \bibinfo {author} {\bibfnamefont {C.}~\bibnamefont {Schmit}},\ }\bibfield
  {title} {\bibinfo {title} {Arithmetical chaos},\ }\href
  {https://www.sciencedirect.com/science/article/abs/pii/S0370157397000161}
  {\bibfield  {journal} {\bibinfo  {journal} {Physics Reports}\ }\textbf
  {\bibinfo {volume} {291}},\ \bibinfo {pages} {219} (\bibinfo {year}
  {1997})}\BibitemShut {NoStop}%
\bibitem [{\citenamefont {Bogomolny}(2003)}]{BogAri}%
  \BibitemOpen
  \bibfield  {author} {\bibinfo {author} {\bibfnamefont {E.}~\bibnamefont
  {Bogomolny}},\ }\bibfield  {title} {\bibinfo {title} {Quantum and
  arithmetical chaos},\ }\href {https://arxiv.org/abs/nlin/0312061} {\bibfield
  {journal} {\bibinfo  {journal} {arXiv:0312061}\ } (\bibinfo {year}
  {2003})}\BibitemShut {NoStop}%
\bibitem [{\citenamefont {Selberg}(1956)}]{selberg}%
  \BibitemOpen
  \bibfield  {author} {\bibinfo {author} {\bibfnamefont {A.}~\bibnamefont
  {Selberg}},\ }\bibfield  {title} {\bibinfo {title} {{Harmonic analysis and
  discontinuous groups in weakly symmetric Riemannian spaces with applications
  to Dirichlet series}},\ }\href@noop {} {\bibfield  {journal} {\bibinfo
  {journal} {J. Indian Math. Soc.}\ }\textbf {\bibinfo {volume} {20}},\
  \bibinfo {pages} {47} (\bibinfo {year} {1956})}\BibitemShut {NoStop}%
\bibitem [{\citenamefont {Iwaniec}(2002)}]{BookAuto}%
  \BibitemOpen
  \bibfield  {author} {\bibinfo {author} {\bibfnamefont {H.}~\bibnamefont
  {Iwaniec}},\ }\href@noop {} {\emph {\bibinfo {title} {{Spectral Methods of
  Automorphic Forms}}}},\ \bibinfo {edition} {2nd}\ ed.,\ \bibinfo {series}
  {Graduate Studies in Mathematics}, Vol.~\bibinfo {volume} {53}\ (\bibinfo
  {publisher} {American Mathematical Society, Providence, Rhode Island},\
  \bibinfo {year} {2002})\BibitemShut {NoStop}%
\bibitem [{\citenamefont {Marklof}(2012)}]{marklof2012ii}%
  \BibitemOpen
  \bibfield  {author} {\bibinfo {author} {\bibfnamefont {J.}~\bibnamefont
  {Marklof}},\ }\bibfield  {title} {\bibinfo {title} {{Selberg's Trace Formula:
  An Introduction}},\ }in\ \href@noop {} {\emph {\bibinfo {booktitle}
  {Hyperbolic geometry and applications in quantum chaos and cosmology}}},\
  Vol.\ \bibinfo {volume} {397},\ \bibinfo {editor} {edited by\ \bibinfo
  {editor} {\bibfnamefont {J.}~\bibnamefont {Bolte}}\ and\ \bibinfo {editor}
  {\bibfnamefont {F.}~\bibnamefont {Steiner}}}\ (\bibinfo  {publisher}
  {Cambridge University Press},\ \bibinfo {year} {2012})\ p.~\bibinfo {pages}
  {83}\BibitemShut {NoStop}%
\bibitem [{\citenamefont {Grosche}(2013)}]{BookGrosche}%
  \BibitemOpen
  \bibfield  {author} {\bibinfo {author} {\bibfnamefont {C.}~\bibnamefont
  {Grosche}},\ }\href@noop {} {\emph {\bibinfo {title} {{Path Integrals,
  Hyperbolic Spaces and Selberg Trace Formulae}}}}\ (\bibinfo  {publisher}
  {Springer, New York},\ \bibinfo {year} {2013})\BibitemShut {NoStop}%
\bibitem [{\citenamefont {Comtet}\ \emph {et~al.}(1993)\citenamefont {Comtet},
  \citenamefont {Georgeot},\ and\ \citenamefont {Ouvry}}]{PhysRevLett.71.3786}%
  \BibitemOpen
  \bibfield  {author} {\bibinfo {author} {\bibfnamefont {A.}~\bibnamefont
  {Comtet}}, \bibinfo {author} {\bibfnamefont {B.}~\bibnamefont {Georgeot}},\
  and\ \bibinfo {author} {\bibfnamefont {S.}~\bibnamefont {Ouvry}},\ }\bibfield
   {title} {\bibinfo {title} {Trace formula for riemann surfaces with magnetic
  field},\ }\href {https://doi.org/10.1103/PhysRevLett.71.3786} {\bibfield
  {journal} {\bibinfo  {journal} {Phys. Rev. Lett.}\ }\textbf {\bibinfo
  {volume} {71}},\ \bibinfo {pages} {3786} (\bibinfo {year}
  {1993})}\BibitemShut {NoStop}%
\bibitem [{\citenamefont {Bytsenko}\ \emph {et~al.}(1996)\citenamefont
  {Bytsenko}, \citenamefont {Cognola}, \citenamefont {Vanzo},\ and\
  \citenamefont {Zerbini}}]{BYTSENKO19961}%
  \BibitemOpen
  \bibfield  {author} {\bibinfo {author} {\bibfnamefont {A.~A.}\ \bibnamefont
  {Bytsenko}}, \bibinfo {author} {\bibfnamefont {G.}~\bibnamefont {Cognola}},
  \bibinfo {author} {\bibfnamefont {L.}~\bibnamefont {Vanzo}},\ and\ \bibinfo
  {author} {\bibfnamefont {S.}~\bibnamefont {Zerbini}},\ }\bibfield  {title}
  {\bibinfo {title} {Quantum fields and extended objects in space-times with
  constant curvature spatial section},\ }\href
  {https://doi.org/https://doi.org/10.1016/0370-1573(95)00053-4} {\bibfield
  {journal} {\bibinfo  {journal} {Physics Reports}\ }\textbf {\bibinfo {volume}
  {266}},\ \bibinfo {pages} {1} (\bibinfo {year} {1996})}\BibitemShut {NoStop}%
\bibitem [{\citenamefont {Chen}\ \emph {et~al.}(2022)\citenamefont {Chen},
  \citenamefont {Brand}, \citenamefont {Helbig}, \citenamefont {Hofmann},
  \citenamefont {Imhof}, \citenamefont {Fritzsche}, \citenamefont
  {Kie{\ss}ling}, \citenamefont {Stegmaier}, \citenamefont {Upreti},
  \citenamefont {Neupert}, , \citenamefont {Bzdusek}, \citenamefont {Greiter},
  \citenamefont {Thomale},\ and\ \citenamefont {Boettcher}}]{AlbertaWu2021}%
  \BibitemOpen
  \bibfield  {author} {\bibinfo {author} {\bibfnamefont {A.}~\bibnamefont
  {Chen}}, \bibinfo {author} {\bibfnamefont {H.}~\bibnamefont {Brand}},
  \bibinfo {author} {\bibfnamefont {T.}~\bibnamefont {Helbig}}, \bibinfo
  {author} {\bibfnamefont {T.}~\bibnamefont {Hofmann}}, \bibinfo {author}
  {\bibfnamefont {S.}~\bibnamefont {Imhof}}, \bibinfo {author} {\bibfnamefont
  {A.}~\bibnamefont {Fritzsche}}, \bibinfo {author} {\bibfnamefont
  {T.}~\bibnamefont {Kie{\ss}ling}}, \bibinfo {author} {\bibfnamefont
  {A.}~\bibnamefont {Stegmaier}}, \bibinfo {author} {\bibfnamefont {L.~K.}\
  \bibnamefont {Upreti}}, \bibinfo {author} {\bibfnamefont {T.}~\bibnamefont
  {Neupert}}, , \bibinfo {author} {\bibfnamefont {T.}~\bibnamefont {Bzdusek}},
  \bibinfo {author} {\bibfnamefont {M.}~\bibnamefont {Greiter}}, \bibinfo
  {author} {\bibfnamefont {R.}~\bibnamefont {Thomale}},\ and\ \bibinfo {author}
  {\bibfnamefont {I.}~\bibnamefont {Boettcher}},\ }\bibfield  {title} {\bibinfo
  {title} {Hyperbolic matter in electrical circuits with tunable complex
  phases},\ }\href {https://arxiv.org/abs/2205.05106} {\bibfield  {journal}
  {\bibinfo  {journal} {arXiv:2205.05106}\ } (\bibinfo {year}
  {2022})}\BibitemShut {NoStop}%
\bibitem [{\citenamefont {Kienzle}\ and\ \citenamefont
  {Rayan}(2022)}]{RayanHyp}%
  \BibitemOpen
  \bibfield  {author} {\bibinfo {author} {\bibfnamefont {E.}~\bibnamefont
  {Kienzle}}\ and\ \bibinfo {author} {\bibfnamefont {S.}~\bibnamefont
  {Rayan}},\ }\bibfield  {title} {\bibinfo {title} {Hyperbolic band theory
  through higgs bundles},\ }\href {https://arxiv.org/abs/2201.12689} {\bibfield
   {journal} {\bibinfo  {journal} {arXiv:2201.12689}\ } (\bibinfo {year}
  {2022})}\BibitemShut {NoStop}%
\bibitem [{\citenamefont {Edmonds}\ \emph {et~al.}(1982)\citenamefont
  {Edmonds}, \citenamefont {Ewing},\ and\ \citenamefont {Kulkarni}}]{Edmonds}%
  \BibitemOpen
  \bibfield  {author} {\bibinfo {author} {\bibfnamefont {A.~L.}\ \bibnamefont
  {Edmonds}}, \bibinfo {author} {\bibfnamefont {J.~H.}\ \bibnamefont {Ewing}},\
  and\ \bibinfo {author} {\bibfnamefont {R.~S.}\ \bibnamefont {Kulkarni}},\
  }\bibfield  {title} {\bibinfo {title} {{Regular Tessellations of Surfaces and
  (p, q, 2)-Triangle Groups}},\ }\href {http://www.jstor.org/stable/2007049}
  {\bibfield  {journal} {\bibinfo  {journal} {Ann. Math.}\ }\textbf {\bibinfo
  {volume} {116}},\ \bibinfo {pages} {113} (\bibinfo {year}
  {1982})}\BibitemShut {NoStop}%
\bibitem [{\citenamefont {Sausset}\ and\ \citenamefont
  {Tarjus}(2007)}]{sausset2007periodic}%
  \BibitemOpen
  \bibfield  {author} {\bibinfo {author} {\bibfnamefont {F.}~\bibnamefont
  {Sausset}}\ and\ \bibinfo {author} {\bibfnamefont {G.}~\bibnamefont
  {Tarjus}},\ }\bibfield  {title} {\bibinfo {title} {{Periodic boundary
  conditions on the pseudosphere}},\ }\href
  {https://doi.org/10.1088/1751-8113/40/43/004} {\bibfield  {journal} {\bibinfo
   {journal} {J. Phys. A}\ }\textbf {\bibinfo {volume} {40}},\ \bibinfo {pages}
  {12873} (\bibinfo {year} {2007})}\BibitemShut {NoStop}%
\bibitem [{\citenamefont {de~Oliveira~Benedito}\ \emph
  {et~al.}(2016)\citenamefont {de~Oliveira~Benedito}, \citenamefont {Palazzo},\
  and\ \citenamefont {Interlando}}]{BENEDITO20161902}%
  \BibitemOpen
  \bibfield  {author} {\bibinfo {author} {\bibfnamefont {C.~W.}\ \bibnamefont
  {de~Oliveira~Benedito}}, \bibinfo {author} {\bibfnamefont {R.}~\bibnamefont
  {Palazzo}},\ and\ \bibinfo {author} {\bibfnamefont {J.~C.}\ \bibnamefont
  {Interlando}},\ }\bibfield  {title} {\bibinfo {title} {An algorithm to
  construct arithmetic fuchsian groups derived from quaternion algebras and the
  corresponding hyperbolic lattices},\ }\href
  {https://www.sciencedirect.com/science/article/pii/S0022404915002868}
  {\bibfield  {journal} {\bibinfo  {journal} {Journal of Pure and Applied
  Algebra}\ }\textbf {\bibinfo {volume} {220}},\ \bibinfo {pages} {1902}
  (\bibinfo {year} {2016})}\BibitemShut {NoStop}%
\bibitem [{\citenamefont {Strohmaier}\ and\ \citenamefont
  {Uski}(2013)}]{Strohmeier}%
  \BibitemOpen
  \bibfield  {author} {\bibinfo {author} {\bibfnamefont {A.}~\bibnamefont
  {Strohmaier}}\ and\ \bibinfo {author} {\bibfnamefont {V.}~\bibnamefont
  {Uski}},\ }\bibfield  {title} {\bibinfo {title} {{An Algorithm for the
  Computation of Eigenvalues, Spectral Zeta Functions and Zeta-Determinants on
  Hyperbolic Surfaces}},\ }\href {https://doi.org/10.1007/s00220-012-1557-1}
  {\bibfield  {journal} {\bibinfo  {journal} {Commun. Math. Phys.}\ }\textbf
  {\bibinfo {volume} {317}},\ \bibinfo {pages} {827} (\bibinfo {year}
  {2013})}\BibitemShut {NoStop}%
\bibitem [{\citenamefont {Helgason}(2004)}]{Helgason}%
  \BibitemOpen
  \bibfield  {author} {\bibinfo {author} {\bibfnamefont {S.}~\bibnamefont
  {Helgason}},\ }\bibfield  {title} {\bibinfo {title} {{Non-Euclidean
  Analysis}},\ }\href {https://arxiv.org/abs/math/0411411} {\bibfield
  {journal} {\bibinfo  {journal} {arXiv:math/0411411}\ } (\bibinfo {year}
  {2004})}\BibitemShut {NoStop}%
\bibitem [{\citenamefont {Ashcroft}\ and\ \citenamefont
  {Mermin}(1976)}]{BookAshcroft}%
  \BibitemOpen
  \bibfield  {author} {\bibinfo {author} {\bibfnamefont {N.~W.}\ \bibnamefont
  {Ashcroft}}\ and\ \bibinfo {author} {\bibfnamefont {N.~D.}\ \bibnamefont
  {Mermin}},\ }\href@noop {} {\emph {\bibinfo {title} {{Solid State
  Physics}}}}\ (\bibinfo  {publisher} {Saunders College, Philadelphia},\
  \bibinfo {year} {1976})\BibitemShut {NoStop}%
\bibitem [{\citenamefont {Voight}(2021)}]{BookQuat}%
  \BibitemOpen
  \bibfield  {author} {\bibinfo {author} {\bibfnamefont {J.}~\bibnamefont
  {Voight}},\ }\href {https://doi.org/10.1007/978-3-030-56694-4} {\emph
  {\bibinfo {title} {{Quaternion Algebras}}}},\ \bibinfo {series} {Graduate
  Texts Mathematics}, Vol.\ \bibinfo {volume} {288}\ (\bibinfo  {publisher}
  {Springer, Cham},\ \bibinfo {year} {2021})\BibitemShut {NoStop}%
\bibitem [{\citenamefont {Maciejko}()}]{JosephPrivate}%
  \BibitemOpen
  \bibfield  {author} {\bibinfo {author} {\bibfnamefont {J.}~\bibnamefont
  {Maciejko}},\ }\href@noop {} {\bibinfo  {journal} {\textit{Private
  Communication}}\ }\BibitemShut {NoStop}%
\bibitem [{\citenamefont {Ihara}(1966)}]{Ihara}%
  \BibitemOpen
\bibfield  {journal} {  }\bibfield  {author} {\bibinfo {author} {\bibfnamefont
  {Y.}~\bibnamefont {Ihara}},\ }\bibfield  {title} {\bibinfo {title} {{On
  discrete subgroups of the two by two projective linear group over p-adic
  fields}},\ }\href {http://10.2969/jmsj/01830219} {\bibfield  {journal}
  {\bibinfo  {journal} {J. Math. Soc. Japan}\ }\textbf {\bibinfo {volume}
  {18(3)}},\ \bibinfo {pages} {219} (\bibinfo {year} {1966})}\BibitemShut
  {NoStop}%
\bibitem [{\citenamefont {Bass}(1992)}]{Bass}%
  \BibitemOpen
  \bibfield  {author} {\bibinfo {author} {\bibfnamefont {H.}~\bibnamefont
  {Bass}},\ }\bibfield  {title} {\bibinfo {title} {{The Ihara-Selberg zeta
  function of a tree lattice}},\ }\href
  {https://doi.org/10.1142/S0129167X92000357} {\bibfield  {journal} {\bibinfo
  {journal} {Int. J. Math}\ }\textbf {\bibinfo {volume} {03}},\ \bibinfo
  {pages} {717} (\bibinfo {year} {1992})}\BibitemShut {NoStop}%
\bibitem [{\citenamefont {Ikeda}\ \emph
  {et~al.}(2021{\natexlab{b}})\citenamefont {Ikeda}, \citenamefont {Aoki},\
  and\ \citenamefont {Matsuki}}]{Ikeda2}%
  \BibitemOpen
  \bibfield  {author} {\bibinfo {author} {\bibfnamefont {K.}~\bibnamefont
  {Ikeda}}, \bibinfo {author} {\bibfnamefont {S.}~\bibnamefont {Aoki}},\ and\
  \bibinfo {author} {\bibfnamefont {Y.}~\bibnamefont {Matsuki}},\ }\bibfield
  {title} {\bibinfo {title} {Hyperbolic band theory under magnetic field and
  dirac cones on a higher genus surface},\ }\href
  {https://doi.org/10.1088/1361-648x/ac24c4} {\bibfield  {journal} {\bibinfo
  {journal} {J. Phys.: Condens. Matter}\ }\textbf {\bibinfo {volume} {33}},\
  \bibinfo {pages} {485602} (\bibinfo {year} {2021}{\natexlab{b}})}\BibitemShut
  {NoStop}%
\bibitem [{\citenamefont {Comtet}\ and\ \citenamefont
  {Houston}(1985)}]{Comtet1985}%
  \BibitemOpen
  \bibfield  {author} {\bibinfo {author} {\bibfnamefont {A.}~\bibnamefont
  {Comtet}}\ and\ \bibinfo {author} {\bibfnamefont {P.~J.}\ \bibnamefont
  {Houston}},\ }\bibfield  {title} {\bibinfo {title} {{Effective action on the
  hyperbolic plane in a constant external field}},\ }\href
  {https://doi.org/10.1063/1.526781} {\bibfield  {journal} {\bibinfo  {journal}
  {J. Math. Phys.}\ }\textbf {\bibinfo {volume} {26}},\ \bibinfo {pages} {185}
  (\bibinfo {year} {1985})}\BibitemShut {NoStop}%
\bibitem [{\citenamefont {Carey}\ \emph {et~al.}(1998)\citenamefont {Carey},
  \citenamefont {Hannabuss}, \citenamefont {Mathai},\ and\ \citenamefont
  {McCann}}]{Carey:1997zvh}%
  \BibitemOpen
  \bibfield  {author} {\bibinfo {author} {\bibfnamefont {A.~L.}\ \bibnamefont
  {Carey}}, \bibinfo {author} {\bibfnamefont {K.~C.}\ \bibnamefont
  {Hannabuss}}, \bibinfo {author} {\bibfnamefont {V.}~\bibnamefont {Mathai}},\
  and\ \bibinfo {author} {\bibfnamefont {P.}~\bibnamefont {McCann}},\
  }\bibfield  {title} {\bibinfo {title} {{Quantum Hall effect on the hyperbolic
  plane}},\ }\href {https://doi.org/https://doi.org/10.1007/s002200050255}
  {\bibfield  {journal} {\bibinfo  {journal} {Commun. Math. Phys.}\ }\textbf
  {\bibinfo {volume} {190}},\ \bibinfo {pages} {629} (\bibinfo {year}
  {1998})}\BibitemShut {NoStop}%
\bibitem [{\citenamefont {Ikeda}(2018)}]{Ikeda3}%
  \BibitemOpen
  \bibfield  {author} {\bibinfo {author} {\bibfnamefont {K.}~\bibnamefont
  {Ikeda}},\ }\bibfield  {title} {\bibinfo {title} {Quantum hall effect and
  langlands program},\ }\href
  {https://doi.org/https://doi.org/10.1016/j.aop.2018.08.002} {\bibfield
  {journal} {\bibinfo  {journal} {Ann. Phys.}\ }\textbf {\bibinfo {volume}
  {397}},\ \bibinfo {pages} {136} (\bibinfo {year} {2018})}\BibitemShut
  {NoStop}%
\bibitem [{\citenamefont {Kravchuk}\ \emph {et~al.}(2021)\citenamefont
  {Kravchuk}, \citenamefont {Mazac},\ and\ \citenamefont {Pal}}]{Kravchuk}%
  \BibitemOpen
  \bibfield  {author} {\bibinfo {author} {\bibfnamefont {P.}~\bibnamefont
  {Kravchuk}}, \bibinfo {author} {\bibfnamefont {D.}~\bibnamefont {Mazac}},\
  and\ \bibinfo {author} {\bibfnamefont {S.}~\bibnamefont {Pal}},\ }\bibfield
  {title} {\bibinfo {title} {{Automorphic Spectra and the Conformal
  Bootstrap}},\ }\href {https://arxiv.org/abs/2111.12716} {\bibfield  {journal}
  {\bibinfo  {journal} {arXiv:2111.12716}\ } (\bibinfo {year}
  {2021})}\BibitemShut {NoStop}%
\bibitem [{\citenamefont {Bonifacio}(2021)}]{Bonifacio}%
  \BibitemOpen
  \bibfield  {author} {\bibinfo {author} {\bibfnamefont {J.}~\bibnamefont
  {Bonifacio}},\ }\bibfield  {title} {\bibinfo {title} {{Bootstrapping Closed
  Hyperbolic Surfaces}},\ }\href {https://arxiv.org/abs/2111.13215} {\bibfield
  {journal} {\bibinfo  {journal} {arXiv:2111.13215}\ } (\bibinfo {year}
  {2021})}\BibitemShut {NoStop}%
\end{thebibliography}%

\end{document}